%% file: arxiv_quantumfriction_20151217.tex
\documentclass[11pt,reqno]{amsart}

\usepackage{amsmath}
\usepackage{amssymb}
\usepackage{amsfonts}
\usepackage{amsthm}
\usepackage{a4wide}
\usepackage{bbm}
\usepackage{graphics}
\usepackage{color}

\usepackage[applemac]{inputenc}
\usepackage[sans]{dsfont}
\usepackage{enumerate}
\usepackage{enumitem}
\usepackage{tikz}
\usepackage{float}
\usetikzlibrary{decorations.pathmorphing}

\newtheorem{theorem}{Theorem}[section]
\newtheorem{lemma}[theorem]{Lemma}
\newtheorem{proposition}[theorem]{Proposition}

\newtheorem{definition}[theorem]{Definition}

\newtheorem{remark}[theorem]{Remark}

\numberwithin{equation}{section}

\newcommand{\ii}{\mathrm{i}}
\renewcommand{\d}{\mathrm{d}}

\def\R{{\mathbb R}}

\newcommand{\rd}{\mathrm{d}}

\def\<{\langle}
\def\>{\rangle}

\newcommand{\im}{\operatorname{Im}}

\newcommand{\DETAILS}[1]{}

\title{Spectral Analysis of a Model for Quantum Friction}

\author[S. {De Bi\`{e}vre}]{Stephan {De Bi\`{e}vre}}
\address[S. {De Bi\`{e}vre}]{Univ. Lille, CNRS, UMR 8524 - Laboratoire Paul Painlevé, F-59000 Lille, France.}
\email{Stephan.De-Bievre@math.univ-lille1.fr}
\address[S. {De Bi\`{e}vre}]{Equipe-Projet MEPHYSTO,
Centre de Recherche INRIA Futurs,
Parc Scientifique de la Haute Borne, 40, avenue Halley B.P. 70478,
F-59658 Villeneuve d'Ascq cedex, France.}

\author[J. Faupin]{J{\'e}r{\'e}my Faupin}
\address[J. Faupin]{Institut Elie Cartan de Lorraine \\
Universit{\'e} de Lorraine, 
57045 Metz Cedex 1, France}
\email{jeremy.faupin@univ-lorraine.fr}

\author[B. Schubnel]{Baptiste Schubnel}
\address[B. Schubnel]{Institut Elie Cartan de Lorraine \\
Universit{\'e} de Lorraine, 
57045 Metz Cedex 1, France}
\email{baptiste.schubnel@univ-lorraine.fr}

\begin{document}

\begin{abstract}
An otherwise free classical particle moving through an extended spatially homogeneous medium with which it may exchange energy and momentum will undergo a frictional drag force in the direction opposite to its velocity with a magnitude which is typically proportional to a power of its speed.  We study here the quantum equivalent of a classical Hamiltonian model for this friction phenomenon that was proposed in \cite{BrDe02_01}.
More precisely, we study the spectral properties of the quantum Hamiltonian and compare the quantum and classical situations. Under suitable conditions on the infrared behaviour of the model, we prove that the Hamiltonian at fixed total momentum has no ground state except when the total momentum vanishes, and that its spectrum is otherwise absolutely continuous.
\end{abstract}

\maketitle

\section{Introduction}
Many classical systems (e.g. a ball in a (viscous) fluid, a classical charged particle emitting Cerenkov radiation, or an electron interacting with its own radiation field) experience a phenomenon of energy dissipation due to a drag force exerted on the system as a result of its interaction  with its environment. A typical example is the phenomenon of linear friction, in which the center of mass $q(t) \in \mathbb{R}^d$ of the system obeys the effective dynamical equation 
\begin{equation} \label{clacla}
m \ddot q(t)= - \gamma \dot q(t) - \nabla V(q(t)), 
\end{equation}
where $V\in \mathrm{C}^1(\R^d)$ is an exterior potential, $m$ is the mass of the system, and $\gamma>0$ is a phenomenological friction coefficient that finds its origin in the interaction between the system and its environment.  In a confining potential with a unique global minimum, the particle will come to rest at this minimum. If the potential is linear, $V(q)=-F\cdot q$, the particle reaches a limiting velocity which is proportional to the applied field, a phenomenon directly related to Ohm's law. If the exterior potential vanishes identically, the center of mass of such a system  will come to rest exponentially fast with rate $\gamma/m$ at some point in space. Note that in these situations, the energy lost by the particle is transferred to the environment, but the phenomenological equation above does not describe this energy transfer since it does not deal with the dynamical variables of the medium. Similar phenomena occur when the friction force is of the form $-\gamma|\dot q(t)|^k\frac{\dot q(t)}{|\dot q(t)|}$ for some $k\geq 1$. The situation is very different with radiation damping. In that case, the drag force vanishes unless the particle accelerates~\cite{KoSpo2000} and in absence of an external potential $V$ the particle will reach a non-vanishing asymptotic velocity. 

It is of interest to understand under what circumstances the interaction with an environment will lead to a (linear) friction force, as opposed to, notably, radiation dissipation. For a classical particle moving through a liquid or a gas, aspects of this question are addressed in~\cite{BuCaMa}.  In \cite{BrDe02_01}, a classical polaron-type Hamiltonian model was introduced to describe linear friction. In this model a particle is coupled to local vibrational modes of an environment modeled by independent real fields $\psi(y,x,t)$ at each point of space $x \in \mathbb{R}^d$ (here $y\in \R^3$ is a coordinate used to label the degrees of freedom of the vibration fields). We shall occasionally refer to these vibration fields as ``membranes''.
\input{modele}

The equations of motion of the particle-field system are given by 
  \begin{align} \label{eqcla}
   \partial_t^2 & \psi(y,x,t) -c^2 \Delta_y \psi(y,x,t)= -\rho_1(x) \sigma_2(y), \\ \label{eqcla2}
 m   \ddot q(t)&=   -\nabla V(q(t))- \int_{\mathbb{R}^{3+d}}  \d y \d x    \rho_{1}(x-q(t)) \sigma_2(y) (\nabla_x \psi)(y,x,t),
  \end{align}
where $\rho_1$ and $\sigma_2$ are ``form factors'' that describe the interaction between the particle and the fields, with $x \in \mathbb{R}^d$, $y \in \mathbb{R}^3$. Precise conditions on $\rho_1$ and $\sigma_2$ will be given below.  The constant $c$ is the phase velocity of the field waves. Note that the Laplacian in equation~\eqref{eqcla} only acts on the $y$ variables, implying that the vibration fields $\psi(y, x, t)$ for two different values of $x$ are \emph{not} coupled. This is a crucial feature of the model, distinguishing it from models such as the classical Nelson model or the Maxwell-Lorentz system that describe radiation damping rather than friction \cite{KoKuSp97_01, KoKuSp99_01, KoSp98_01, KoSpo2000}. We will come back to this point below.

Introducing an appropriate phase space of initial conditions (see Section~\ref{classmod} below), it was shown in~\cite{BrDe02_01} that the system \eqref{eqcla}--\eqref{eqcla2} is a well-posed Hamiltonian dynamical system. In addition, the asymptotic behavior of the solutions is studied  for various potentials $V$. It is in particular proven that, when $V=0$, the case we will address in this paper,  and provided the propagation speed $c$ is large enough, $\dot q(t)\to 0$ and $q(t)\to q_{\infty}\in\R^d$ as the time $t$ tends to infinity. In other words, the particle comes to rest asymptotically and the fields acquire a limiting configuration consisting of a Coulomb-type static field surrounding the particle and a radiation field carrying of all the momentum of the system.  This is in sharp contrast to the situation in the classical Nelson model or the Maxwell-Lorentz system, where, when $V=0$, the particle acquires a non-vanishing limiting velocity that depends on the initial condition of the system, and in which it carries along a soliton-type field configuration~\cite{KoKuSp97_01, KoKuSp99_01, KoSp98_01, KoSpo2000}.

The purpose of our work (present and future) is to analyze the quantum version of the above model with $V=0$,  which is obtained by replacing the classical fields  by quantized fields, and the classical particle by a quantum particle. 
We expect that, asymptotically in time, the quantum particle-field system will now converge to a ``ground state'' in which the expectation value of the particle momentum vanishes, plus a radiation field carrying the total momentum of the system.   While a full proof of such a result is well beyond the present work,  the spectral results obtained in this paper can be seen as a first step towards a proof of such a statement, as we explain in Section~\ref{presmod}.

The paper is organised as follows. In Section~\ref{classmod} we give some further information on the dynamical properties of the classical model~\eqref{eqcla}-\eqref{eqcla2}. In Section~\ref{presmod} we present the quantum model we study as well as our main results, summarized in Theorem~\ref{thm:main}, and discuss their interpretation. The rest of our paper is devoted to the proof of Theorem~\ref{thm:main}. \\

\noindent{\bf Acknowledgments.}  S.D.B. acknowledges the support of the Labex CEMPI (ANR-11-LABX-0007-01). The research of J.F. is supported in part by ANR grant ANR-12-JS01-0008-01. The research of B.S. is supported by ``Region Lorraine''. The authors thank the GDR Dynqua for its support.

\section{The classical model}\label{classmod}
To better understand the relevance of the quantum mechanical results obtained in this paper, it is helpful to relate them to the properties of the classical system~\eqref{eqcla}-\eqref{eqcla2} which has Hamiltonian 
 \begin{align} 
  H(q, \psi, p, \pi)&=\frac{p^2}{2m} + V(q)+\frac{1}{2} \int_{\mathbb{R}^{3+d}} \d y \d x  (c^2 \vert \nabla_y \psi(y,x) \vert^2 + \vert \pi(y,x) \vert^2) \notag \\
  & \quad + \int_{\mathbb{R}^{3+d}} \d y \d x  \rho_{1}(x-q) \sigma_2(y) \psi(y,x). \label{Hcl}
  \end{align}
In Fourier variables, this same Hamiltonian reads
\begin{align}
H(\psi,q,\pi,p)&= \frac{p^2}{2m} + V(q)+\frac{1}{2} \int_{\mathbb{R}^{3+d}}  \d k  \d\xi  (c^2  \vert k \vert^2 \vert \hat{\psi}(k,\xi) \vert^2 + \vert \hat{\pi}(k,\xi) \vert^2) \nonumber\\
&\qquad +  \int_{\mathbb{R}^{3+d}}   \d k  \d\xi   e^{\ii \xi \cdot q} \hat{\rho}_1(\vert \xi \vert) \hat{\sigma}_2(\vert k \vert)  \hat{\psi}(k,\xi).\label{HcII}
\end{align}
In what follows, we will put $V=0$, $c=1$. We consider furthermore $\rho_1\in \mathrm{C}_0^\infty(\R^d)$ and $\sigma_2$ of the form 
\begin{equation}\label{eq:hatsigma2}
\hat\sigma_2(k)=|k|^{\mu+\frac12}\hat\rho_2(k), \quad\hat \rho_2(0)\not=0,
\end{equation}
with $\mu\in\R$ and  $\rho_2\in \mathrm{L}^2(\R^3)$. It was shown in~\cite{BrDe02_01} that, provided 
$$
(1+\frac1{|k|})\hat\sigma_2\in \mathrm{L}^2(\R^3, \rd k), \quad \mu>-1,
$$ 
the system \eqref{eqcla}--\eqref{eqcla2} is well-posed in the sense of Hadamard. Namely, if we introduce $p=\dot q$, $\pi=\dot \psi$, and the finite energy space $\mathcal{E} = E \times \mathbb{R}^d \times \mathrm{L}^{2}(\mathbb{R}^{3+d}) \times \mathbb{R}^d$, where $E=\overline{(\mathrm{C}_{0}^{\infty}(\mathbb{R}^{3+d}), \| \nabla_y \cdot \|_{2})}^c$ is the completion of $\mathrm{C}_{0}^{\infty}(\mathbb{R}^{3+d})$ under the norm $\| \nabla_y \cdot \|_{2}$, and where the norm of any   $Y=(\psi,q, \pi, p) \in \mathcal{E}$ is given by  $\|Y\|=\|\nabla \psi \|_2 +|q| + \| \pi \|_2 + |p|$, then for any $Y_0 \in \mathcal{E}$, the system \eqref{eqcla}--\eqref{eqcla2} has a unique solution $Y(\cdot) \in \mathrm{C}_0(\mathbb{R},\mathcal{E})$. Moreover, the solutions depend continuously on the initial data $Y_0$, and the  Hamiltonian $H$ in~\eqref{Hcl}  is a constant of the motion.  

It is clear from the expression of the interaction Hamiltonian in~\eqref{HcII} that the exponent $\mu$ in~\eqref{eq:hatsigma2} controls the strength of the coupling of the particle to the low frequency modes of the vibration fields.  The larger $\mu$, the weaker the coupling to these modes. Note that the extra term $\frac12$  in the exponent,which simply shifts the $\mu$ axis, is notationally convenient in the quantum model. Clearly, $\mu= - \frac12$ corresponds to a critical value. Indeed, if $\mu = - \frac12$, $\sigma_2$ has a non-vanishing ``charge'' $\int_{\R^3} \sigma_2(y)\ \rd y.$ If $\mu>-\frac12$, this total charge vanishes, and the coupling to the low frequency modes of the field is cut off by the factor $|k|^{\mu+\frac12}$. If $\mu<-\frac12$, this total charge is infinite and the coupling to the low frequency modes is enhanced.

We are now in a position to discuss the effect of $\mu$ on the dynamics of the system.  For that purpose, it is instructive to first consider the situation where the particle moves through the medium at constant velocity and to explicitly compute the drag force exerted by the medium on the particle in that situation, following~\cite{BrDe02_01, Br02}.  We write $f(v)=-f_r(|v|)\frac{v}{|v|}$ for this drag force, with
$$
f_r(|v|)=(2\pi)^{\frac32}|v|\int_0^{+\infty}\rd \omega\ |\hat \sigma_2|^2(|v|\omega)\ \hat K(\omega),
$$
where
$$
\hat K(\omega)=\sqrt{2\pi}\int_{\R^{d-1}} \rd\eta \left[\widehat{\partial_1\sigma_1}(\omega, \eta)\right]^2,\quad \hat K(0)=0,\quad \hat K(\omega)\leq 0,
$$
is a smooth function of rapid decrease. Now suppose that, for some $\alpha\in\R$,
$$
C_\alpha:=\lim_{\omega\to0}\frac{|\hat \sigma_2(\omega)|^2(\omega)}{\omega^\alpha}>0.
$$
Then
$$
\lim_{v\to0} \frac{f_r(|v|)}{|v|^{1+\alpha}}=(2\pi)^{\frac32}C_\alpha\int_0^{+\infty}\rd \omega\ \omega^{\alpha}\ \hat K(\omega)=-\gamma_\alpha<0.
$$
For the integral to converge (around $\omega=0$) we need $\alpha>-1$, which is also a necessary and sufficient condition to guarantee for the drag force to vanish for $v=0$.  In the case where $\hat\sigma_2$ is given by~\eqref{eq:hatsigma2}, we find that 
$$
\alpha=2\mu+1,\quad \alpha+1=2(\mu+1).
$$
So in that case 
$$
f_r(|v|)\sim -\gamma_\alpha |v|^{2(\mu+1)}.
$$
Note that $\mu=-\frac12$ corresponds to linear friction, and hence Ohmic behaviour of the system. For higher $\mu$, the drag force is weaker at small speeds, for lower $\mu$ it is stronger, a reflection of the stronger coupling to low frequency modes. Provided the drag force is not too small, we expect the particle to slow down and come to a complete stop as $t\to+\infty$. This intuition comes from the study of the solutions to $\dot v(t)=-v(t)^k$, $v(0)=v_0>0$ for $t>0$, and with $k=2(\mu+1)$. If $ -\frac12\leq \mu$, one easily checks the solutions satisfy $v(t)\to 0,$ as $t\to+\infty$. In addition, if $-\frac12\leq \mu<0$,  $q(t)\to q_\infty\in\R$. And if $0\leq \mu$, $q(t)\to+\infty$: in these cases the particle's velocity tends so slowly to zero that it does not come to a stop at a finite distance from its starting point. In~\cite{BrDe02_01}, it was proven for the full model described by \eqref{eqcla}--\eqref{eqcla2} that, provided $\mu=-\frac12$ and some additional technical conditions are satisfied, the drag force will slow the particle down until it stops: in other words $\dot q(t)\to 0, q(t)\to q_{\infty}$. It is clear that the low velocity behaviour of the drag force is essential here and it would be of interest to investigate if the result continues to hold for other values of $\mu$. 

To put the spectral results on the quantum model obtained in this paper in perspective, it is of interest to study the ``momentum shells'' of the classical model. The total momentum of the system is
$$
P_\mathrm{tot}(Y)=p-\int_{\R^{d+3}} \rd x\rd y \ \pi(x,y)\nabla_x \psi(x,y).
$$
The translational invariance of the model guarantees it is a constant of the motion. Let us write $\Sigma_P$ for the surface 
$P_{\mathrm{tot}}(Y)=P$. It is proven in~\cite{Br02} that, for all $P\in\R^d$, and provided $-1<\mu$,
\begin{equation}\label{eq:classicalinfimum}
\inf_{Y\in \Sigma_P} H(Y)=E_0,\quad\mathrm{where} \ E_0=\min_{Y\in\mathcal E}H(Y)=\min_{Y\in \Sigma_0} H(Y).
\end{equation}
Here
$$
E_0=-g^2\frac{\|\sigma_1\|^2}2 \int_{\R^3}\rd k \frac{|\hat\sigma_2|^2(|k|)}{|k|^2} <0.
$$
In other words, the global minimum of the Hamiltonian is reached on the zero momentum shell. Moreover, the energy infimum is the same on all momentum shells, but not reached on any other one. This phenomenon is directly linked to the fact that the dispersion relation $\omega(k)=|k|$ of the vibration fields does not depend on $\xi$. In other words, to the fact that the membranes at different $x$ are not coupled. It implies that one can create vibrational states that carry arbitrary momentum, but very little energy. 

The energy minimizers of the model are explicitly given by $Y=(q, \psi_q, 0,0)\in\mathcal E$ $(-1<\mu)$ where $\psi_q$ is the static field surrounding the particle at rest at position $q$:
$$
\hat\psi_q(k, \xi)=-\exp(i\xi\cdot q)\frac{\hat\rho_1(|\xi|)\hat\sigma_2(|k|)}{\omega^2(k)}.
$$
These minimizers serve a asymptotic attractors of the dynamics. Note that they do not only all have vanishing total momentum, but also vanishing particle and field momentum. They are furthermore unique up to translational invariance.

A formal quantization of the above model can be readily performed using the usual second quantization approach. First, one introduces the normal field variables 
\begin{align*}
\alpha(k,\xi)&=\frac{1}{\sqrt{2}} \Big( \sqrt{c \vert k \vert} \hat \psi(k,\xi) + \frac{\ii}{\sqrt{\vert k\vert c}} \hat \pi(k,\xi)\Big),\\
\alpha^{*}(k,\xi)&=\overline{\alpha}(k,\xi),
\end{align*}
and one easily verifies that 
\begin{align*}
H(\alpha,q,\alpha^*,p)&= \frac{p^2}{2m} + \frac{1}{2} \int_{\mathbb{R}^{3+d}}   \d k  \d\xi  c \vert k \vert \alpha^{*}(k,\xi) \alpha(k,\xi) \\
&\qquad + \frac{1}{\sqrt{2c}} \int_{\mathbb{R}^{3+d}} \d k  \d\xi  \vert k \vert^{-\frac12} e^{\ii \xi \cdot q} \hat{\rho}_1(\vert \xi \vert) \hat{\sigma}_2(\vert k \vert)  \alpha(k,\xi)+  \mathrm{h.c.},
\end{align*}
where we have used that  $\overline{\hat \pi(k,\xi)}=\hat \pi(-\xi,-k)$, $\overline{\hat \psi(k,\xi)}=\hat \psi(-\xi,-k)$ because  the fields are real-valued.  The normal field variables satisfy the poisson bracket equations $\{ \alpha(k,\xi), \alpha^{*}(k',\xi') \}= -\ii \delta(k-k') \delta(\xi-\xi')$,  $\{ \alpha^{\sharp}(k,\xi), \alpha^{\sharp}(k',\xi') \}=0$, where the poisson bracket comes from the symplectic form 
\begin{equation*}
\sigma((\hat \psi, \hat \pi),(\hat \psi', \hat \pi')):= \int_{\mathbb{R}^{3+d}} \d k  \d\xi  [\hat \psi(k,\xi) \hat  \pi'(k,\xi)  -\hat \pi(k,\xi) \hat  \psi'(k,\xi)]. 
\end{equation*}
To formally quantize $H$, one then just makes the substitutions
\begin{equation*}
q \rightarrow Q, \qquad p \rightarrow - \ii  \nabla_q, \qquad \alpha(k,\xi) \rightarrow a(k,\xi), \qquad  \alpha^{*}(k,\xi) \rightarrow a^{*}(k,\xi),
\end{equation*} 
where  $a(k,\xi)$ et $ a^{*}(k,\xi)$  are annihilation and creation ``operators'' on the symmetric Fock space over $\mathrm{L}^{2}(\mathbb{R}^{3+d})$. To make this into a rigorous procedure, one chooses in the usual manner a new phase space for the field (different from the finite energy space $E\times \mathrm{L}^2(\R^{d+3})$ used above), given by $\mathcal D(\omega^{1/2})\times \mathcal D(\omega^{-1/2})$, on which 
$$
J=
\begin{pmatrix}
0&-\omega^{-1}\\
\omega&0
\end{pmatrix}
$$
defines a dynamics-invariant complex structure and on which the map $(\psi,\pi)\to \alpha$ is unitary \cite{DB}. The quantum Hilbert space of the quantized vibration field,  on which the quantum Hamiltonian is defined as a self-adjoint operator, is then the Fock space over this space. We note for later reference that the minima $\psi_q$ defined above belong to this space iff $\mu>-1/2$.

\begin{remark} (i) We have taken three dimensional membranes, but one could also consider membranes in $\R^n$. For $n>3$, there are fewer low frequency modes in the membranes, which is similar to taking $n=3$ and increasing $\mu$. Conversely, taking $n<3$, one increases the number of low frequency modes in the membranes, which corresponds to lowering $\mu$. The treatment of the spectral problems in this paper could be readily adapted to these situations.

(ii) We mention that the above friction phenomenon is related to Cerenkov radiation, which occurs when a particle moves through a wave field at a speed higher than the propagation speed of the waves.  Such a particle is expected to loose energy through radiation and slow down to the wave propagation speed. In the model considered here, the fact that the vibrational fields are not coupled for different $x$ means that the propagation speed in the $x$-direction vanishes identically. Hence the particle always moves faster than the waves in this model, even when going very slowly.  Furthermore, if one couples the different membranes by adding a term $-c'^2|\partial_x\psi(x,y)|^2$ in the Hamiltonian, where $c'$ could be different from $c$, one can notice that the above drag force vanishes identically as long as the particle moves slower than the wave propagation speed $c'$ in the $x$-direction. We finally note that the slowing down of classical particles due to a Cerenkov radiation type phenomenon has been investigated also in a model for a heavy tracer particle moving through a Bose-Einstein condensate in~\cite{FrGSo_12}. 
\end{remark}

\section{The quantum model}\label{presmod}
\subsection{The total Hamiltonian}
The Hilbert space associated with the physical system we consider is given by
\begin{equation*}
\mathcal{H} := \mathcal{H}_{\mathrm{p}} \otimes \mathcal{H}_{\mathrm{f}} ,
\end{equation*}
where $\mathcal{H}_{\mathrm{p}} := \mathrm{L}^2( \mathbb{R}^d, \d q )$  is the Hilbert space for the particle, and
\begin{equation*}
\mathcal{H}_{\mathrm{f}} := \Gamma_s( \mathrm{L}^2 ( \mathbb{R}^{ 3 + d } ) ) \equiv \mathbb{C} \oplus \bigoplus_{l=1}^{\infty} \otimes_s^l \mathrm{L}^2 ( \mathbb{R}^{ 3 + d } ) \simeq \mathbb{C} \oplus \bigoplus_{l=1}^{\infty} \mathrm{L}^2_s ( \mathbb{R}^{ (3 + d)l } ) ,
\end{equation*}
is the symmetric Fock space over $\mathrm{L}^2 ( \mathbb{R}^{ 3 + d } )$. Here $\otimes_s$ denotes the symmetrized tensor product and $\mathrm{L}^2_s$ denotes the subspace of $\mathrm{L}^2$ composed of totally symmetric functions.
It will often be convenient to identify $\mathcal H$ with $\mathrm{L}^2(\mathbb{R}^d, \mathcal{H}_\mathrm{f})$. 

We work in units such that the Planck constant $\hbar$ and the velocity of light $c$ are equal to $1$. In $\mathcal{H}$, we consider the Hamiltonian
\begin{eqnarray}
H &:=& H_{\mathrm p} \otimes \mathds{1}_{ \mathcal{H}_{\mathrm{f}} } + \mathds{1}_{ \mathcal{H}_{\mathrm{p}} } \otimes H_{\mathrm f} + H_I ,\quad \mathrm{with}\ H_{\mathrm p} := - \Delta_q  \ \mathrm{and}\label{eq:def_H1}\\
H_\mathrm{f} &:=& \d \Gamma( \omega( k ) ) \equiv \int_{ \mathbb{R}^{3 + d} } \omega( k ) a^*( k , \xi ) a( k , \xi ) \d k \d \xi , \quad \omega(k) := |k| ,
\label{eq:def_H2}
\end{eqnarray}
where $H_\mathrm{p}$
is the kinetic Hamiltonian for the particle (whose mass is set equal to $1/2$), with $q$ the position variable of the particle, and $H_\mathrm{f}$
is the free Hamiltonian for the field. We work in the momentum representation, with $k \in \mathbb{R}^3$ the momentum variable of the field bosons and $\xi \in \mathbb{R}^d$ 
the variable dual to $x\in\R^d$. 

The interaction term in the quantum Hamiltonian is of the form
\begin{equation*}
H_I := g \Phi( h_q ) \equiv g ( a^*( h_q ) + a( h_q ) ) = g \int_{ \mathbb{R}^{3 + d} } \big ( h_q( k , \xi ) a^*( k , \xi ) + \overline{h_q}( k , \xi ) a( k , \xi ) \big ) \d k \d \xi ,
\end{equation*}
in analogy with the classical expression above. Here $g \in \mathbb{R}$ is a coupling constant and $h_q \in \mathrm{L}^2( \mathbb{R}^{ 3 + d } )$ is a coupling function given by 
\begin{equation}\label{eq:coupling}
h_q( \xi , k ) := e^{- \ii q \cdot \xi } | k |^\mu \hat{ \rho }_1 ( \xi ) \hat{\rho}_2( k ) .
\end{equation}
In what follows, we suppose $\rho_1 \in \mathcal{S}( \mathbb{R}^d )$, where $\mathcal{S}( \mathbb{R}^d )$ denotes the Schwartz space of rapidly decreasing functions on $\mathbb{R}^d$. In addition, $\rho_2 \in \mathcal{S}( \mathbb{R}^3 )$. Moreover we suppose that $\rho_1$ and $\rho_2$ are radial,
\begin{equation*}
\rho_1( x ) = \rho_1( |x|), \quad \rho_2( y ) = \rho_2( |y| ).
\end{equation*}
We will always assume that $\mu> - 1 $; depending on the result, more restrictive constraints on $\mu$ will be imposed.   

Using standard estimates on creation and annihilation operators (see Lemma \ref{lm:appFock1}) together with the Kato-Rellich theorem, it is easy to verify that for any $g \in \mathbb{R}$, $H$ defines a self-adjoint operator with domain
\begin{equation*}
\mathcal{D}( H ) = \mathcal{D}( H_{\mathrm p} \otimes \mathds{1}_{ \mathcal{H}_{\mathrm{f}} } + \mathds{1}_{ \mathcal{H}_{\mathrm{p}} } \otimes H_{\mathrm f} ).
\end{equation*}
In the sequel, to simplify notations, we will drop the tensor products with identities, writing for instance $H_{\mathrm p} = H_{\mathrm p} \otimes \mathds{1}_{ \mathcal{H}_{\mathrm{f}} }$ and $H_{\mathrm f} = \mathds{1}_{ \mathcal{H}_{\mathrm{p}} } \otimes H_{\mathrm f}$.

Note that the Hamiltonian~\eqref{eq:def_H1}--\eqref{eq:def_H2} was studied in~\cite{Br07_01}, when a confining potential $V$ is added to the particle Hamiltonian. It is proven in~\cite{Br07_01} that the Hamiltonian then has a ground state if $\mu>-\frac12$. The confinement of the particle provides a form of compactness that is exploited in the proof. In this paper, we put $V=0$, and very different techniques are needed to establish the results we obtain. {Let us also point out that, using the argument in Appendix~\ref{app:absence}, it can be shown readily that in the model of~\cite{Br07_01} there is no ground state when $-1<\mu\leq -\frac12$. }

\subsection{Translation invariance and fiber decomposition}\label{subsec:trans}

We define the total momentum operator acting on $\mathcal{H}$ by 
\begin{equation*}
P_{ \mathrm{tot} } := - \ii \nabla_q + \d \Gamma( \xi ).
\end{equation*}
The physical system is translation invariant in the sense that $H$ commutes with each component of $P_{ \mathrm{tot} }$,
\begin{equation*} 
[ H , P_{ \mathrm{tot}, j } ] = 0, \quad j = 1 , \dots , d.
\end{equation*}
This implies that $H$ admits a direct integral decomposition,
\begin{equation}\label{eq:dirint}
\mathcal{H} = \int^\oplus_{ \mathbb{R}^d }\mathcal H(P) \d P =\int^\oplus_{ \mathbb{R}^d } \mathcal{H}_{\mathrm{f}} \d P , \quad H = \int^\oplus_{ \mathbb{R}^d } H( P ) \d P ,
\end{equation}
where for all $P \in \mathbb{R}^d$, 
\begin{equation*}
H( P ) := ( P - \d\Gamma( \xi ) )^2 + \d\Gamma( |k| ) + g \Phi( h_0 ).
\end{equation*}
More precisely, defining the operator $U : \mathcal{H} \to \int^\oplus_{ \mathbb{R}^d } \mathcal{H}_{\mathrm{f}} \d P$ by
\begin{equation*}
( U \psi ) ( P ) := \mathcal{F}_{ q \to P } \big ( e^{ \ii \d \Gamma( \xi ) \cdot q } \psi( q ) \big ) \equiv ( 2 \pi )^{-\frac{d}{2}} \int_{ \mathbb{R}^d } e^{ - \ii ( P - \d \Gamma( \xi ) ) \cdot q } \psi( q ) \d q ,
\end{equation*}
where $\mathcal{F}_{ q \to P }$ denotes the usual Fourier transform in $\mathcal H=\mathrm{L}^2( \mathbb{R}^d ; \mathcal{H}_{\mathrm{f}} )$, one verifies that $U$ is unitary and that
\begin{equation*}
U H U^* = \int^\oplus_{ \mathbb{R}^d } H( P ) \d P.
\end{equation*}
The relative bounds of Lemma \ref{lm:appFock1} together with the Kato-Rellich theorem show that for any $g \in \mathbb{R}$, $H(P)$ is a semi-bounded self-adjoint operator with domain $\mathcal{D}( H(P) ) = \mathcal{D} ( \d \Gamma( \xi )^2 + \d\Gamma( \omega(k) ) )$. Moreover any core for $\d \Gamma( \xi )^2 + \d \Gamma( \omega ( k ) )$ is a core for $H(P)$.

Our aim is to analyse the spectrum of $H(P)$ for any $P \in \mathbb{R}^d$.

\subsection{The uncoupled Hamiltonian at a fixed total momentum}

Our analysis will be perturbative. Without coupling, the spectrum of the Hamiltonian at a fixed total momentum $P \in \mathbb{R}^d$ is explicit. Let 
\begin{equation*}
H_{0}( P ) := ( P - \d\Gamma( \xi ) )^2 + \d \Gamma( |k| ) , \quad P \in \mathbb{R}^d.
\end{equation*}
It is not difficult to verify that
\begin{align*}
& \sigma( H_{0}( P ) ) = \sigma_{ \mathrm{ess} }( H_{0}( P ) ) = \sigma_{ \mathrm{ac} }( H_{0}( P ) ) = [ 0 , \infty ) , \quad \sigma_{ \mathrm{pp} }( H_{0}( P ) ) = \{ P^2 \} , \quad \sigma_{ \mathrm{sc} }( H_{0}( P ) ) = \emptyset ,
\end{align*}
for any $P \in \mathbb{R}^d$. The eigenvalue $P^2$ is simple, associated to the Fock vaccuum $\Omega \in \mathcal{H}_{\mathrm{f}}$. In particular we observe that the  infinimum of the spectrum  of $H_0(P)$, equal to $0$, is independent of $P$, and that $H(P)$ has a ground state if and only if $P = 0$.

\subsection{Main results--Discussion}

Our main results are summarized in Theorem \ref{thm:main}. We recall that we assume throughout that  $\rho_1 \in \mathcal{S}( \mathbb{R}^d )$ and that $\sigma_2$ is of the form~\eqref{eq:hatsigma2}, with $\mu\in\R$ and $\rho_2\in \mathcal{S}( \mathbb{R}^3 )$, $\hat{\rho}_2( 0 ) \neq 0$.
\begin{theorem}\label{thm:main}
The spectra of the family of Hamiltonians $H(P)$  satisfy the following properties:
\begin{itemize}
\item[(i)] Suppose that $-1<\mu$. For all $g \in \mathbb{R}$, there exists $E_g \le 0$ such that, for all $P \in \mathbb{R}^d$,
\begin{equation}
\sigma( H(P) ) = \sigma_{ \mathrm{ess} } ( H( P ) ) = [ E_g , \infty ), \label{eq:thm0}
\end{equation}
In particular, $E_g = \inf \sigma ( H(P) )$ does not depend on $P$. 
\item[(ii)] Suppose that $-1/2<\mu $. There exists $g_c = g_c( \mu ) > 0$ such that, for all $0 \le |g| \le g_c$, 
\begin{equation}
H(0) \text{ admits a unique ground state}. \label{eq:thm1}
\end{equation}
In other words,  $E_g$ is a simple eigenvalue of $H(0)$.
\item[(iii)] Suppose that $1/2<\mu$. There exists $g_c = g_c( \mu ) > 0$ such that, for all $0 \le |g| \le g_c$,
\begin{equation}
\sigma_{ \mathrm{pp} } ( H (0) ) = \{ E_g \}, \quad \sigma_{ \mathrm{ac} }( H( 0 ) ) = [ E_g , \infty ) , \quad \sigma_{ \mathrm{sc} }( H( 0 ) ) = \emptyset. \label{eq:thm3-1}
\end{equation}
Suppose in addition that $\hat{\rho}_1$ and $\hat{\rho}_2$ never vanish and let $\nu_1$, $\nu_2$ be such that $0 < \nu_1 < \nu_2$. Then there exists $g_c = g_c( \mu , \nu_1 , \nu_2) > 0$ such that, for all $0 < |g| \le g_c$ and $P \in \mathbb{R}^d$, $|P| \in (\nu_1, \nu_2)$,
\begin{equation}
\sigma_{ \mathrm{pp} } ( H (P) ) = \emptyset, \quad \sigma_{ \mathrm{ac} }( H( P ) ) = [ E_g , \infty ) , \quad \sigma_{ \mathrm{sc} }( H( P ) ) = \emptyset .  \label{eq:thm3-2}
\end{equation}
In particular, for $|P| \in (\nu_1, \nu_2)$, the unperturbed eigenvalue $P^2$ disappears as the coupling is turned on.
\end{itemize}
\end{theorem}

\input{spectre}

We make the following remarks on the statement of Theorem \ref{thm:main}. 
Part~(i) of the theorem is the quantum analog of statement~\eqref{eq:classicalinfimum} about the classical model; both hold for all $-1<\mu$. Note that it is valid for all $g\in\R$.  Part~(ii) of the theorem, which only holds for larger $\mu$, namely $-\frac12<\mu$, is completed with the following result. 
\begin{proposition}\label{prop:nogroundstate}
Suppose that $-1 < \mu \le - 1/2$ and that $\hat{\rho}_1( 0 ) \neq 0$. For all $P \in \mathbb{R}^d$ and $g \in \mathbb{R}{\setminus\{0\}}$,
\begin{equation}
H(P) \text{ does \emph{not} have a ground state}. \label{eq:thm2}
\end{equation}
\end{proposition}
\noindent This should be compared to the observation made at the end of Section~\ref{classmod}:  the classical ground states $\psi_q$ belong to the phase space $\mathcal D(\omega^{1/2})\times\mathcal D(\omega^{-1/2})$ iff $\mu>-\frac12$. This is therefore precisely the same regime under which the quantum ground state exists in Fock space.

Note that we need to impose a smallness condition on the coupling constant to prove (ii). This is due to the method that we employ, which is based on a suitable version of the spectral renormalization group (see Section \ref{subsec:strategy} below for more details).

Part~(iii) of the theorem finally gives much more detailed information on the spectrum of $H(P)$. For technical reasons, we need $\frac12<\mu$ here. Under that condition, $H(0)$ has no other point spectrum than its ground state energy, and has otherwise absolutely continuous spectrum. In addition, $H(P)$ has no eigenvalues at all when $P\not=0$, not even at the bottom of its spectrum. Again, this is analogous to the classical situation, where the Hamiltonian does not reach a minimum on the momentum shells $\Sigma_P$ when $P\not=0$. The condition $\frac12<\mu$  we impose here, which is quite strong, comes from the need to control two commutators of $H(P)$ with the conjugate operator of the Mourre theory used in the proof. Exploiting interpolation arguments, it might be possible to relax this infrared condition to $-\frac12<\mu $ at the cost of some technical complications. The proof of \eqref{eq:thm3-2}, more specifically the absence of eigenvalues for $H(P)$, $P \neq 0$, is based on second order perturbation theory in the coupling constant $g$, and in particular on Fermi's Golden Rule. As will appear in our proof, the second order term given by Fermi's Golden Rule vanishes if $|P| \to 0$ or $|P| \to \infty$. This is the reason for the restriction $|P| \in (\nu_1, \nu_2)$ with $0 < \nu_1 < \nu_2$ in \eqref{eq:thm3-2}.

The above spectral properties of the Hamiltonian $H(P)$ give an indication of the dynamical behavior of the system at total momentum $P$. Indeed, if the interaction between the field and the particle is turned off, the total energy of the system can be made arbitrary small but positive by choosing a state in which the particle has arbitrary small momentum and hence kinetic energy, whereas the field carries all the momentum $P$ of the combined system, while having arbitrary low energy.  The state in which -- on the contrary -- the particle carries all momentum and the field is in its ground state (the vacuum) corresponds to an embedded eigenvalue of $H_0(P)$, but not at all to a state of minimal energy.  If now the interaction is turned on, one expects that -- in analogy with the corresponding classical model --  starting from any initial state with total momentum $P$, the particle will slow down  as time goes to infinity, while transferring all its momentum to the field. This is indicated  by the fact that the spectrum of $H(P)$ is purely absolutely continuous and in particular has no embedded eigenvalue close to $P^2$, the unperturbed eigenvalue. A complete understanding of the asymptotic behaviour of the system at fixed total momentum would require developing a full scattering theory for the system, a task well beyond the present paper. Some dynamical information can nevertheless be obtained from our work here.

Indeed, the spectral statements \eqref{eq:thm3-1} and \eqref{eq:thm3-2} are consequences of a limiting absorption principle that will be proven in Section \ref{section:Mourre}. As another consequence of the latter, one can prove local decay properties, precisely stated in~\eqref{eq:locdecay1} and~\eqref{eq:locdecay2} below. Loosely speaking, such statements mean that, for initial states suitably localized in energy and total momentum, some energy is carried away in the $y$-direction by the field.

Let us point out that the situation for this model is very different from what happens in the Nelson model where the minimal energy state of the uncoupled system at fixed total and not too large momentum $P$ is an eigenstate of the system in which the particle carries all the momentum, while the field is in its vacuum state and carries no momentum. In that case, when the perturbation is turned on, the eigenstate survives and continues to represent a state in which the particle moves ballistically with a non-zero momentum. It is then expected that all initial states converge asymptotically to this state together with a radiation field~\cite{Fr73_01, FrGrSc04_01}. Nevertheless, Cerenkov radiation phenomena do occur in the Nelson model (as well as in the Pauli-Fierz model of non-relativistic QED) in the energy-momentum region where the particle propagates faster than the speed of light. In that region, the existence of Cerenkov radiation is expected to manifest itself through similar spectral properties as the ones we obtained here. Proving them remains however an open problem (but see \cite{DeFrPi10_01} for preliminary results in this direction).

\subsection{Strategy of the proof}\label{subsec:strategy}

The proofs of (i), (ii) and (iii) in Theorem \ref{thm:main} are independent. They rely on different mathematical tools.

To prove (i), we adapt localization techniques introduced by Derezi{\'n}ski and Gérard in the framework of scattering theory for quantum field theory models \cite{DeGe99_01}. Roughly speaking, we exploit the fact that to any state $\varphi$ of the coupled particle-field system with total momentum $P$, sufficiently localized in $x$-space, we can add one-quantum with wave function $f$ localized near infinity in $x$-space and with a momentum close to $\xi = - P$ and an energy close to $0$. Estimating various localization errors, we will then show that the state $a^*(f) \varphi$ (which can be defined in a proper sense) has an energy arbitrary close to the energy of $\varphi$ and a momentum arbitrary close to $0$. From this one can deduce that $\inf \sigma ( H(P) ) = \inf \sigma ( H(0) )$.

To prove (ii) (existence of a ground state for $H(0)$), we use a modified version of the original spectral renormalization group method of \cite{BaFrSi98_01,BaFrSi98_02} that has been introduced recently in \cite{BaFaFrSc15_01}. It is based on an iterative application of the so-called smooth Feshbach-Schur map (see Section \ref{subsec:Feshbach} for mode details) that allows one to construct an isospectral sequence of effective Hamiltonians acting on a Hilbert space with fewer and fewer degrees of freedom. At each step of the iterative procedure, a Neumann series decomposition combined with Wick ordering is performed, in order to rewrite the corresponding effective Hamiltonians as convergent series of (generalized) Wick monomials. In this respect, it is important to be able to control suitable norms of the derivatives of the Wick monomial kernels. To that end, we have developed in this paper a further substantial modification of the method of~\cite{BaFaFrSc15_01} since we need to control not only the first derivatives, but also the second derivatives of the Wick monomial kernels with respect to the variable associated with the total momentum operator $\d \Gamma( \xi )$. This is made possible thanks to the rotation invariance of the original Hamiltonian $H(0)$ and the fact that the Feshbach-Schur map preserves rotation invariance. We mention that rotation invariance in combination with the spectral renormalization group approach is also an important ingredient in the work of Hasler and Herbst \cite{HaHe11_01}. Yet both the purpose of \cite{HaHe11_01} and the way rotation invariance is used there are very different from the situation studied here.

We note that simpler methods for proving the existence of a ground state in various quantum field theory models exist. The method of~\cite{Ge20_01} and~ \cite{GrLiLo01_01} uses a compactness criterion that is not satisfied in the present model.  Another method, due to Pizzo \cite{Pi03_01}, has proven to be efficient in several different contexts but, again, it does not seem to be applicable in our setting, at least in any straightforward way.

The proof of (iii) is based on the conjugate operator theory of Mourre \cite{Mo81_01}. More precisely, we use an extension of the Mourre theory developed in \cite{GeGeMo04_01,GeGeMo04_02} that allows for a non self-adjoint conjugate operator and a commutator between the Hamiltonian and the conjugate operator that is not controllable by the Hamiltonian itself. We obtain absence of singular continuous spectrum in this manner and in addition, use the techniques of \cite{FaMoSk11_01} to show that the unperturbed eigenvalue of $H(P)$ at $P^2$ is unstable because Fermi's Golden Rule holds.

To prove Proposition~\ref{prop:nogroundstate} we adapt a simple argument due to Derezi{\'n}ski and Gérard \cite{DeGe04_01} (see Appendix \ref{app:absence}).

\section{Proof of $(\mathrm{i})$: Localization of the spectrum of $H(P)$}\label{sec:ess_spectrum}
In this section we prove \eqref{eq:thm0}. We use a partition of unity in Fock space as introduced in \cite{DeGe99_01}. Here we define it as follows. Let $j_0 \in \mathrm{C}_0^\infty( \mathbb{R} ; [ 0 , 1 ] )$, $j_0 \equiv 1$ on $[0,1]$, and $j_\infty \in \mathrm{C}^\infty( \mathbb{R} ; [ 0 , 1 ] )$ be defined by the relation $j_0^2 + j_\infty^2 = 1$. Let $x = \ii \nabla_\xi$. We define the (bounded) operators $\mathbf{j}_0$ and $\mathbf{j}_\infty$ in $\mathrm{L}^2( \mathbb{R}^{3+d} {,\rd k\rd \xi} )$ by $\mathbf{j}_\# := j_\#( | x |/ R )$, where $\#$ stands for $0$ or $\infty$. Here $R > 1$ is a (large) parameter. {The operators   $\mathbf{j}_\#$ should be thought of as operating in the one-boson sector of the quantized field: they approximately localize the field excitations into the corresponding regions. We will need their second quantized versions, that we now construct. For that purpose,} we introduce
\begin{equation*}
\mathbf{j} := \mathbf{j}_0 \oplus \mathbf{j}_\infty : \mathrm{L}^2( \mathbb{R}^{3+d} {,\rd k\rd \xi} ) \to \mathrm{L}^2( \mathbb{R}^{3+d} {,\rd k\rd \xi}) \oplus \mathrm{L}^2( \mathbb{R}^{3+d} {,\rd k\rd \xi} ),
\end{equation*}
which is easily seen to be an isometry. This induces an operator
\begin{equation*}
\Gamma( \mathbf{j} ) : \mathcal{H}_{\mathrm{f}} = \Gamma_s( \mathrm{L}^2( \mathbb{R}^{3+d} {,\rd k\rd \xi} ) ) \to \Gamma_s( \mathrm{L}^2( \mathbb{R}^{3+d} {,\rd k\rd \xi} ) \oplus \mathrm{L}^2( \mathbb{R}^{3+d} {,\rd k\rd \xi}) ),
\end{equation*}
where we recall that for any operator $b$ from a Hilbert space $\mathfrak{h}_1$ to another Hilbert space $\mathfrak{h}_2$, the second quantization of $b$, $\Gamma( b ) : \Gamma_s( \mathfrak{h}_1 ) \to \Gamma_s( \mathfrak{h}_2 )$, is defined by
\begin{equation*}
\Gamma ( b ) \Omega = \Omega , \quad \Gamma( b ) |_{ \otimes_s^l \mathfrak{h}_1 } := b \otimes \cdots \otimes b , \ l \ge 1.
\end{equation*}
The operator
\begin{equation*}
U : \Gamma_s( \mathrm{L}^2( \mathbb{R}^{3+d} {,\rd k\rd \xi} ) \oplus \mathrm{L}^2( \mathbb{R}^{3+d} {,\rd k\rd \xi}) ) \to  \mathcal{H}_{\mathrm{f}} \otimes \mathcal{H}_{\mathrm{f}} ,
\end{equation*}
defined by $U \Omega = \Omega \otimes \Omega$ and $U a^\sharp ( f \oplus g ) U^*= a^\sharp ( f ) \otimes \mathds{1} + \mathds{1} \otimes a^\sharp( g )$ (where $a^\sharp$ stands for $a$ or $a^*$) is unitary.  Composing with $\Gamma( \mathbf{j} )$, we arrive at the the following definition
\begin{equation*}
\check{\Gamma}( \mathbf{j} ) := U \Gamma( \mathbf{j} ) : \mathcal{H}_{\mathrm{f}} \to \mathcal{H}_{\mathrm{f}} \otimes \mathcal{H}_{\mathrm{f}}.
\end{equation*}
It is not difficult to verify that $\check{\Gamma}( \mathbf{j} )$ is an isometry.

Using standard properties of the operators $\check{\Gamma}( \mathbf{j} )$ we obtain the decomposition stated in the following lemma. The proof is deferred to Appendix \ref{app:Fock}.
\begin{lemma}\label{lm:decomp_in_Fock}
Let $\mu > - 1$, $g \in \mathbb{R}$, $P \in \mathbb{R}^d$. Then 
\begin{align*}
H( P ) &= \big ( \check{\Gamma}( \mathbf{j} )^* \big ( - \d\Gamma( \xi ) \otimes \mathds{1} + \mathds{1} \otimes ( P -  \d\Gamma( \xi ) ) \big ) \check{\Gamma}( \mathbf{j} ) \big )^2 \notag \\
&  + \check{\Gamma}( \mathbf{j} )^* \big ( \d\Gamma( |k| ) \otimes \mathds{1} + \mathds{1} \otimes \d\Gamma( |k| )  \big ) \check{\Gamma}( \mathbf{j} ) + g \check{\Gamma}( \mathbf{j} )^* \big ( \Phi( \mathbf{j}_0 h_0 ) \otimes \mathds{1} + \mathds{1} \otimes \Phi( \mathbf{j}_\infty h_0 ) \big ) \check{\Gamma}( \mathbf{j} ) . 
\end{align*}
\end{lemma}
Now we proceed to the proof of \eqref{eq:thm0}.
\begin{theorem}\label{prop:loc_spectrum}
Let $\mu > -1$ and $g \in \mathbb{R}$. There exists $E_g \le 0$ such that
\begin{equation*}
\sigma( H(P) ) = [ E_g , \infty ) ,
\end{equation*}
for all $P \in \mathbb{R}^d$.
\end{theorem}
\begin{proof}
Let $E_g(P) := \inf \sigma ( H(P) )$. We prove that $E_g(P) = E_g(0)$ for all $P \in \mathbb{R}^d$. 
Let $\varepsilon > 0$, and let $\varphi_\varepsilon \in \mathcal{D}( H( 0 ) )  \subset \mathcal{H}_{\mathrm{f}}$, $\| \varphi_\varepsilon \| = 1$ be such that
\begin{equation}
\big \|( H( 0 ) - E_g(0) ) \varphi_\varepsilon \big\| \le  \varepsilon. \label{eq:a3}
\end{equation}
Let $\mathcal{C} := \mathcal{D} ( \xi^2 ) \cap \mathcal{D}( |x| ) \cap \mathcal{D}( |x|^{-1} ) \subset \mathrm{L}^2( \mathbb{R}^{3+d} , \d k \d \xi )$. Since one can verify that
\begin{equation*}
\Gamma_{ \mathrm{fin} }( \mathcal{C} ) := \big \{ ( \varphi^{(0)} , \varphi^{(1)} , \cdots ) \in \Gamma_s( \mathcal{C} ) , \, \varphi^{(n)} = 0 \text{ for all but finitely many } n \big \} 
\end{equation*}
is a core for $H_{0}(0)$, and hence also for $H(0)$, we can assume without loss of generality that $\varphi_\varepsilon$ in \eqref{eq:a3} is such that $\varphi_\varepsilon \in \Gamma_{ \mathrm{fin} }( \mathcal{C} )$. The latter insures that $\varphi_\varepsilon \in \mathcal{D}( \d \Gamma( \xi )^2 ) \cap \mathcal{D}( \d \Gamma( |x| ) ) \cap \mathcal{D}( \d \Gamma( |x|^{-1} ) ) \cap \mathcal{D}( N )$, which we will need in the proof. Here $N$ is the number operator on Fock space.

Let $\tilde{j}_0 \in \mathrm{C}_0^\infty( \mathbb{R} ; [ 0 , 1 ] )$ be such that $\mathrm{supp} ( \tilde{j}_0 ) \subset [0,1]$ and $\tilde j_0 \equiv 1$ on $[0,1/2]$. As above the notation $\tilde{\mathbf{j}}_0$ stands for the bounded operator $\tilde{j}_0( | x |/ R )$. We claim that
\begin{equation}
\big \|( H( 0 ) - E_g(0) ) \Gamma( \tilde{\mathbf{j}}_0 ) \varphi_\varepsilon \big\| \le \varepsilon + \frac{ C }{ R } \| ( \d\Gamma( \xi )^2 + N + \d \Gamma( |x|^{-1} ) + \mathds{1} ) \varphi_\varepsilon \| , \label{eq:a4}
\end{equation}
where $C$ is a positive constant. To prove \eqref{eq:a4}, we commute $\Gamma( \tilde{\mathbf{j}}_0 )$ through $H(0)$, using Appendix \ref{app:Fock}. Since $\Gamma( \tilde{\mathbf{j}}_0 )$ is a contraction, it follows from \eqref{eq:a3} and Lemma \ref{lm:appFock3} that
\begin{align*}
\big \|( H( 0 ) - E_g(0) ) \Gamma( \tilde{\mathbf{j}}_0 ) \varphi_\varepsilon \big\| &\le \varepsilon + \big \| [ H( 0 ) , \Gamma( \tilde{\mathbf{j}}_0 ) ] \varphi_\varepsilon \big \| \notag \\
&\le \varepsilon + \frac{ C }{ R } \| ( \d\Gamma( \xi )^2 + N + \d \Gamma( |x|^{-1} ) + \mathds{1} ) \varphi_\varepsilon \| .  
\end{align*}
Now let $\tilde \delta_{ \xi = P } \in \mathrm{L}^2 ( \mathbb{R}^d , \d \xi )$ be an approximation of the delta function $\delta( \xi - P )$ given by $\tilde \delta_{ \xi = P } := \mathcal{F}_{ x \to \xi } ( e^{ i P \cdot x } R^{-d/2} \chi( | x |/ R ) )$ where $\chi \in \mathrm{C}_0^\infty( \mathbb{R} )$, $\| \chi \|_{ \mathrm{L}^2 } = 1$ and $\mathrm{supp}( \chi ) \cap \mathrm{supp}( j_0 ) = \emptyset$.  We define similarly $\tilde \delta_{ k = 0 } := R^{3/2} \eta( R |k| ) \in \mathrm{L}^2( \mathbb{R}^3 , \d k )$, with $\mathrm{supp} ( \eta ) \subset [0,1] $ and $\| \eta \|_{ \mathrm{L}^2} = 1$.  Letting $\tilde \delta_\infty := a^*( \tilde \delta_{ \xi = P } \tilde \delta_{ k = 0 } ) \Omega$, one verifies using the previous definitions that
\begin{equation}
\big \| ( P - \d \Gamma( \xi ) ) \tilde \delta_\infty \big \| \lesssim R^{-1}, \quad \big \| \d \Gamma( \omega( k ) ) \tilde \delta_\infty \| \le R^{-1} . \label{eq:a7}
\end{equation}
{In other words, $\tilde \delta_\infty$ is a one-boson state of the field, which, viewed as an element of $\mathcal{H}(P)=\mathcal H_{\mathrm f}$ has very small particle kinetic energy, as well as field energy, provided $R$ is large. In addition, it lives in an annular region far from the origin in $x$-space.} 
By construction, since all the excitations in the state $\Gamma( \tilde{\mathbf{j}}_0 ) \varphi_\varepsilon$ are localized in the region where $\mathbf{j}_0 \equiv \mathds{1}$ and since those in the state $\tilde \delta_\infty$ are localized in the region where $\mathbf{j}_\infty \equiv \mathds{1}$, we have that $( \Gamma( \tilde{\mathbf{j}}_0 ) \varphi_\varepsilon ) \otimes \tilde \delta_\infty \in \mathrm{Ran}( \check{ \Gamma }( \mathbf{j} ) )$. Therefore, applying Lemma \ref{lm:decomp_in_Fock}, we deduce that
\begin{align}
E_g( P ) \| \Gamma( \tilde{\mathbf{j}}_0 ) \varphi_\varepsilon  \|^2 &\le \big \langle \check{ \Gamma }( \mathbf{j} )^* ( ( \Gamma( \tilde{\mathbf{j}}_0 ) \varphi_\varepsilon ) \otimes \tilde \delta_\infty ) , H ( P ) \check{ \Gamma }( \mathbf{j} )^* ( ( \Gamma( \tilde{\mathbf{j}}_0 ) \varphi_\varepsilon ) \otimes \tilde \delta_\infty ) \big \rangle \notag \\
&= \big \langle \check{ \Gamma }( \mathbf{j} )^* ( ( \Gamma( \tilde{\mathbf{j}}_0 ) \varphi_\varepsilon ) \otimes \tilde \delta_\infty ) , \big ( \check{\Gamma}( \mathbf{j} )^* \big ( - \d\Gamma( \xi ) \otimes \mathds{1} + \mathds{1} \otimes ( P -  \d\Gamma( \xi ) ) \big ) \check{\Gamma}( \mathbf{j} ) \big )^2 \notag \\
&\qquad \phantom{ = \langle \check{ \Gamma }( \mathbf{j} )^* ( ( \Gamma( j_0 ) \varphi_\varepsilon ) \otimes \tilde \delta_\infty ) } \check{ \Gamma }( \mathbf{j} )^* ( ( \Gamma( \tilde{\mathbf{j}}_0 ) \varphi_\varepsilon ) \otimes \tilde \delta_\infty ) \big \rangle \notag \\
&  + \big \langle ( \Gamma( \tilde{\mathbf{j}}_0 ) \varphi_\varepsilon ) \otimes \tilde \delta_\infty , \big ( \d\Gamma( |k| ) \otimes \mathds{1} + \mathds{1} \otimes \d\Gamma( |k| )  \big ) ( ( \Gamma( \tilde{\mathbf{j}}_0 ) \varphi_\varepsilon ) \otimes \tilde \delta_\infty ) \big \rangle \notag \\
&  + g \big \langle ( \Gamma( \tilde{\mathbf{j}}_0 ) \varphi_\varepsilon ) \otimes \tilde \delta_\infty , \big ( \Phi( \mathbf{j}_0 h_0 ) \otimes \mathds{1} + \mathds{1} \otimes \Phi( \mathbf{j}_\infty h_0 ) \big ) ( ( \Gamma( \tilde{\mathbf{j}}_0 ) \varphi_\varepsilon ) \otimes \tilde \delta_\infty ) \big \rangle , \label{eq:a11}
\end{align}
where we used that $\check{ \Gamma }( \mathbf{j} ) \check{ \Gamma }( \mathbf{j} )^*$ is the projection onto the closure of $\mathrm{Ran} ( \check{ \Gamma }( \mathbf{j} ) )$. Likewise, since all the excitations in the state $\d \Gamma( \xi ) \Gamma( \tilde{\mathbf{j}}_0 ) \varphi_\varepsilon$ remain localized in the region where $\mathbf{j}_0 \equiv \mathds{1}$ and since those in $\d \Gamma( \xi ) \tilde \delta_\infty$ remain localized in the region where $\mathbf{j}_\infty \equiv \mathds{1}$, we have as before that $( - \d\Gamma( \xi ) \otimes \mathds{1} + \mathds{1} \otimes ( P -  \d\Gamma( \xi ) ) ) ( \Gamma( \tilde{\mathbf{j}}_0 ) \varphi_\varepsilon ) \otimes \tilde \delta_\infty \in \mathrm{Ran}( \check{ \Gamma }( \mathbf{j} ) )$. Therefore the first term in the right-hand side of the previous equality reduces to
\begin{align*}
& \big \langle \check{ \Gamma }( \mathbf{j} )^* ( ( \Gamma( \tilde{\mathbf{j}}_0 ) \varphi_\varepsilon ) \otimes \tilde \delta_\infty ) , \big ( \check{\Gamma}( \mathbf{j} )^* \big ( - \d\Gamma( \xi ) \otimes \mathds{1} + \mathds{1} \otimes ( P -  \d\Gamma( \xi ) ) \big ) \check{\Gamma}( \mathbf{j} ) \big )^2  \check{ \Gamma }( \mathbf{j} )^* ( ( \Gamma( \tilde{\mathbf{j}}_0 ) \varphi_\varepsilon ) \otimes \tilde \delta_\infty ) \big \rangle \notag \\
& = \big \langle ( \Gamma( \tilde{\mathbf{j}}_0 ) \varphi_\varepsilon ) \otimes \tilde \delta_\infty , \big ( - \d\Gamma( \xi ) \otimes \mathds{1} + \mathds{1} \otimes ( P -  \d\Gamma( \xi ) ) \big )^2 ( ( \Gamma( \tilde{\mathbf{j}}_0 ) \varphi_\varepsilon ) \otimes \tilde \delta_\infty ) \big \rangle.
\end{align*}
Introducing this into \eqref{eq:a11} and reorganizing terms, we arrive at
\begin{align}
E_g( P )  \| \Gamma( \tilde{\mathbf{j}}_0 ) \varphi_\varepsilon  \|^2 &\le \langle \Gamma( \tilde{\mathbf{j}}_0 ) \varphi_\varepsilon , H( 0 ) \Gamma( \tilde{\mathbf{j}}_0 ) \varphi_\varepsilon \rangle - 2 \langle \Gamma( \tilde{\mathbf{j}}_0 ) \varphi_\varepsilon , \d \Gamma( \xi ) \Gamma( \tilde{\mathbf{j}}_0 ) \varphi_\varepsilon \rangle \cdot \langle \tilde \delta_\infty , ( P - \d \Gamma( \xi ) ) \tilde \delta_\infty \rangle  \notag \\
& + \| \Gamma( \tilde{\mathbf{j}}_0 ) \varphi_\varepsilon \|^2 \| ( P - \d \Gamma( \xi ) ) \tilde \delta_\infty \|^2 + \| \Gamma( \tilde{\mathbf{j}}_0 ) \varphi_\varepsilon \|^2 \langle \tilde \delta_\infty , \d \Gamma( |k| ) \tilde \delta_\infty \rangle \notag \\
& + g \langle \Gamma( \tilde{\mathbf{j}}_0 ) \varphi_\varepsilon , \Phi( (\mathbf{j}_0 - \mathds{1} ) h_0 ) \Gamma( \tilde{\mathbf{j}}_0 ) \varphi_\varepsilon \rangle+ g \big \| \Gamma( \tilde{\mathbf{j}}_0 ) \varphi_\varepsilon \big \|^2 \langle \tilde \delta_\infty , \Phi( \mathbf{j}_\infty h_0 ) \tilde \delta_\infty \rangle. \label{eq:a6}
\end{align}
The first term on the right-hand side of the previous equation is controlled as follows: From \eqref{eq:a4} and the fact that $\Gamma( \tilde{\mathbf{j}}_0 )$ is a contraction, we obtain
\begin{align*}
\langle \Gamma( \tilde{\mathbf{j}}_0 ) \varphi_\varepsilon , H( 0 ) \Gamma( \tilde{\mathbf{j}}_0 ) \varphi_\varepsilon \rangle \le E_g( 0 ) \| \Gamma( \tilde{\mathbf{j}}_0 ) \varphi_\varepsilon \|^2 + \varepsilon + \frac{ C }{ R } \| ( \d\Gamma( \xi )^2 + N + \d \Gamma( |x|^{-1} ) + \mathds{1} ) \varphi_\varepsilon \|.
\end{align*}
Moreover, since $\varphi_\varepsilon \in \mathcal{D}( \d \Gamma ( |x| ) )$, we can write
\begin{align*}
\| ( \mathds{1} - \Gamma( \tilde{\mathbf{j}}_0 ) ) \varphi_\varepsilon \| \le \| \d \Gamma( \mathds{1} - \tilde{\mathbf{j}}_0 ) \varphi_\varepsilon \| \le \frac{C}{R} \| \d \Gamma( |x| ) \varphi_\varepsilon \| ,
\end{align*}
and hence, since $E_g(0)\leq 0$ (because $\langle \Omega , H_g ( 0 ) \Omega \rangle = 0$ with $\Omega$ the Fock vacuum), we deduce that
\begin{align*}
& \langle \Gamma( \tilde{\mathbf{j}}_0 ) \varphi_\varepsilon , H( 0 ) \Gamma( \tilde{\mathbf{j}}_0 ) \varphi_\varepsilon \rangle \notag \\
&\le E_g( 0 ) (1- \frac{ C }{ R } \| \d \Gamma( |x| ) \varphi_\varepsilon \|)^2+ \varepsilon + \frac{ C }{ R } \left( \| ( \d\Gamma( \xi )^2 + N + \d \Gamma( |x|^{-1} ) + \mathds{1} ) \varphi_\varepsilon \| \right). 
\end{align*}
To estimate the second term in the right-hand side of \eqref{eq:a6}, we use again Lemma \ref{lm:appFock3}, which shows that
\begin{align*}
 \big |\langle \Gamma( \tilde{\mathbf{j}}_0 ) \varphi_\varepsilon , \d \Gamma( \xi_j ) \Gamma( \tilde{\mathbf{j}}_0 ) \varphi_\varepsilon \rangle \big | &\le \| \d \Gamma ( \xi_j ) \varphi_\varepsilon \| + \| [ \d \Gamma( \xi_j ) , \Gamma( \tilde{\mathbf{j}}_0 ) ] \varphi_\varepsilon \| \\
& \le \| \d \Gamma ( \xi_j ) \varphi_\varepsilon \| + \frac{C}{R} \| ( N + \mathds{1} ) \varphi_\varepsilon \| ,
\end{align*}
for $j = 1 , \dots , d$. Together with \eqref{eq:a7}, this gives
\begin{align*}
&\big |\langle \Gamma( \tilde{\mathbf{j}}_0 ) \varphi_\varepsilon , \d \Gamma( \xi ) \Gamma( \tilde{\mathbf{j}}_0 ) \varphi_\varepsilon \rangle \langle \tilde \delta_\infty , ( P - \d \Gamma( \xi ) ) \tilde \delta_\infty \rangle \big |  \le \frac{C}{R} \big ( \sum_j \| \d \Gamma ( \xi_j ) \varphi_\varepsilon \| + \frac{1}{R} \| ( N + \mathds{1} ) \varphi_\varepsilon \| \big ). 
\end{align*}
The third and fourth term in the right-hand side of \eqref{eq:a6} are directly estimated by \eqref{eq:a7}. More precisely, 
\begin{align}
\| \Gamma( \tilde{\mathbf{j}}_0 ) \varphi_\varepsilon \|^2 \| ( P - \d \Gamma( \xi ) ) \tilde \delta_\infty \|^2 \le \frac{ C }{ R^2 } \quad \text{and}\quad \| \Gamma( \tilde{\mathbf{j}}_0 ) \varphi_\varepsilon \|^2 \langle \tilde \delta_\infty , \d \Gamma( |k| ) \tilde \delta_\infty \rangle \le R^{-1} .\label{eq:a13}
\end{align}
Finally, since $\mathrm{supp} ( j_0 - \mathds{1} ) \subset [ 1 , \infty )$ and $\mathrm{supp}( j_\infty ) \subset [ 1 , \infty )$, the expression \eqref{eq:coupling} of the coupling function $h_0$ shows that
\begin{equation}
\| (\mathbf{j}_0 - \mathds{1} ) h_0 \|_{ \mathrm{L}^2 } \le \frac{ C_n }{ R^n } , \quad \| \mathbf{j}_\infty h_0 \|_{ \mathrm{L}^2 } \le \frac{ C_n }{ R^n } , \quad n \in \mathbb{N}. \label{eq:a18}
\end{equation}
Together with Lemma \ref{lm:appFock1} and the fact that $N$ commutes with $\Gamma( \tilde{\mathbf{j}}_0 )$, this yields
\begin{align}
& \big | \langle \Gamma( \tilde{\mathbf{j}}_0 ) \varphi_\varepsilon , \Phi( (\mathbf{j}_0 - \mathds{1} ) h_0 ) \Gamma( \tilde{\mathbf{j}}_0 ) \varphi_\varepsilon \rangle \big |\le \frac{C}{R} \| ( N^{\frac12} + \mathds{1} ) \varphi_\varepsilon\| \label{eq:a15}
\end{align}
and
\begin{align}
 \big \| \Gamma( \tilde{\mathbf{j}}_0 ) \varphi_\varepsilon \big \|^2 \big |\langle \tilde \delta_\infty , \Phi( \mathbf{j}_\infty h_0 ) \tilde \delta_\infty \rangle \big | \le \frac{C}{R}. \label{eq:a16}
\end{align}

Putting together \eqref{eq:a6}--\eqref{eq:a16}, we obtain
\begin{align*}
 E_g( P ) (1- \frac{ C }{ R } \| \d \Gamma( |x| ) \varphi_\varepsilon \|)^2 & \le E_g(0)  (1- \frac{ C }{ R } \| \d \Gamma( |x| ) \varphi_\varepsilon \|)^2+ \varepsilon \\
 &+ \frac{ C }{ R } \big ( 1 + \| ( \d\Gamma( \xi )^2 + N + \d \Gamma( |x|^{-1} ) + \mathds{1} ) \varphi_\varepsilon \| \big ).
\end{align*}
Letting $R \to \infty$, next $\varepsilon \to 0$, we conclude that $E_g(P) \le E_g(0)$. The proof that $E_g(0) \le E_g(P)$ is analogous.  

Denoting $E_g := E_g(0) = E_g(P)$, for any $P \in \mathbb{R}^d$, we have proven that $\sigma ( H(P) ) \subset [ E_g , \infty )$. In order to show that $[ E_g , \infty ) \subset \sigma_{ \mathrm{ess} }( H( P ) )$, it suffices to construct a Weyl sequence associated to $\lambda$, for any $\lambda \in [ E_g , \infty )$. The construction being standard (see e.g. \cite{DeGe99_01}), the details are left to the reader.
\end{proof}

\section{Proof of $\mathrm{ii)}$: Existence of a ground state for $P=0$} \label{S4}
\subsection{Preliminaries}\label{subsec:Feshbach}
\noindent We now look at the particular case where $P=0$. Then
\begin{equation} \label{H(0)}
H( 0 )=  \d\Gamma( \xi )^2 + \d \Gamma( |k| ) + g \Phi( h_0 ).
\end{equation}
We prove that $H(0)$ has a ground state if $\vert g \vert$ is small enough  and if the coupling is infrared regular. To do so, we use a variant of the spectral renormalisation group method as developed in \cite{BaFaFrSc15_01}. The central result of this section is: 
\begin{theorem}\label{thm:GS}
Let $\mu > - 1/2$. There is $g_c>0$ such that for all $\vert g \vert \leq  g_c$, $H(0)$ has a ground state.
\end{theorem}
The main ingredient of our proof is the smooth Feshbach-Schur map.  If $z$ belongs to a well-chosen open complex set, the Feshbach-Schur map maps $H(0)-z$ to a new operator,  $F_{\chi}(H(0)-z,\d\Gamma( \xi )^2 + \d \Gamma( |k| )-z)$, which acts on a subspace of $\mathcal{H}_{\mathrm{f}}$ with less degrees of freedom. By restricting further the  possible values of $z$ at each step, it is possible to iterate this transformation infinitely many times (see Section~\ref{sec:iteration}). The eigenvectors with eigenvalue zero, if any,  of the limiting operator are then easy to study. They can be used to reconstruct the original eigenvectors of the operator $H(0)$.  Even if this technique is conceptually simple, it necessitates a long exposition if one carries out   all details. We will concentrate here on presenting the key steps, emphasizing the new ingredients needed in order to adapt the general construction to the model under study (see Section \ref{subsec:strategy}). We will present some additional details in Appendix \ref{extras} (or refer to the literature if they can be inferred directly from existing works).  

The Feshbach-Schur map is constructed from the so-called ``Feshbach-Schur'' pairs.

\begin{definition} \label{def:fesh} 
Let $ \chi $ be a positive operator on a complex, separable Hilbert space $ \mathcal{H} $,  with $0 \leq  \chi   \leq 1$. 
Assume that $ \chi $ and $\overline  \chi  : = \sqrt{1 -  \chi ^2}$ are both non-zero. Let $H$ and $T$ be two closed operators on $\mathcal{H}$
with domains $\mathcal{D}(H) = \mathcal{D}(T)$. Assume that $ \chi $ and $\overline  \chi $ commute with $T$. Let $W : = H - T$ and define 
$
H_{ \chi } : = T +  \chi W  \chi $, $  H_{\overline  \chi } : = T + \overline{  \chi  } W \overline{ \chi }.
$
 The pair $(H, T)$ is called a Feshbach-Schur pair associated with $ \chi $ if
\begin{itemize}
\item[(i)] $ H_{\overline  \chi } , T : \mathcal{D}( T ) \cap \mathrm{Ran} ( \overline  \chi ) \to \mathrm{Ran}( \bar \chi )$ are bijections with bounded inverses,
\item[(ii)]
$ H_{\overline  \chi }^{-1} \overline  \chi  W  \chi   $
extends to a bounded operator on $\mathcal{H}$.
\end{itemize}
If $(H,T)$ is a Feshbach-Schur pair associated with $ \chi $, the smooth Feshbach-Schur map is defined by 
\begin{equation}\label{eq:def-fesh}
F_{ \chi }(\cdot,T): H \mapsto F_ \chi (H, T) : =  T +  \chi W \chi  -  \chi  W\overline  \chi  H_{\overline  \chi }^{-1}\overline  \chi  W  \chi .
\end{equation}
 \end{definition}

The following result is taken from \cite{BaChFrSi07_01,GrHa08}.

\begin{theorem} \label{thm:fesh}
Let $0 \leq \chi \leq 1$, and let $(H, T)$ be a Feshbach-Schur pair  associated with $  \chi $.  Let $V$ be a closed subspace of $\mathcal{H}$ such that $\mathrm{Ran}( \chi ) \subset V$. Suppose in addition that 
$T: \mathcal{D}(T) \cap V \rightarrow V$ and $\overline{ \chi}T^{-1} \overline{ \chi} (V) \subset V$. Let $ Q_\chi(H, T) : =  \chi - \overline  \chi H_{\overline  \chi}^{-1}\overline  \chi W  \chi$.Then the following holds true:  
\begin{itemize}
\item[(i)] $H : \mathcal{D}( H ) \to \mathcal{H}$ is a bijection with bounded inverse iff $ F_ \chi(H, T): \mathcal{D}(T) \cap V \rightarrow V $ is a bijection with bounded inverse.
\item[(ii)] $H$ is injective iff $ F_ \chi(H, T) $ is injective. More precisely, 
\begin{align*}
& H \psi = 0, \, \psi \in \mathcal{D}( T ) ,  \, \psi \neq 0 \Longrightarrow  F_ \chi(H, T)  \chi \psi = 0,\,  \chi \psi \neq 0, \\
& F_ \chi(H, T) \varphi = 0,\, \varphi \in \mathcal{D}( T ) \cap V , \varphi \neq 0 \Longrightarrow  H Q_ \chi(H, T)\varphi = 0,\, Q_ \chi(H, T)\varphi \neq 0.
\end{align*}
\end{itemize}
\end{theorem}
Theorem  \ref{thm:fesh} describes what is often presented as the ``isospectral" property of the Feshbach-Schur map. Note that the spectra of $H$ and of $ F_\chi(H, T) $ are however not identical.  We remark that if  $T : \mathcal{D}( T ) \cap \mathrm{Ran} ( \overline  \chi ) \to \mathrm{Ran}( \bar \chi )$ is a bijection with bounded inverse and if $\| T^{-1} \overline{\chi} W \overline{\chi} \| < 1$, $\| \overline{\chi} W T^{-1} \overline{\chi} \| < 1$ and $\|T^{-1} \overline{\chi} W \chi \|< \infty$, then $(H,T)$ is a Feshbach-Schur pair associated with $\chi$ (see \cite{GrHa08}). We will use this criterion below (see the proof of Lemma \ref{lm:fesh-pair}).
  
 We now concentrate on operators $H,T$ and $\chi$ acting on $\mathcal{H}_{\mathrm{f}}$. Note that the Feshbach-Schur map preserves  unitary invariance in the sense of the following Lemma.
\begin{lemma} \label{eq:rotinv}
Let $u : \mathrm{L}^{2}(\mathbb{R}^{3+d}) \rightarrow  \mathrm{L}^{2}(\mathbb{R}^{3+d})$ be a unitary operator and let $\mathcal{U} = \Gamma( u )$ be the lifting of $u$ to $\mathcal{H}_{\mathrm{f}}$. Let $(H,T)$ be a Feshbach-Schur pair of operators in $\mathcal{H}_{\mathrm{f}}$ associated to $\chi$. We assume that $ \mathcal{U}^* \chi \mathcal{U} = \chi$,  $ \mathcal{U}^* T \mathcal{U}=T$, and $ \mathcal{U}^* H \mathcal{U}=H$. Then 
\begin{equation*}
\mathcal{U}^* F_{\chi}(H,T) \mathcal{U} =F_{\chi}(H,T).
\end{equation*}
\end{lemma}
The proof  Lemma  \ref{eq:rotinv} is straightforward. We will use it in the following manner. If $H,T$ and $\chi$ are invariant under rotations in $\xi$-space, then $ F_{\chi}(H,T)$ is invariant under rotations in $\xi$-space  as well. Indeed, if $\mathcal{R}: \mathbb{R}^{d} \rightarrow \mathbb{R}^{d}$ is a rotation, one can consider the unitary map $u_{\mathcal{R}} : \mathrm{L}^{2}(\mathbb{R}^{3+d}) \rightarrow  \mathrm{L}^{2}(\mathbb{R}^{3+d})$ defined by
\begin{equation*}
( u_{\mathcal{R}} f) (k,\xi)=f(k, \mathcal{R} \xi), \qquad \forall (k,\xi) \in \mathbb{R}^{3+d}, 
\end{equation*}
for all $f \in \mathrm{L}^{2}(\mathbb{R}^{3+d})$. Clearly $H(0)$ and  $h_0$ in \eqref{H(0)}  (see also \eqref{eq:coupling}) are invariant under rotations in $\xi$-space in the sense that $u_{\mathcal{R}} h_0 = h_0$ and $\mathcal{U}_{\mathcal{R}} H(0) \mathcal{U}^*_{\mathcal{R}} = H(0)$, where $\mathcal{U}_{\mathcal{R}} := \Gamma( u_{\mathcal{R}} )$. Therefore, if we use operators $\chi$ that share this property, the sequence of operators obtained by iterative applications of the Feshbach-Schur map will all be invariant under rotations in $\xi$-space. This property is essential in Section~\ref{sec:iteration} since it will permit us to control the leading part of the operators $H^{(n)}(z)$ that we obtain by the iterative application of the Feshbach-Schur map.  Remark that the rotation invariance in $\xi$-space is not satisfied when $P \neq 0$ since in that case the original fiber Hamiltonian is not rotationally invariant. This explains why the proof below breaks down when $P\not=0$. 

We now specify the operators $\chi$ that are used in our analysis. Let $\chi: \mathbb{R} \rightarrow [0,1]$ be a smooth function such that $\chi(x)=1$ if $x\leq 3/4$, and $\chi(x)=0$ if $x\geq1$. We introduce 
\begin{equation*}
\chi_{\rho}:= \chi(\d \Gamma( |k| )/\rho),
\end{equation*}
for a suitably chosen parameter $0<\rho<1$.  We will first prove that $(H(0)-z,\d\Gamma( \xi )^2 + \d \Gamma( |k| )-z)$ is a Feshbach-Schur pair associated with $\chi_{\rho}$ if $\vert g \vert$ is  small enough and if $z$ lies in a small open ball centered at the origin. 
Then we explain how to recast the operator $F_{\chi_{\rho}}(H(0)-z,\d\Gamma( \xi )^2 + \d \Gamma( |k| )-z)$ into a Wick ordered form using Wick ordering. To shorten notations, we set
\begin{equation}\label{eq:Hchibarrho}
H_{\overline\chi_\rho}(z)=\rd\Gamma(\xi)^2+\rd\Gamma(|k|)-z+\overline\chi_\rho H_I\overline\chi_\rho.
\end{equation}

\subsection{First decimation step}
We introduce 
\begin{align*}
& B_{\rho}:= \{  (k,\xi)  \in \mathbb{R}^{3+d} \mid  \vert k \vert \leq \rho \}, \ B_{\rho}^{(M)}:= \{ ((k_1,\xi_1),...,(k_M,\xi_M)) \in \mathbb{R}^{(3+d)M} \mid \sum_{i=1}^{M} \vert k_i \vert \leq \rho \} , \\
& B_{\rho}^{(M,N)}:=B_{\rho}^{(M)} \times B_{\rho}^{(N)}.
\end{align*}

We begin with proving that $(H(0)-z,\d\Gamma( \xi )^2 + \d \Gamma( |k| )-z)$  is a Feshbach-Schur pair associated to $\chi_\rho$. 
\begin{lemma}
\label{lm:fesh-pair}
There exist $\mathrm{C}_0 > 0$ (only depending on the coupling function $h_0$) and $g_c > 0$ such that, for all $| g | \le g_c$, $\rho$ satisfying $\mathrm{C}_0 g^2 \le \rho < 1$ and $z \in D(0,\rho/4)$, the pair $(H(0)-z,\d\Gamma( \xi )^2 + \d \Gamma( |k| )-z)$ is a Feshbach-Schur pair associated with $ \chi_{\rho}$.
\end{lemma}
\begin{proof}
It is easy to prove that  $[\d\Gamma( \xi )^2 + \d \Gamma( |k| )-z]_{| \mathrm{Ran}(\overline{\chi}_{\rho})}$ is invertible with bounded inverse using spectral calculus. Indeed, 
\begin{equation*}
\vert r + l^2 - z  \vert  \geq r- \rho/4  \geq \rho/2,
\end{equation*}
if $r \geq 3 \rho/4$, and hence $\| [\d\Gamma( \xi )^2 + \d \Gamma( |k| )-z]^{-1}_{| \mathrm{Ran}(\overline{\chi}_{\rho})} \| \leq 2/\rho.$ The bounded invertibility of $[H(0)-z]_{| \mathrm{Ran}(\overline{\chi}_{\rho})}$ is clear using the Neumann series expansion of its inverse. Indeed, 
\begin{equation*} 
\big \| [H(0)-z]^{-1}_{| \mathrm{Ran}(\overline{\chi}_{\rho})} \big \|  
 \leq  \rho^{-1}  \sum_{n=0}^{\infty} ( C g  \rho^{-\frac12} \big )^{n}
\end{equation*}
for some constant $C>0$ independent of $g$ and $\rho$, where we have used the bounds (see Lemma \ref{lm:appFock1})
\begin{align} 
\| ( \d \Gamma( |k| ) + \rho)  [\d\Gamma( \xi )^2 + \d \Gamma( |k| )-z]^{-1}_{| \mathrm{Ran}(\overline{\chi}_{\rho})}\| &= \mathcal{O}(1), \notag\\
\|( \d \Gamma( |k| ) + \rho)^{-\frac12} \Phi(h_0) ( \d \Gamma( |k| ) + \rho)^{-\frac12} \|&= \mathcal{O}(\rho^{-\frac12}), \label{eq:mu2}
\end{align}
and where we have inserted the operators $(\d \Gamma( |k| )+\rho)^{-1/2}$ on the right and on the left of the interaction $\Phi(h_0)$ in the Neumann series expansion; see e.g. \cite{BaFaFrSc15_01} for more details. That $ [H(0)-z]^{-1}_{| \mathrm{Ran}(\overline{\chi}_{\rho})} \overline{\chi}_{\rho} \Phi(h_0) \chi_{\rho}$ extends to a bounded operator is immediate, using \eqref{eq:mu2} once more.
\end{proof}
 To verify that the Feshbach-Schur map can be applied again to $F_{\chi_{\rho}}(H(0)-z,\d\Gamma( \xi )^2 + \d \Gamma( |k| )-z)$ (see~\eqref{eq:def-fesh}) 
after restricting the possible values of $z$, it is useful to Wick order 
 \begin{align*}
 H^{(0)}(z):&=F_{\chi_{\rho}}(H(0)-z,\d\Gamma( \xi )^2 + \d \Gamma( |k| )-z)_{ |  \mathcal{H}^{(0)}}\\
& =\Big (  \d \Gamma(\vert k \vert) +  \d \Gamma(\xi)^2- z \Big ) \mathds{1}_{\d \Gamma(\vert k \vert) \leq \rho} +  {\chi}_{\rho} H_{I} {\chi}_{\rho} -   {\chi}_{\rho} H_{I}  {\overline{\chi}}_{\rho}  [H_{{\overline{\chi}}_{\rho} }(z)]^{-1}_{| \text{Ran}({\overline{\chi}}_{\rho})}  {\overline{\chi}}_{\rho} H_{I}  {\chi}_{\rho} \nonumber
 \end{align*}
acting on $$\mathcal{H}^{(0)}= \mathds{1}_{\d \Gamma( |k| ) \leq \rho} (\mathcal{H}_{\mathrm{f}}). $$ 
More precisely the resolvent $[H_{{\overline{\chi}}_{\rho} }(z)]^{-1}$ is expanded into a Neumann series, then the obtained expression is Wick ordered using the pull-through formula
\begin{equation} \label{eq:pullthrough}
a(k_1,\xi_1) f(\d\Gamma( |k| ) , \d\Gamma(\xi))=f( \d\Gamma( |k| ) + \vert k_1 \vert , \d\Gamma(\xi)+ \xi_1) a(k_1,\xi_1),
\end{equation}
which holds for any measurable function $f: \mathbb{R}^4 \rightarrow \mathbb{C}$. Lemma \ref{lm:b0} below shows that  $H^{(0)}(z)$ can indeed be rewritten as a convergent series of (generalized) ``Wick monomials" on $\mathcal{H}^{(0)}$ (see equation~\eqref{eq:wickexpansion}). 

To that end, we first need to recall the precise definition of the Wick monomials (see \cite{BaFaFrSc15_01} for more details). They are operators on $\mathcal{H}^{(0)}$ (or on $\mathcal{H}^{(j)}$, see below) denoted by  $W_{M,N}(h)$, that  are  defined in the sense of  quadratic forms, for all $M+N\geq 1$,  by 
\begin{align}
 W_{M,N}(h)  : =  \mathds{1}_{\d \Gamma( |k| ) \leq \rho} & \int_{\mathbb{R}  ^{(3+d) M} \times\mathbb{R}  ^{(3+d)N}} 
a^*(k_1,\xi_1)\cdots a^*(k_m,\xi_M) \notag \\
& h( \d \Gamma( |k| ), \d \Gamma(\xi), (k_1,\xi_1), \cdots,   (k_M,\xi_M), (\tilde{k}_1,\tilde{\xi}_1), \cdots,(\tilde{k}_N,\tilde{\xi}_N))  \notag \\
& a  (\tilde{k}_1,\tilde{\xi}_1) \cdots a (\tilde{k}_N,\tilde{\xi}_n)  \prod_{i=1}^{M} \d k_i \d\xi_i    \prod_{j=1}^{N} \d \tilde{k}_j \d\tilde{\xi}_j \mathds{1}_{\d \Gamma( |k| ) \leq \rho}. \label{eq:Wmn}
\end{align}
 The functions  $ h : \mathbb{R} \times \mathbb{R}^3  \times\mathbb{R}  ^{(3+d)M} \times 
\mathbb{R}  ^{(3+d) N} \to \mathbb{C}$  in \eqref{eq:Wmn}, $M+N \geq 1$ that will appear below, 
are bounded measurable, symmetric in the $M$ variables in  $\mathbb{R}  ^{(3+d)M}$ and the $N$ variables in $\mathbb{R}  ^{(3+d)N}$.   An easy calculation shows that the operator norm of a Wick monomial can be controlled  by the norm of its kernel  $h$, meaning that 
\begin{equation} \label{eq:lambda1}
\| W_{M,N}(h) \| \leq \| h\|_{\mu}  \rho^{ (M+N)(\frac{3}{2}+ \mu)} (4 \pi)^{\frac{M+N}{2}} ,
\end{equation}
 if $\| h\|_{\mu}<\infty$, where  $\| h\|_{\mu} $ is defined by 
\begin{equation} \label{mu}
\|h \|_{\mu} := \Big \| \sup_{(r,l ) \in [0,\rho] \times \mathbb{R}^d}  \underset{ K^{(M,N)}  \in B_{\rho }^{(M,N)}}{\text{ ess sup}}  \frac{ \vert h\left(r, l,  (k,\xi)^{(M)}, (\tilde k, \tilde \xi)^{(N)} \right) \vert }{ \vert  k_1 \vert ^{\mu} ...  \vert  k_M \vert ^{\mu} \vert \tilde{k}_1  \vert^{\mu} ...  \vert \tilde{k}_M  \vert^{\mu}}  \Big \|_{\mathrm{L}^2(\mathbb{R}^{d(M+N)})},
\end{equation}
with the shorthands 
\begin{equation*}
(k,\xi)^{(N)}:=((k_1,\xi_1),...,(k_n,\xi_N)), \quad (K,\Xi)^{(M,N)}: =((k,\xi)^{(M)},(\tilde k, \tilde \xi)^{(N)}).
\end{equation*}
We observe that the norm introduced in \eqref{mu} differs from the one considered in \cite{BaChFrSi03_01} or \cite{BaFaFrSc15_01}, mainly because we have to control the kernels in the $\xi$ variable using an $L^2$-norm, whereas it is convenient to use an $L^\infty$-norm in the $k$ variable. Nevertheless the proof of \eqref{eq:lambda1} is a straightforward modification of \cite[Theorem 3.1]{BaChFrSi03_01} (see also \cite[Lemma 3.1]{BaFaFrSc15_01}). Hence we do not give the details.

Note that the Wick monomials depend on the parameter $\rho$. To state the next lemma, we will furthermore need
the following class of functions:
\begin{definition}
A function $f: [0,\rho] \times \mathbb{R}^d \to \mathbb{C}$ is said to be of class $\mathcal{C}^{1,2}(\rho)$ if 
\begin{itemize}
\item $f$ is continuous on $[0,\rho] \times \mathbb{R}^d$,
\item $f(\cdot,l)$  is $\mathcal{C}^1$ on $[0,\rho]$ for all $l \in \mathbb{R}^d$, 
\item $f(r, \cdot)$ is $\mathcal{C}^2$ on $ \mathbb{R}^d$  for all $r \in [0,\rho]$. 
\end{itemize}
\end{definition}
\begin{lemma}
\label{lm:b0}
There exists a positive constant $\mathrm{C}_0$ such that the following holds. Let $\gamma > 0$. There exists $g_c > 0$ such that, for all $| g | \le g_c$, $\rho$ such that $\mathrm{C}_0 g^2 \le \rho < 1$ and $z \in D(0,\rho/4)$, 
\begin{equation}\label{eq:wickexpansion}
H^{(0)}(z)= \sum_{M+N \geq 0} W_{M,N}^{(0)}(z)  + \mathcal{E}^{(0)}(z),
\end{equation}
where the series converges uniformly on $D( 0 , \rho /  4 )$, $W_{M,N}^{(0)}(z):=W_{M,N}(w_{M,N}^{(0)}(z,\cdot, \cdot))$ and
\begin{equation*}
\mathcal{E}^{(0)}(z):=\langle\Omega, H^{(0)}(z)\Omega\rangle.
\end{equation*}
The kernels 
\begin{equation*}
w_{M,N}^{(0)}: D(0,\rho/4) \times [0,\rho] \times \mathbb{R}^d \times B_{\rho}^{(M,N)} \rightarrow \mathbb{C} ,
\end{equation*}
and the function $\mathcal{E}^{(0)}: D(0,\rho/4) \rightarrow \mathbb{C}$ satisfy the following properties: 
\begin{list}{\labelitemi}{\leftmargin=1em}
\item For all $z \in D( 0 , \rho / 4 )$, $w^{(0)}_{0,0}(z,\cdot,\cdot) \in \mathcal{C}^{1,2}(\rho)$, and $w_{0,0}^{(0)}(z,0,0)=0$.
\item  For all $M+N \ge 1$, $z  \in D(0,\rho/4)$ and for a.e. $ (K,\Xi)^{(M,N)} \in B_{\rho}^{(M,N)}$, 
\begin{equation*}
w^{(0)}_{M,N}(z, \cdot, \cdot , (K,\Xi)^{(M,N)}) \in \mathcal{C}^{1,2}(\rho).
\end{equation*}
\item  For all $M+N \ge 1$ and $z \in D(0,\rho/4)$,
\begin{align}
\label{eq:gamma1}
\| w^{(0)}_{M,N} (z) \|_{ \mu }& \le  \gamma  \rho^{1 - \frac12 (M+N) } ,\\ 
\| \partial_{\sharp} w^{(0)}_{M,N}(z) \|_{\mu}& \le \gamma \rho^{ - \frac12 (M+N)} , \label{eq:gamma2} \\ 
\| \partial_{l_q l_q'}^{2} w^{(0)}_{M,N}(z) \|_{\mu}& \le \gamma  \rho^{- \frac12 (M+N)}, \label{eq:gamma3}
\end{align}
where $\partial_\sharp$ stands for $\partial_r$ or $\partial_{l_q}$, and
\begin{align}
\vert z + \mathcal{E}^{(0)} (z) \vert &\le \gamma \rho ,\\ 
\label{eq:gamma5}
\| \partial_r w_{0,0}^{(0)}(z) -1  \|_{\infty}   +  \sum_{q \neq q'} \|  \partial^2_{l_q l_{q'}} w_{0,0}^{(0)}(z) -2 \delta_{q,q'}  \|_{\infty}  &\le \gamma .
\end{align}
\end{list}
Furthermore, the map $z \mapsto H^{(0)}(z) - \d\Gamma(\xi)^{2}_{\mid \mathcal{H}^{(0)}} \in \mathcal{L}(\mathcal{H}^{(0)})$ is analytic on $D(0,\rho/4)$.
\end{lemma}
The proof of Lemma~\ref{lm:b0}  follows closely the lines of \cite[Lemma 3.4]{BaFaFrSc15_01} and is therefore omitted. The main new results are the estimates on the second derivatives with respect to $l_q$, $l_{q'}$ appearing in \eqref{eq:gamma3}. These estimates may look surprising at a first glance because when we differentiate $w^{(0)}_{M,N}(z)$ (with respect to $l$ or to $r$) we have in particular to differentiate the ``free resolvents'' $( r + l^2 - z )^{-1} \bar{\chi}_\rho( r )$ appearing in the Neumann series expansion. This produces an extra resolvent and hence we obtain an estimate which is worse by a power of $\rho$. This explains the difference between \eqref{eq:gamma1} and \eqref{eq:gamma2}. With such a naive estimate we would thus obtain an exponent $\rho^{-1- \frac12 (M+N)}$ in the right-hand side of \eqref{eq:gamma3}, which would be too singular for our purpose. In order to prove \eqref{eq:gamma3}, the key property is the rotation invariance in the $\xi$-variable, as we explain in Appendix \ref{app:rotation}.

Note that the Wick monomials $W_{M,N}^{(0)}(z)$ are bounded operators for $M+N \geq 1$ as follows from \eqref{eq:lambda1} and \eqref{eq:gamma1}, but that $W_{0,0}^{(0)}(z)$ is unbounded because of the term $\d\Gamma(\xi)^2$ in $H(0)$. More precisely, one can verify that $W_{0,0}^{(0)}( z )$ is of the form
\begin{equation*}
W_{0,0}^{(0)}( z ) = \chi_\rho \big ( \d\Gamma( \xi )^2 + \d \Gamma( |k| ) \big ) \chi_\rho + \tilde{W}_{0,0}^{(0)}(z),
\end{equation*}
where $\tilde{W}_{0,0}^{(0)}(z) = \tilde{w}_{0,0}^{(0)}( z ,  \d\Gamma ( |k| ) , \d \Gamma( \xi ) )$ and the function $\tilde{w}_{0,0}^{(0)}$ satisfies the estimates.
\begin{equation*}
\| \tilde{w}_{0,0}^{(0)}( z ) \|_\infty \le \gamma \rho , \quad \| \partial_\sharp \tilde{w}_{0,0}^{(0)}( z ) \|_\infty \le \gamma \rho, \quad \| \partial^2_{l_ql_q'} \tilde{w}_{0,0}^{(0)}( z ) \|_\infty \le \gamma \rho.
\end{equation*}
Therefore $W_{0,0}^{(0)}( z )$, and hence $H^{(0)}(z)$, are closed, unbounded operators on $\mathcal{H}^{(0)}$, with domains $\mathcal{D}(\d\Gamma(\xi)^2) \cap \mathcal{H}^{(0)}$. This is a particular feature of the current model: the cutoff in $|k|\leq \rho$ does not control $\xi$. It is the source of some technical difficulties, as we will see below, but is also at the origin of the physical features of the model. The fact that $z \mapsto H^{(0)}(z) - \d\Gamma(\xi)^{2}_{\mid \mathcal{H}^{(0)}}$ is analytic on $D(0,\rho/4)$ is proven in Appendix \ref{app:analyticity}.

In the estimates \eqref{eq:gamma1}--\eqref{eq:gamma5} the parameter $\gamma$ can be chosen arbitrarily small. Naturally, the smaller $\gamma$ is chosen, the smaller we have to fix the critical value $g_c$. In the sequel we choose to fix $\gamma = 1/8$ to simplify the exposition.

We can   establish that $ z \mapsto \mathcal{E}^{(0)}(z)$ has a unique zero in $D(0,\rho/4)$ using Rouch\'{e}'s Theorem. We refer to \cite[Lemma 3.5]{BaFaFrSc15_01} for the proof.
\begin{lemma}
\label{lm:zeros}
 There exists $g_c > 0$ such that, for all $| g | \le g_c$, $\rho$ such that $\mathrm{C}_0 g^2 \le \rho < 1$ (where $\mathrm{C}_0$ is the constant of Lemma \ref{lm:b0}) and $z \in D(0,\rho/4)$, the map $z \mapsto   \mathcal{E}^{(0)}(z) \in \mathbb{C}$ has a unique zero $z^{(0)} \in D(0,\rho/4)$. Moreover, for any $0 < \eta < 1/8$, we have that $D(z^{(0)}, \rho \eta/4  ) \subset D(0,\rho/4)$, and
\begin{align*}
\vert \mathcal{E}^{(0)}(z)\vert &\le  ( 1 + \eta ) \rho /4, \qquad \vert z^{(0)} \vert \le \rho / 8 ,
 \end{align*}
for all $z \in  D(z^{(0)}, \rho \eta  /4 )$.
\end{lemma}
Lemma \ref{lm:zeros} is an important ingredient in the  iterative construction below. It shows that the modulus  of  $\mathcal{E}^{(0)}(z)$ can be controlled on a smaller disk included in $ D(0, \rho/4 )$. This estimate is used to prove that the Feshbach-Schur map can be applied once more to $H^{(0)}(z)$ upon restricting $z$ to $ D(z^{(0)}, \rho_1/4)$, giving rise to a new operator $H^{(1)}(z)$ acting on $$\mathcal{H}^{(1)}:=\mathds{1}_{\d \Gamma( |k| ) \leq \rho_1} (\mathcal{H}_{\mathrm{f}}),$$
where $\rho_1 = \rho^{3/2+ \mu -\varepsilon}$ and $\varepsilon \in (0,1/2+ \mu)$.  To simplify matters, we choose $\varepsilon =1/4 + \mu/2$. The modulus of $\mathcal{E}^{(1)}(z)$ can be itself controlled on a small disk included in $D(z^{(0)}, \rho_1/4 )$.  Applying a similar argumentation at each iterative step,  one can construct a sequence $H^{(j)}(z)$ of operator-valued functions defined on shrinking complex open sets $D^{(j)}$ and shrinking subspaces   $\mathcal{H}^{(j)} \subset \mathcal{H}_{\mathrm{f}}$ (Lemma~\ref{lm:induc}). An important property of this iterative construction is that  the  norm $\| \cdot \|_{\mu}$  of the kernels $w_{M,N}^{(j)}$, $M+N \geq1$, stays finite (but grows) at each iteration step. On the other hand, thanks to \eqref{eq:lambda1}, the norm of the Wick monomials $W_{M,N}^{(j)}$ goes quickly to zero as we iterate the process.  

We clarify this construction in the next subsection.

\subsection{Iterative construction} \label{sec:iteration}
\subsubsection{Statement of the main lemma} 

We introduce 
\begin{align*}
\rho_{j}&:=\rho^{(\frac{5}{4}+ \frac{\mu}{2})^j}, \qquad D^{(j)}:= D(z^{(j-1)},\rho_{j}/4), \qquad \mathcal{H}^{(j)}:=\mathds{1}_{\d \Gamma( |k| ) \leq \rho_{j}} (\mathcal{H}_{\mathrm{f}}),
\end{align*}
where $(z^{(j)})_{j=-1,0,\dots}$ is a sequence of complex numbers which is defined recursively below as the zeros of the applications $\mathcal{E}^{(j)}$. The first two values of this sequence, $z^{(-1)}=0$, and $z^{(0)}$, have been already introduced above. Note that $\rho_j$ decays super exponentially fast to $0$ (because $\mu > -1/2$), but the decay speed of $\rho_{j}$ has been chosen to simplify notations. More generally, one can choose any decay of the type $\rho^{(3/2+ \mu-  \varepsilon)^j}$, with $\varepsilon \in (0,1/2 + \mu)$ (see \cite{BaFaFrSc15_01} for more details). We also set  
$$H^{(-1)}(z):=H(0)-z \qquad \text{and} \qquad W_{0,0}^{(-1)}(z) + \mathcal{E}^{(-1)}(z):=\d\Gamma( |k| ) + \d\Gamma(\xi)^2 -z.$$ We construct operator-valued functions $D^{(j)} \ni z \mapsto H^{(j)}(z)$ iteratively, where the operators $H^{(j)}(z)$ are closed with domain $\mathcal{D}(\d\Gamma(\xi)^2) \cap \mathcal{H}^{(j)}$. 
\begin{lemma} \label{lm:induc} 
Suppose that $\mu > -1/2$. There exist $g_c > 0$, $\rho_c > 0$, $\mathrm{C}_0 > 0$ and $\mathbf{C}>1$ such that, for all $| g | \le g_c$ and $\rho$ such that $\mathrm{C}_0 g^2 \le \rho < \rho_c$, the following is satisfied:
For all  $j \in \mathbb{N} \cup \{0\}$, there is a sequence of kernels $w_{M,N}^{(j)} : D^{(j)} \times [0,\rho_j] \times \mathbb{R}^d \times B_{\rho_{j}}^{(M,N)} \rightarrow \mathbb{C}$,    a  function $\mathcal{E}^{(j)}:D^{(j)} \rightarrow \mathbb{C}$, and an operator-valued function $ D^{(j)} \ni z \mapsto H^{(j)}(z)$ defined recursively by 
 \begin{equation} \label{eq:defHj}
 H^{(j)}(z):=F_{\chi_{\rho_j}}(H^{(j-1)}(z),W_{0,0}^{(j-1)}(z) + \mathcal{E}^{(j-1)}(z))_{| \mathcal{H}^{(j)}} =: \sum_{M+N \geq 0} W_{M,N}^{(j)}(z) + \mathcal{E}^{(j)}(z),
 \end{equation}
where the series converges uniformly on $D^{(j)}$, $W_{M,N}^{(j)}(z) := W_{M,N}[w_{M,N}^{(j)}(z, \cdot, \cdot)]$ and
\begin{equation*}
\mathcal{E}^{(j)}(z):=\langle\Omega, H^{(j)}(z)\Omega\rangle.
\end{equation*}
For all $z \in D^{(j)}$, the operator $H^{(j)}(z)$ is closed with domain $\mathcal{H}^{(j)} \cap \mathcal{D} ( \d\Gamma(\xi)^2 )$, and  $H^{(j)}(z)- \d\Gamma(\xi)^{2}_{\mid \mathcal{H}^{(j)}} $ extends to a bounded operator on $\mathcal{H}^{(j)}$. The kernels  $w_{M,N}^{(j)}$ satisfy the following properties.

\begin{itemize}[leftmargin=30pt, itemsep=5pt]
\item[(a)]  For all $z \in D^{(j)}$, $w_{0,0}^{(j)}(z,\cdot,\cdot) \in \mathcal{C}^{1,2}(\rho_j)$ and  $w_{0,0}^{(j)}(z,0,0)=0$.
\item[(b)]  For all $M+N \geq 1$, $z  \in D^{(j)}$ and a.e. $(K,\Xi)^{(M,N)} \in B_{\rho_j}^{(M,N)}$, $w_{M,N}^{(j)}(z, \cdot, \cdot, (K,\Xi)^{(M,N)}) \in \mathcal{C}^{1,2}(\rho_j)$.   Moreover, $w_{M,N}^{(j)}(z, \cdot, \cdot, (K,\Xi)^{(M,N)})$ is symmetric in $(k^{(M)},\xi^{(M)})$ and $(\tilde{k}^{(N)},\tilde{\xi}^{(N)})$.
\item[(c)] For all $M+N \ge 1$ and $z \in D^{(j)}$, 
\begin{align}
\label{eq:AA1}
\| w_{M,N}^{(j)} (z) \|_{\mu}& \le \mathbf{C}^{j(M+N)} \rho_{j}^{- (M+N) + 1} ,\\
\| \partial_{\sharp} w^{(j)}_{M,N}(z) \|_{\mu}& \le  \mathbf{C}^{j(M+N)}    \rho_{j}^{- (M+N) } , \label{eq:AA2} \\
\| \partial^2_{l_q l_{q'}} w^{(j)}_{M,N}(z) \|_{\mu}& \le  \mathbf{C}^{j(M+N)}    \rho_{j}^{- (M+N) } , \label{eq:AA3}
\end{align}
where $\partial_\sharp$ stands for $\partial_r$ or $\partial_{l_q}$. Moreover, for all $z \in D^{(j)}$
\begin{align} \label{eq:AA4}
\| \partial_r w^{(j)}_{0,0}(z) - 1 \|_{\infty} +  \sum_{q\neq q'}   \|  \partial^2_{l_q l_{q'}} w^{(j)}_{0,0}(z)  -2 \delta_{q,q'}  \|_{\infty} & \le \frac{1 }{ 4  } .
\end{align}
\item[(d)] The maps $\mathcal{E}^{(j)}  :D^{(j)} \rightarrow \mathbb{C}$ and $z \mapsto H^{(j)}(z)-\d \Gamma(\xi)^{2}_{\mid \mathcal{H}^{(j)}}   \in \mathcal{L}(\mathcal{H}^{(j)})$ are analytic on  $D^{(j)}$, and for all $z \in D(z^{(j-1)} , \rho_j / 6)$,
 \begin{equation}
 \big \vert \partial_z \mathcal{E}^{(j)}(z)  + 1 \big \vert < \frac{1}{4}. \label{eq:AA5}
 \end{equation}
\item[(e)] The map $z \mapsto   \mathcal{E}^{(j)} (z) \in \mathbb{C}$ has a unique zero $  z^{(j)}  \in  D^{(j)}$.   One has that  $ D^{(j+1)}=D(z^{(j)},\rho_{j+1}/4) \subset  D^{(j)}$, and for all $z  \in D^{(j+1)}$,
\begin{align}
\label{eq:alpha1}
\vert \mathcal{E}^{(j)}(z)\vert &\le \frac{   \rho_{j+1} }{ 2} ,\\
 \label{eq:alpha2}
 \vert z^{(j)}  -z^{(j-1)}  \vert   &< \frac{\rho_j}{8} .
\end{align}
\end{itemize}
 \end{lemma}

\subsubsection{Outline  of  the proof of Lemma \ref{lm:induc}}
We outline the proof which closely follows \cite[Section 4]{BaFaFrSc15_01}, and stress those passages that need to be altered for the model under study. The main differences with \cite{BaFaFrSc15_01} are that the operators $H^{(j)}( z )$ are not bounded (but this can be compensated by the fact that $H^{(j)}(z)-\d \Gamma(\xi)^{2}_{\mid \mathcal{H}^{(j)}}$ is bounded) and more importantly that we need to control the second derivatives of the kernels with respect to $l_q$, $l_{q'}$ as stated in \eqref{eq:AA3}. This can be done thanks to the rotation invariance in the $\xi$-variable as we explain below.

It is clear from Lemmas \ref{lm:b0} and \ref{lm:zeros} that the properties (a)-(e) are satisfied for $j=0$. The rest of the  proof is  done by induction and is divided into three steps: knowing that properties (a)-(e) are satisfied at all  steps $ k \leq j$, for $j \geq 0$,  we first show that the Feshbach-Schur map can be applied to $H^{(j)}(z)$ for all $z \in D^{(j+1)}$. This defines  the operator-valued function $z \mapsto H^{(j+1)}(z)$ after restriction to $\mathcal{H}^{(j+1)}$.  In the second step, we explain how to normal order $H^{(j+1)}(z)$ and we prove the upper bounds \eqref{eq:AA1}-\eqref{eq:AA4} at step $j+1$. In the third step, we prove properties (d) and (e).

\vspace{0,2cm}

\textbf{Step 1.} We begin with the applicability of the Feshbach-Schur map. Let $z \in D^{(j+1)}$.  The main new idea we need here is to use the invariance under rotations in $\xi$-space to control the absolute value of $ w^{(j)}_{0,0} ( z,r, l )  + \mathcal{E}^{(j)} ( z)$, and therefore, to show that the restriction of $W_{0,0}^{(j)}(z)$ to $\text{Ran}(\overline{\chi}_{ \rho_{j+1} })$ is bounded invertible if $z \in D^{(j+1)}$.  We write $l=(l_1,l_2,\dots,l_d) \in \mathbb{R}^d$ and  we get
\begin{align*}
 w_{0,0}^{(j)} ( z,r, l ) &=  w_{0,0}^{(j)} ( z,r, l ) - w_{0,0}^{(j)} ( z,0, l )   \\
 & + \sum_{i=1}^{d} [w_{0,0}^{(j)} ( z,0, (0,\dots,l_i,l_{i+1},\dots,l_d )) -   w_{0,0}^{(j)} (z,0,(0,\dots,0,l_{i+1},\dots,l_d )) ]\\
 &=\int_{0}^{r} (\partial_{r'}  w_{0,0}^{(j)}) ( z,r', l ) \d r'  + \sum_{i=1}^{d} \int_{0}^{l_i}(\partial_{l'_i} w_{0,0}^{(j)}) ( z,0, (0,\dots,l'_i,l_{i+1},\dots,l_d )) \d l'_i\\
 &= \int_{0}^{r} (\partial_{r'}  w_{0,0}^{(j)}) ( z,r', l ) \d r'  + \sum_{i=1}^{d} \int_{0}^{l_i} \int_{0}^{l'_i}(\partial_{l''_i} w_{0,0}^{(j)}) ( z,0, (0,\dots,l''_i,l_{i+1},\dots,l_d )) \d l''_i \d l'_i.
\end{align*}
In the last line, we have used that $(\partial_{l''_i} w_{0,0}^{(j)}) ( z,0, (0,\dots,0,l_{i+1},\dots,l_d ))=0$, a property that follows directly from the fact that the $\mathrm{C}^2$-function $l \mapsto  w_{0,0}^{(j)} ( z,r, l )$ is invariant under rotations, and therefore, $(\partial_{l_i} w_{0,0}^{(j)} ( z,r, l ))= (\partial_{l_i}(\vert l \vert))  (\partial_{\vert l \vert } w_{0,0}^{(j)} ( z,r, \vert l \vert ))$ is zero in $l_i=0$. The invariance under rotation in $\xi$-space of the kernel $ w_{0,0}^{(j)} $ follows from the invariance under rotation in $\xi$-space of the effective operator $H^{(j)}(z)$; see Appendix \ref{extrac}.  Using \eqref{eq:AA4} and \eqref{eq:alpha1}, it follows that 
\begin{align*}
\vert w_{0,0}^{(j)} ( z,r, l )  + \mathcal{E}^{(j)} ( z) \vert &\geq  \vert w_{0,0}^{(j)}  (z,r,l) \vert -  \vert  \mathcal{E}^{(j)} (z) \vert  \geq    r  +l^2  - \frac{ r + l^2 }{ 4 }  - \frac12 \rho_{j+1} \\
&   \geq \frac{3 r}{4}  - \frac12 \rho_{j+1} \ge \frac{ 1 }{16} \rho_{j+1},
\end{align*}
for all $ z \in D^{(j+1)}$ and $r \ge 3 \rho_{j+1} /4$.  We deduce that  the restriction of  $w_{0,0}^{(j)} ( z)  + \mathcal{E}^{(j)}(z)$ to $\mathrm{Ran}(\overline{\chi}_{  \rho_{j+1} })$ is invertible and that its inverse has a norm of order $\rho_{j+1}^{-1}$ for all $z \in D^{(j+1)}$.    Moreover, using \eqref{eq:lambda1} and  \eqref{eq:AA1}, we obtain that
\begin{equation}
\label{eq:delta0}
\|W^{(j)}_{M,N} (z) \| \leq  (4 \pi)^{\frac{M+N}{2}} \rho_j^{(M+N) (\frac{3}{2} + \mu)  } \|w_{M,N} (z) \|_{\mu}  \le (4 \pi)^{\frac{M+N}{2}}\mathbf{C}^{j(M+N)} \rho_{j}^{(\frac{1}{2}+ \mu)(M+N) + 1}.
\end{equation}
Summing \eqref{eq:delta0} over $M+N \geq 1$, we  deduce that 
\begin{equation*}
 \sum_{M+N \ge 1 }\|W^{(j)}_{M,N} (z) \| \le \mathbf{C}^{j+1}  \rho_j^{\frac{3}{2}+ \mu} ,
\end{equation*}
provided that $\mathbf{C}$ is large enough, which yields 
\begin{align*}
&\big \|    [W^{(j)}_{0,0}(z) + \mathcal{E}^{(j)} (z)]^{-1}_{\mathrm{Ran}(\overline{\chi}_{\rho})}  \overline{\chi}_{\rho_{j+1}}   \Big[\sum_{M+N \ge 1 }W^{(j)}_{M,N} (z) \Big] \big \| \le  \mathbf{C}^{j+2}  \rho_j^{\frac{1}{4}+ \frac{\mu}{2}} .
\end{align*} 
This concludes the first step of the proof and shows that $(H^{(j)}(z),W^{(j)}_{0,0}(z)+ \mathcal{E}^{(j)}(z))$ is a Feshbach-Schur pair associated to $\chi_{\rho_{j+1}}$ for all $z \in  D^{(j+1)}$. It follows that  $ H^{(j+1)} (z)$ (see \eqref{eq:defHj}) is well-defined for all $z \in D^{(j+1)}$.

\vspace{0,2cm}

\textbf{Step 2.} The next step is to rewrite the operator $ H^{(j+1)} (z)$ using a Neumann series expansion and Wick ordering. This allows us to  express $H^{(j+1)}$ in the form 
 \begin{equation*} 
 H^{(j+1)}(z) = F_{\chi_{\rho_{j+1}}}(H^{(j)}(z),W_{0,0}^{(j)}(z) + \mathcal{E}^{(j)}(z))_{| \mathcal{H}^{(j+1)}} = \sum_{M+N \geq 0} W_{M,N}^{(j+1)}(z) + \mathcal{E}^{(j+1)}(z) ,
 \end{equation*}
with  $W_{M,N}^{(j+1)}(z)=W_{M,N}[w_{M,N}^{(j+1)}(z, \cdot, \cdot)]$. The algebraic computations leading to such an expression are very similar to what is done in \cite{BaChFrSi03_01} of  \cite{BaFaFrSc15_01}. The computations are therefore omitted here and we refer the reader to those papers for more details. Moreover, adapting \cite[Section 4]{BaFaFrSc15_01} in a straightforward way, one can verify that, for $M + N \ge 1$,
\begin{align*}
\| w_{M,N}^{(j+1)} ( z) \|_{\mu} & \leq \rho_{j+1}^{1-M-N} \mathbf{C}^{(j+1)(M+N)} ,
\end{align*}
provided that $\mathbf{C}$ is chosen large enough and $\rho_c$ is chosen small enough (depending on $\mathbf{C}$). Likewise, one verifies that the bounds \eqref{eq:AA2}--\eqref{eq:AA3} are satisfied by the partial derivatives with respect to $r$ and $l_q$, $q=1,2,3$, with $j$ replaced by $j+1$. Here again we need to use rotation invariance in $\xi$-space in a crucial way (see Appendix \ref{app:rotation}).  Similar calculations (see again \cite[Section 4]{BaFaFrSc15_01}) also lead to 
\begin{align}
\|  w^{(j+1)}_{0,0} (z) +   \mathcal{E}^{(j+1)} (z) - w^{(j)}_{0,0} (z) -\mathcal{E}^{(j)} (z) \|_{\infty} &  \leq \mathbf{C}^{2j+1} \rho_{j}^{3+2 \mu -(\frac{5}{4}+ \frac{\mu}{2}) }, \label{eq:lambda2}
\end{align}
and, for the derivatives,
\begin{align*}
\|  \partial_{\sharp} w^{(j+1)}_{0,0} (z) - \partial_{\sharp} w^{(j)}_{0,0} (z) \|_{\infty} & \leq  \mathbf{C}^{2j+1} \rho_{j}^{\frac{1}{2}+\mu},\\
\|  \partial_{\flat}^2 w^{(j+1)}_{0,0} (z) - \partial_{\flat}^2 w^{(j)}_{0,0} (z) \|_{\infty} & \leq \mathbf{C}^{2j+1} \rho_{j}^{\frac{1}{2} + \mu},
\end{align*}
where $\sharp=r$ or $l_q$, and $\flat=l_ql_q'$. The supremum is taken on the smallest  domain of definition of the functions. In particular, since one can show that theses inequalities are valid for any $k \leq j$ by the induction hypothesis, one deduces using Lemma \ref{lm:b0} (with $\gamma = 1/8$) that 
\begin{align*}
& \| \partial_r w^{(j+1)}_{0,0}(z) - 1 \|_{\infty}  + \sum_{q,q'=1}^{3}  \|  \partial^2_{l_q l_{q'}} w^{(j+1)}_{0,0}(z) -  2 \delta_{q,q'}   \|_{\infty}  \leq    \frac18 + 7 \sum_{k=0}^{j} \mathbf{C}^{2k+1} \rho_{k}^{\frac{1}{2}  + \mu } \leq \frac{1}{4}  ,
\end{align*}
provided that $\rho_c$ is sufficiently small. This establishes \eqref{eq:AA4}. Similarly, one shows that
\begin{equation*}
\| w^{(j+1)}_{0,0}(z) + \mathcal{E}^{(j+1)}(z)- l^2-r+z \|_{\infty} \le \mathbf{C} \Big (  \sum_{k=0}^{j} \mathbf{C}^{2k} \rho_{k}^{ \frac{7}{4}+ \frac{3\mu}{2}}  + \frac18 \rho \Big ) ,
\end{equation*}
and hence the operators $W_{0,0}^{(j+1)}(z)+ \mathcal{E}^{(j+1)}(z) -\d\Gamma(\xi)^{2}_{\mid \mathcal{H}^{(j+1)}}$  and   $H^{(j+1)}(z)-\d\Gamma(\xi)^{2}_{\mid \mathcal{H}^{(j+1)}}$ extend to  bounded operators on $\mathcal{H}^{(j+1)}$, as claimed.

\vspace{0,2cm}

\textbf{Step 3.} It remains to verify the statements $(d)$ and $(e)$ concerning analyticity. In Appendix \ref{app:analyticity} we show that $ z \mapsto H^{(j)}(z) - \d\Gamma(\xi)^{2}_{\mid \mathcal{H}^{(j)}}$  is analytic on $D^{(j)}$ for all $j$. Therefore $z \mapsto  \mathcal{E}^{(j)} (z)  = \langle \Omega \vert ( H^{(j)}(z) - \d \Gamma(\xi)^{2}_{\mid \mathcal{H}^{(j)}} ) \Omega \rangle$ is analytic in $z$, too. We  set $r_{j}= \rho_{j}/4$ and $r_{j+1}= \rho_{j+1}/4$. Let  $z = z^{(j)}+ \beta r_{j+1}  \in D( z^{(j)} , r_{j+1} )$, with $0 \leq \vert  \beta  \vert \leq 2/3$. The triangle inequality and the inequality \eqref{eq:alpha2}  imply that $z \in  \overline{D}( z^{(j-1)}, 2 r_j/3)$ if $\rho_c$ is sufficiently small.   We consider the  circular  contour $\mathcal{C}$ centered at $z^{(j)}$ and with radius $r_{\mathcal{C}}= 3 r_{j+1}/4$. We have that $ \mathcal{C}  \subset D^{(j+1)} \subset   D^{(j)} $, and, by Cauchy's formula,
\begin{equation*}
\begin{split}
\vert \partial_z \mathcal{E}^{(j+1)}(z) - \partial_z \mathcal{E}^{(j)}(z) \vert& \leq \frac{1}{2\pi} \Big \vert  \int_{\mathcal{C}} \d z' \frac{\mathcal{E}^{(j+1)}(z') -\mathcal{E}^{(j)}(z') }{ (z-z')^2} \Big \vert  \leq \frac{3 }{ 4 (3/4-\vert\beta\vert)^2 } \frac{ \mathbf{C}^{2j+1}  \rho_{j}^{\frac{7}{4}+ \frac{3\mu}{2}}}{r_{j+1} } ,
\end{split}
\end{equation*}
where we have used \eqref{eq:lambda2} and the equality $w_{0,0}^{(j)}(z,0,0) =  w_{0,0}^{(j+1)}(z,0,0)=0$. We deduce that 
\begin{equation*}
\vert \partial_z \mathcal{E}^{(j+1)}(z) - \partial_z \mathcal{E}^{(j)}(z) \vert  \leq   432 \mathbf{C}^{2j+1}   \rho_j^{\frac{1}{2}+\mu}.
\end{equation*}
Since this property is true for all $0 \leq k \leq j$, we can sum up over $k$, using Lemma \ref{lm:b0} (with $\gamma = 1/8$), to obtain 
\begin{equation*} 
\vert \partial_z \mathcal{E}^{(j+1)}(z) +  1  \vert  \leq   432 \sum_{k=0}^{j} \mathbf{C}^{2k+1}   \rho_k^{\frac{1}{2}+ \mu} + \frac18 < \frac{1}{4},
\end{equation*}
for all  $z \in \overline{D}( z^{(j)},2 r_{j+1} / 3 )$ if $\rho_c$ is small enough. Using similar calculations as in \cite[Section 4]{BaFaFrSc15_01}, it is then easy to show that $\mathcal{E}^{(j+1)}$ has a unique zero in $D(z^{(j)} , \rho_{j+1} / 6 )$ using Rouch\'{e}'s theorem. The proof of \eqref{eq:alpha1} and \eqref{eq:alpha2} at step $j+1$ follows closely \cite{BaFaFrSc15_01}.

\subsection{Existence of a ground state} 
A direct consequence of Lemma \ref{lm:induc}, and in particular of \eqref{eq:alpha2}, is that the sequence $z^{(j)}$ is Cauchy and converges to some non-zero limit $z^{(\infty)} \in \mathbb{C}$. The limit $z^{(\infty)}$ is actually real and is a non-degenerate eigenvalue of $H(0)$. This can be proven completely similarly as in \cite[Section 5]{BaFaFrSc15_01}. Here we prove that $z^{(\infty)}= \inf \sigma(H(0))$. To do so we  just need to show that $H(0)-z$ is bounded invertible for all $z \in (-\infty,z^{(\infty)})$. We first remark that $H^{(j)}(z)$ is self-adjoint for all $z \in D^{(j)} \cap \mathbb{R}$ because the Feshbach-Schur map preserves self-adjointness. Let $\psi \in \mathcal{D}(\d\Gamma(\xi)^2) \cap \mathcal{H}^{(j)}$.
One has that 
\begin{align}
  \langle  \psi  \vert  H^{(j)}(z) \psi \rangle & =  \langle  \psi \vert W_{0,0}^{(j)}(z)  \psi \rangle + \| \psi\|^2  \mathcal{E}^{(j)}(z) + \sum_{m+n \geq 1} \langle \psi \vert W_{M,N}^{(j)}(z) \psi \rangle \notag \\
& \geq  \langle  \psi \vert W_{0,0}^{(j)}(z)  \psi \rangle + \| \psi\|^2  \mathcal{E}^{(j)}(z )   -  \| \psi\|^2  \mathbf{C}^{j+1}  \rho_j^{\frac{3}{2}+ \mu},  \label{eq:beta1}
\end{align}
for all $z \in D^{(j)} \cap \mathbb{R}$. Moreover $\langle \psi \vert W_{0,0}^{(j)}(z)  \psi \rangle \geq 0$ because $z$ is real and because $w_{0,0}^{(j)}(z,r,l) \geq \frac{3}{4}(r + l^2)$ by \eqref{eq:AA4}. Using \eqref{eq:AA5}, we deduce that 
\begin{equation*}
\vert \mathcal{E}^{(j)}(z) - \mathcal{E}^{(j)}(z^{(\infty)}) -z^{(\infty)}+ z \vert < \frac{1}{4} \vert z-z^{(\infty)} \vert,
\end{equation*}
for all $z \in D^{(j)} \cap \mathbb{R}$ (because $D^{(j)} \cap \mathbb{R} \subset D(z^{(j-1)}, \rho_{j}/6)$), where we have used also that $z^{(\infty)} \in D^{(j)} \cap \mathbb{R}$. Therefore,  for all $z \in D^{(j)} \cap \mathbb{R}$,
\begin{equation*}
   \mathcal{E}^{(j)}(z) =    \mathcal{E}^{(j)}(z)   - \mathcal{E}^{(j)}(z^{(\infty)}) +z -z^{(\infty)} -z + z^{(\infty)} + \mathcal{E}^{(j)}(z^{(\infty)})  \geq   \frac{3}{4} (z^{(\infty)}  - z) - \vert \mathcal{E}^{(j)}(z^{(\infty)}) \vert.
\end{equation*}
Moreover, by \eqref{eq:alpha1}, we have that $\vert \mathcal{E}^{(j)}(z^{(\infty)}) \vert \le \rho_{j+1} / 2 = \rho_j^{5/4+\mu/2} / 2$.
Introducing this into \eqref{eq:beta1}, we obtain that
\begin{align*}
\langle  \psi  \vert  H^{(j)}(z) \psi \rangle & \ge  \Big ( \frac{3}{4} (z^{(\infty)}  - z) -  \frac12 \rho_j^{\frac54+\frac{\mu}{2}} - \mathbf{C}^{j+1}  \rho_j^{\frac{3}{2}+ \mu} \Big )  \| \psi\|^2  \\
&\ge \Big ( \frac{3}{4} (z^{(\infty)}  - z) -\rho_j \Big ( \frac12 \rho_j^{\frac14+\frac{\mu}{2}} - \mathbf{C}^{j+1}  \rho_j^{\frac{1}{2}+ \mu} \Big ) \Big )  \| \psi\|^2  ,
\end{align*}
and we deduce that for all $z \in [z^{(\infty)}-\frac{1}{6}\rho_j,z^{(\infty)}-\frac{1}{6}\rho_{j+1}) \subset D^{(j)} \cap \mathbb{R}$, the operator $ H^{(j)}(z)$ is bounded invertible provided that $\rho \le \rho_c$ with $\rho_c$ small enough. This is true for all $j \in \mathbb{N} \cup \{0\}$, and, by isospectrality of the Feshbach-Schur map, this implies that $H(0)-z$ is bounded invertible for all $z \in [z^{(\infty)} -\frac{1}{6} \rho, z^{(\infty)})$.
 
It remains to verify that $ H(0)-z$ is bounded invertible for all $z<z^{(\infty)}-\rho/6$. We write
 \begin{align*}
\langle \psi \vert (H(0)-z) \psi \rangle& = \langle \psi | ( \d \Gamma( |k| ) + \d\Gamma(\xi)^2 + g \Phi( h_0 ) - z ) \psi \rangle \\
& \geq \langle \psi | ( \d \Gamma( |k| ) - z ) \psi \rangle - C | g | \| \psi \|\|\d \Gamma( |k| )^{\frac12} \psi \| \\
& \geq \langle \psi | ( \d \Gamma( |k| ) - z ) \psi \rangle - \frac{1}{2} \langle \psi , \d \Gamma ( | k |) \psi \rangle - \frac12 C^2 g^2 \| \psi \|^2 \\
& \geq - \Big ( z + \frac12 C^2 g^2 \Big ) \| \psi \|^2 \geq \Big ( \frac{ \rho}{6} - z^{(\infty)} - \frac12 C^2 g^2 \Big ) \| \psi \|^2 .
 \end{align*}
 Summing \eqref{eq:alpha1} over $j$, we deduce that $\vert z^{(\infty)} \vert \leq \rho / 7$ provided that $\rho_c$ is small enough. 
Choosing the constant $\mathrm{C}_0$ in the statement of the lemma large enough then shows that $\rho/6 - z^{(\infty)} - C^2 g^2 / 2 > 0$, which concludes the proof.

\section{Proof of $\mathrm{iii})$: Absoute continuity of the spectrum}\label{section:Mourre}
In this section we show that the spectrum of $H(P)$ is purely absolutely continuous if $P \neq 0$ and $g \neq 0$, and that $\sigma( H(0) ) \setminus \{ E_g \}$ is purely absolutely continuous. Our analysis is based on Mourre's positive commutator method \cite{Mo81_01}. We work in the setting of the so-called ``singular Mourre theory'' developed in \cite{GeGeMo04_01,GeGeMo04_02}, using results and methods from \cite{FaMoSk11_01} and \cite{GeGeMo04_01,GeGeMo04_02}.

This section is divided into two subsections. First we define the conjugate operator that we shall use, we establish useful regularity properties of $H(P)$ with respect to auxiliary operators, and we deduce from the abstract results of Appendix \ref{app:Mourre} that a limiting absorption principle for $H(P)$, $P\in \mathbb{R}^d$, holds outside its pure point spectrum. In a second subsection, using Fermi's Golden Rule, we show that $H(P)$ does not have eigenvalues for $P \neq 0$.

\subsection{Limiting absorption principle}
The main result of this section is stated in the following theorem.
\begin{theorem}\label{thm:LAPH(p)}
Suppose that $\mu > 1/2 $. There exists $g_c > 0$ such that, for all $| g | \le g_c$ and $P \in \mathbb{R}^d$, the following holds:
\begin{itemize}
\item[(i)] $H(P)$ as at most $1$ eigenvalue counting multiplicity. Moreover, $H(P)$ does not have eigenvalue in $( - \infty , P^2 - 1 ] \cup [P^2 + 1 , \infty )$.
\item[(ii)] Let $J \subset [ E_g , \infty )$ be a compact interval such that $\sigma_{\mathrm{pp}}(H(P)) \cap J=  \emptyset$. Then 
\begin{equation*} 
    \sup_{z\in S} \| \langle A \rangle^{-s} (H-z)^{-1} \langle A \rangle^{-s}\|<\infty ,
\end{equation*}
for any $1/2<s\le 1$, with $\langle A \rangle = ( 1 + A^* A )^{1/2}$ and $S=\{ z \in \mathbb{C} , \mathrm{Re}(z) \in J , 0 < | \mathrm{Im}(z) | \leq 1\}$. 
Here $A$ is defined in~\eqref{eq:conjugateoperator}.
Moreover, the spectrum of $H(P)$ in $J$ is purely absolutely continuous, and we have that
\begin{equation*}
\big \| \langle A \rangle^{-s} e^{ - \ii t H(P) } \chi( H(P) ) \langle A \rangle^{-s} \big \| \lesssim t^{ - s + \frac12 } , \quad t \to \infty ,
\end{equation*}
for any $1/2 < s \le 1$ and $\chi \in \mathrm{C}_0^\infty( J ; \mathbb{R} )$.
\end{itemize}
\end{theorem}

This theorem is based on the abstract results recalled in Appendix \ref{app:Mourre}. In the following we explain how the hypotheses of Appendix \ref{app:Mourre} can be verified in our context. Note that the second statement of $(i)$ can be replaced by the more precise one that $H(P)$ does not have eigenvalue in $( - \infty , P^2 - C |g| ] \cup [P^2 + C |g| , \infty )$ for some positive constant $C$.
 
Before turning to the proof, let us point out that a consequence of the above result together with Theorem \ref{thm:GS} is the following local decay property:
\begin{equation}\label{eq:locdecay1}
\big \| \langle \d \Gamma( | y |) \rangle^{-s} e^{ - \ii t H(0) } \chi( H(0) ) \langle \d \Gamma( |y| ) \rangle^{-s} \big \| \lesssim t^{ - s + \frac12 } , \quad t \to \infty ,
\end{equation}
for any $1/2 < s \le 1$ and $\chi \in \mathrm{C}_0^\infty( ( E_g , \infty ) ; \mathbb{R} )$, where we used the notation $\langle x \rangle := (1+ x^2)^{1/2}$. Likewise, anticipating Theorem \ref{coi} and defining the set $B := \{ P \in \mathbb{R}^3 , |P| \in ( \nu_1 , \nu_2 ) \}$, with $\nu_1$ and $\nu_2$ as in \emph{(iii)}  of Theorem \ref{thm:main}, we have that
\begin{equation}\label{eq:locdecay2}
\big \| \langle \d \Gamma( | y |) \rangle^{-s} e^{ - \ii t H } \chi( H , P_{ \mathrm{tot} } ) \langle \d \Gamma( |y| ) \rangle^{-s} \big \| \lesssim t^{ - s + \frac12 } , \quad t \to \infty ,
\end{equation}
for any $1/2 < s \le 1$ and $\chi \in \mathrm{C}_0^\infty( \mathbb{R} \times B  ; \mathbb{R} )$. As mentioned in the introduction, estimates \eqref{eq:locdecay1} and \eqref{eq:locdecay2} show that some energy of the system is carried of by the field to infinity.

\subsubsection{The conjugate operator}
The positive commutator method is used, in particular, to extend expectation values of the resolvent from the open upper (or lower) half-plane to parts of the real axis. One of the key of the method consists in choosing a suitable ``conjugate operator'' that allows one to control the extension of the resolvent. We refer the reader to e.g. \cite{Amrein} or \cite{CFKS} for an introduction to this theory, and to \cite{AmBoGe96_01} for an extensive study.

The conjugate operator that we consider here is the generator of radial translations (associated to the variable $k$) in Fock space, \emph{formally} defined by
\begin{equation}\label{eq:conjugateoperator}
A := \d \Gamma \Big ( \frac{ k }{ |k| } \cdot y + y \cdot \frac{ k }{ |k| } \Big ),
\end{equation}
where we remind the reader that $y= \ii \nabla_k$. It is convenient to rewrite $A$ in polar coordinates. We consider the unitary operator 
\begin{equation*}
T : \mathrm{L}^2( \mathbb{R}^3 ) \to \mathrm{L}^2( \mathbb{R}^+ ) \otimes \mathrm{L}^2( S^{2} ) ,
\end{equation*}
defined by $(Tu)( \omega , \theta ) = \omega u ( \omega \theta )$, where $S^2$ denotes the sphere in $\mathbb{R}^3$. Lifting $T$ to Fock space, we obtain a unitary map
\begin{equation*}
\Gamma( T ) : \mathcal{H}_{\mathrm{f}} \equiv \Gamma_s \big ( \mathrm{L}^2( \mathbb{R}^{3+d} ) \big ) \to \Gamma_s \big ( \mathrm{L}^2( \mathbb{R}^d ) \otimes \mathrm{L}^2( \mathbb{R}^+ ) \otimes \mathrm{L}^2( S^2 ) \big ) =: \tilde{ \mathcal{H}_{\mathrm{f}} }.
\end{equation*}
In this representation, we have
\begin{equation*}
\tilde A := \Gamma( T ) A \Gamma( T )^* \equiv \d \Gamma \big ( \tilde a \big ), \quad \tilde a := \ii \partial_\omega .
\end{equation*}

Now, more precisely, we suppose that the operator $\tilde a$ is defined on the dense domain 
\begin{equation*}
\mathcal{D} ( \tilde a ) := \mathrm{L}^2( \mathbb{R}^d ) \otimes \mathrm{H}_0^1( \mathbb{R}^+ ) \otimes \mathrm{L}^2( S^2 ) ,
\end{equation*}
where $\mathrm{H}_0^1( \mathbb{R}^+ )$ denotes the closure of $\mathrm{C}_0^\infty( ( 0 , \infty ) )$ in the usual Sobolev space $\mathrm{H}^1( \mathbb{R} )$. On this domain, one verifies that $\tilde a$ is closed, maximal symmetric, satisfies $\dim \mathrm{Ker} ( \tilde{a}^* - \ii) = 0$ and $\dim \mathrm{Ker} ( \tilde{a}^* + \ii) = 1$, and that
\begin{equation*}
\mathcal{D} ( \tilde{a}^* ) := \mathrm{L}^2( \mathbb{R}^d ) \otimes \mathrm{H}^1( \mathbb{R}^+ ) \otimes \mathrm{L}^2( S^2 ) .
\end{equation*}
In particular, $\tilde{a}$  is not self-adjoint and has no self-adjoint extension. The same properties hold for $\tilde A = \d\Gamma( \tilde a )$, and therefore $\tilde A$ is maximal symmetric but not self-adjoint. The proper definition for the conjugate operator on $\mathcal{H}_{\mathrm{f}}$ that we consider is then $A := \Gamma( T )^* \tilde{A} \Gamma( T )$.  

To manipulate commutators between the Hamiltonian $H(P)$ and the conjugate operator $A$, it is useful to establish regularity properties of $H$ with respect to $A$. The latter, proven in the next paragraph, are defined through the semigroup generated by $A$. Here it is given as follows: Let $\{ \tilde{w}_t \}_{t \ge 0}$ be the $\mathrm{C}_0$-semigroup of isometries generated by $a$. It is not difficult to verify that $( \tilde{w}_t \varphi ) ( \omega, \theta , \xi ) = \varphi( \omega - t , \theta , \xi )$ if $\omega \ge t$ and $( \tilde{w}_t \varphi ) ( \omega, \theta , \xi ) = 0$ if $\omega \le t$. The adjoint semigroup is given by $( \tilde{w}_t^* \varphi ) ( \omega, \theta , \xi ) = \varphi( \omega + t , \theta , \xi )$. It is a $\mathrm{C}_0$-semigroup of contractions with generator $- a^*$. On the Fock space $\tilde{ \mathcal{H} }_f$, we set $\tilde{W}_t := \Gamma( \tilde{w}_t )$, and hence $\tilde{W}_t^* := \Gamma( \tilde{w}_t^* )$. The operator $\tilde{A}$ on $\tilde{ \mathcal{H} }_f$ is the generator of $\{ \tilde{W}_t \}_{ t \ge 0 }$, and the generator of $\{ \tilde{W}_t^* \}$ is $- \tilde{A}^*$. The generators of $A$ and $A^*$ are then
\begin{equation*}
W_t := \Gamma ( T )^* \tilde{W}_t \Gamma( T ) , \quad W_t^* = \Gamma( T )^* \tilde{W}_t^* \Gamma( T ) ,
\end{equation*}
respectively.

\subsubsection{Regularity properties of $H(P)$ with respect to auxiliary operators}
In this section we establish regularity properties of the Hamiltonian $H(P)$ with respect to the number operator $N$ and the conjugate operator $A$ defined above. In the following, those regularity properties will  allow us to manipulate the commutators between $H(P)$ and the unbounded operators $N$ and $A$. The definitions of the class of operators $\mathrm{C}^1( A )$, in the case where $A$ is self-adjoint, and $\mathrm{C}^{m}(A_1 ; A_2)$, $m=1,2$, in the case where $A_1$, $A_2$ are generators of $\mathrm{C}_0$-semigroups, are recalled in Appendix \ref{app:Mourre}.

We begin with the following easy lemma whose proof is left to the reader (see e.g. \cite[Lemma A.6]{BoFaSi12_01} for similar arguments).
\begin{lemma}\label{lm:hyp(1)}
Suppose that $\mu> - 1 $. For all $P \in \mathbb{R}^d$ and $g \in \mathbb{R}$, we have that
$
H(P) \in \mathrm{C}^1 ( N ) .
$
Moreover, the commutator $[ H(P) , \ii N ]$ defined in the sense of quadratic forms on $\mathcal{D}( H(P) ) \cap \mathcal{D}( N )$ extends to an $H(P)$-bounded operator. In particular, the hypothesis (1) of Appendix \ref{app:Mourre} is satisfied with $H = H(P)$ and $M=N$.
\end{lemma}

Our next concern is to establish regularity of $H(P)$ with respect to $A$. Since $\mathcal{D}( H(P) ) = \mathcal{D}( \d \Gamma( \xi )^2 + \d \Gamma( |k| ) )$, an argument of complex interpolation shows that $\mathcal{D}( |H(P)|^{1/2} ) = \mathcal{D}( ( \d \Gamma( \xi )^2 + \d \Gamma( |k| ) )^{1/2} )$. We introduce the subspace $\mathcal{G} \subset \mathcal{H}_{\mathrm{f}}$ defined by 
\begin{equation*}
\mathcal{G} := \mathcal{D}( ( \d \Gamma( \xi )^2 + \d \Gamma( |k| ) )^{1/2} ) \cap \mathcal{D}( N^{1/2} ) ,
\end{equation*}
and equipped with the norm of the intersection topology (see Appendix \ref{app:Mourre}). In view of definitions given in the previous paragraph, the proof of following lemma is straightforward. It is therefore omitted. 
\begin{lemma}\label{lm:Wt}
For all $t \ge 0$, we have that $W_t^\sharp \mathcal{G} \subset \mathcal{G}$, where $W_t^{\sharp}$ stands for $W_t$ or $W_t^*$. Furthermore, for all $0 \le t \le 1$, 
\begin{equation*}
\big \| ( \d \Gamma( \xi )^2 + \d \Gamma( |k| ) + N )^{\frac12} W_t^{\sharp} ( \d \Gamma( \xi )^2 + \d \Gamma( |k| ) + N + \mathds{1} )^{-\frac12} \big \| \le 2 .
\end{equation*}
In particular, the hypothesis (3) of Appendix \ref{app:Mourre} is satisfied with $H = H(P)$ and $M = N$.
\end{lemma}
Let $A_\mathcal{G}$ denote the generator of the $\mathrm{C}_0$-semigroup $W_t |_\mathcal{G}$ (which is well-defined by Lemma \ref{lm:Wt}), and let $A_{\mathcal{G}^*}$ be the generator of the $\mathrm{C}_0$-semigroup given as  the extension of $W_t$ to $\mathcal{G}^*$.
\begin{lemma}\label{lm:regularity_A}
Suppose that $\mu > 1 / 2 $. For all $P \in \mathbb{R}^d$ and $g \in \mathbb{R}$, we have that
$
H(P) \in \mathrm{C}^2 ( A_{ \mathcal{G} } ; A_{ \mathcal{G}^* } ) .
$
The quadratic form $[H(P),\ii A]$ on $\mathcal{D}(H(P)) \cap \mathcal{D}(A)$ extends to an $N$-bounded operator $H'$, with
\begin{equation}\label{H'}
H'=  N - g \Phi( \ii a h_0 ),
\end{equation}
on $\mathcal{D}(N)$ for all $P \in \mathbb{R}^d$ (in particular, $H'$ is independent of $P$). Furthermore, the quadratic form $ [H', \ii A]$ on $\mathcal{D}(N) \cap \mathcal{D}(A)$ extends to an $N$-bounded operator $H''$, with \begin{equation}\label{H''}
H''=  - g \Phi( a^2 h_0 ).
\end{equation}
on $\mathcal{D}( N )$. In particular, the hypothesis (4) of Appendix \ref{app:Mourre} is satisfied with $H= H(P)$ and $M=N$.
\end{lemma}
Note that it is the control on the second commutator of $H(P)$ with $A$ that imposes the condition $\mu>\frac12$ in this lemma. In particular it is not difficult to verify that $h_0 \in \mathcal{D}( a^2 )$ if $\mu > \frac12$, and hence $\Phi( \ii a h_0 )$ and $\Phi ( a^2 h_0 )$ in \eqref{H'} and \eqref{H''} are well-defined.
\begin{proof}
An explicit computation based on the expression of $W_t$ given in the previous subsection shows that
$
W_t H_0( P ) - H_0( P ) W_t = - t N ,
$
for all $t \ge 0$, as an identity on the set $\mathcal{B} ( \mathcal{G} ; \mathcal{G}^* )$ of bounded operators from $\mathcal{G}$ to $\mathcal{G}^*$. This already shows that $H_0(P) \in \mathrm{C}^1 ( A_{ \mathcal{G} } ; A_{ \mathcal{G}^* } )$ and that
\begin{equation}\label{eq:H'1}
[ H_0(P) , \ii A ]^0 = N.
\end{equation}
Since $N$ commutes with $W_t$ for all $t \ge 0$, the latter property actually implies that $H_0(P) \in \mathrm{C}^2 ( A_{ \mathcal{G} } ; A_{ \mathcal{G}^* } )$.

Next we have to consider the field operator $\Phi( h_0 )$. We compute
$
W_t \Phi( h_0 ) - \Phi( h_0 ) W_t = \Phi( ( w_t - \mathds{1} ) h_0 ) W_t.
$
Since $h_0 \in \mathcal{D}( a )$ under our assumptions (in particular because $\mu > - 1/2 $), we deduce that $\| ( w_t - \mathds{1} ) h_0 \|_{\mathrm{L}^2} \le C t$, and hence, by Lemma \ref{lm:appFock1}, that $\| \Phi( ( w_t - \mathds{1} ) h_0 ) ( N + \mathds{1} )^{-1/2} \| \le C t$. This implies that $\Phi( h_0 ) \in \mathrm{C}^1 ( A_{ \mathcal{G} } ; A_{ \mathcal{G}^* } )$. Moreover, using again the commutation relations of Appendix \ref{app:Fock}, we can compute
\begin{equation}\label{eq:H'2}
[ \Phi( h_0 ) , \ii A ]^0 = - \Phi( \ii a h_0 ).
\end{equation}
Equations \eqref{eq:H'1} and \eqref{eq:H'2} prove \eqref{H'}.

Likewise, the assumption that $\mu > 1/2 $ implies that $h_0 \in \mathcal{D}( a^2 )$ and we compute
\begin{equation*}
[ [ \Phi( h_0 ) , \ii A ] , \ii A]  = - \Phi( a^2 h_0 ).
\end{equation*}
This shows that $\Phi( h_0 ) \in \mathrm{C}^2 ( A_{ \mathcal{G} } ; A_{ \mathcal{G}^* } )$ and establishes \eqref{H''}, which concludes the proof of the lemma. 
\end{proof}
%
%

\subsubsection{The Mourre estimate}

We verify that for any $P \in \mathbb{R}^d$, a Mourre estimate holds for $H(P)$ with respect to the conjugate operator $A$.
\begin{lemma}\label{lm:Mourre_estimate}
Suppose that $\mu > - 1/2 $. There exists $g_c > 0$, $\mathrm{c}_0 > 0$ and $C >0$ such that, for all $| g | \le g_c$ and $P \in \mathbb{R}^d$,
\begin{equation}\label{Mourre}
H' = [ H(P) , \ii A ]^0 \ge \mathrm{c}_0 \mathds{1} - C \Pi_\Omega ,
\end{equation}
in the sense of quadratic forms on $\mathcal{D} ( H(P) ) \cap \mathcal{D}( N )$, where $\Pi_\Omega$ denotes the orthogonal projection onto the vacuum in $\mathcal{H}_{\mathrm{f}}$. 
In particular the hypothesis (2) of Appendix \ref{app:Mourre} holds, with $I = \mathbb{R}$, $H = H(P)$, $M = N$ and $K = \Pi_\Omega$.
\end{lemma}
\begin{proof}
From Lemma \ref{lm:regularity_A} we know that $H' = N - g \Phi( \ii a h_0 )$. Using Lemma \ref{lm:appFock1}, we deduce that
\begin{align*}
\langle \varphi , H' \varphi \rangle &\ge \langle \varphi , N \varphi \rangle - 2 |g| \| \varphi \| \| a h_0 \|_{2} \| N^{\frac12} \varphi \| \\
& \ge ( 1 - |g| \| a h_0 \|_{2} ) \langle \varphi , N \varphi \rangle - |g|\| a h_0 \|_{2} \langle \varphi , \varphi \rangle \\
& \ge ( 1 - 2 |g| \| a h_0 \|_{2} ) \langle \varphi , \varphi \rangle - ( 1 - |g| \| a h_0 \|_{2}) \langle \varphi , \Pi_\Omega \varphi \rangle ,
\end{align*}
for any $\varphi \in \mathcal{D}( H(P) ) \cap \mathcal{D}( N )$, where in the last inequality we used that $N \ge \mathds{1} - \Pi_\Omega$. Since $\Pi_\Omega$ is a one-dimensional projection, hence a compact operator, the lemma is proven, with $\mathrm{c}_0 = 1 - 2 |g| \| a h_0 \|_{2}$ and $C = 1 - |g| \| a h_0 \|_{2}$, assuming that $|g|$ is small enough.
\end{proof}
Now we can justify that the limiting absorption principle for $H(P)$ stated at the beginning of this section is indeed satisfied:
\begin{proof}[Proof of Theorem \ref{thm:LAPH(p)}]
$(i)$ From the proof of Lemma \ref{lm:Mourre_estimate}, we see that the constants $\mathrm{c}_0$ and $C$ in the Mourre estimate \eqref{Mourre} can be taken arbitrary close to $1$, provided that $|g|$ is small enough. Since $\mathrm{dim}( \mathrm{Ran} ( \Pi_\Omega ) ) = 1$, the virial theorem together with a standard argument (see e.g. \cite[Lemma 10]{HuSp95_01}) show that $H(P)$ has at most one eigenvalue counting multiplicity. Moreover, using the functional calculus, it is not difficult to verify that $\|\chi ( H(P) ) - \chi ( H_0(P) ) \| \le C_\chi |g|$ for any $\chi$ s.t. $\chi' \in \mathrm{C}_0^\infty( \mathbb{R})$. Next we can consider a function $\chi \in \mathrm{C}_0^\infty( \mathbb{R})$ such that $\chi \equiv 1$ on $\mathbb{R} \setminus ( P^2 - 1 , P^2 + 1 )$, $\chi \equiv 0$ on $[ P^2 - 1/2, P^2 + 1/2 ]$, and $C_\chi = \mathcal{O}(1)$ uniformly in $|g|$. Moreover $\chi( H_0(P) ) \Pi_\Omega = 0$ and therefore \eqref{Mourre} yields
\begin{equation*}
 H' \ge \mathrm{c}'_0 \mathds{1} - C_1 \chi^\perp(H(P))^2 ,
\end{equation*}
where $\mathrm{c}'_0>0$ provided that $|g|$ is small enough, $C_1 > 0$, and $\chi^\perp = \mathds{1} - \chi$. The virial theorem then implies that $H(P)$ does not have eigenvalue in $\mathbb{R} \setminus ( P^2 - 1 , P^2 + 1 )$, as claimed.

$(ii)$ Lemmas \ref{lm:hyp(1)}--\ref{lm:Mourre_estimate} show that the hypotheses (1)--(4) of Appendix \ref{app:Mourre} hold with $H=H(P)$ and $M=N$. Since in addition the Mourre estimate \eqref{Mourre} is uniform in $P \in \mathbb{R}^d$ and does not depend on the spectral interval $J$ that we consider, Theorem \ref{thm:LAP} implies the statement of Theorem \ref{thm:LAPH(p)}.
\end{proof}
%
%

\subsection{Absence of eigenvalues for the operator $H(P)$ if $P \neq 0$ and $g \neq 0$} \label{eigen}
Now we prove that the operator $H(P)$ has no eigenvalue if $P \neq 0$ and $g \neq 0$. 
\begin{theorem} \label{coi}
Let $\mu > 1/2$ and $\nu_1$, $\nu_2$ be such that $0 <\nu_1 < \nu_2$. There exists $g_c = g_c(\mu , \nu_1,\nu_2)>0$ such that, for all $0 < \vert g \vert \le g_c$ and $P \in \mathbb{R}^d$, $\vert P \vert  \in (\nu_1, \nu_2)$, 
\begin{equation*}
\sigma_{\mathrm{pp}}(H(P))=\emptyset.
\end{equation*} 
\end{theorem}
To prove this theorem, we mainly follow the strategy of \cite{FaMoSk11_01} with some simplifications due to the particular setting that we consider here. One of the main ideas is to introduce the self-adjoint operator 
\begin{equation*}
\overline{H}_{\alpha_J}(P):=H(P)+ \alpha_J \Pi_\Omega,
\end{equation*}
for any compact interval $J \subset  \sigma(H(P))=[E, \infty)$, with $\alpha_J \geq \sup J-\inf J$. The two key points are then the use of a limiting absorption principle for $\overline{H}_{\alpha_J}$ and the fact that Fermi's Golden Rule for $H(P)$ holds for all non-zero $P$'s. We begin with establishing these two auxiliary results before going into the proof of Theorem \ref{coi}.

\subsubsection{Fermi's Golden Rule and the limiting absorption principle}
Recall that for all  $P \in \mathbb{R}^d$, $P^2$ is a non-degenerate eigenvalue of the uncoupled Hamiltonian $H_0(P)$ associated with the Fock vacuum eigenvector $\Omega$. 
\begin{lemma}[Fermi Golden rule] \label{Fermi}
Suppose that $\hat{\rho}_1$ and $\hat{\rho}_2$ do not vanish. For all $P \in \mathbb{R}^d \setminus \{ 0 \}$, there exists $\mathrm{c}(P) > 0$ such that 
\begin{equation} \label{golden}
\Pi_\Omega \Phi( h_0 ) \mathrm{Im} \big ( ( H_0(P) - P^2 - \ii 0^+)^{-1}\bar \Pi_\Omega \big ) \Phi( h_0 ) \Pi_\Omega \ge \mathrm{c}(P) \Pi_\Omega ,
\end{equation}
where $\Pi_\Omega$ is the projection onto the Fock vacuum and $\bar \Pi_\Omega = \mathds{1} - \Pi_\Omega$.
\end{lemma}
\begin{center}
\begin{figure}[t]
\includegraphics[width=9cm, keepaspectratio]{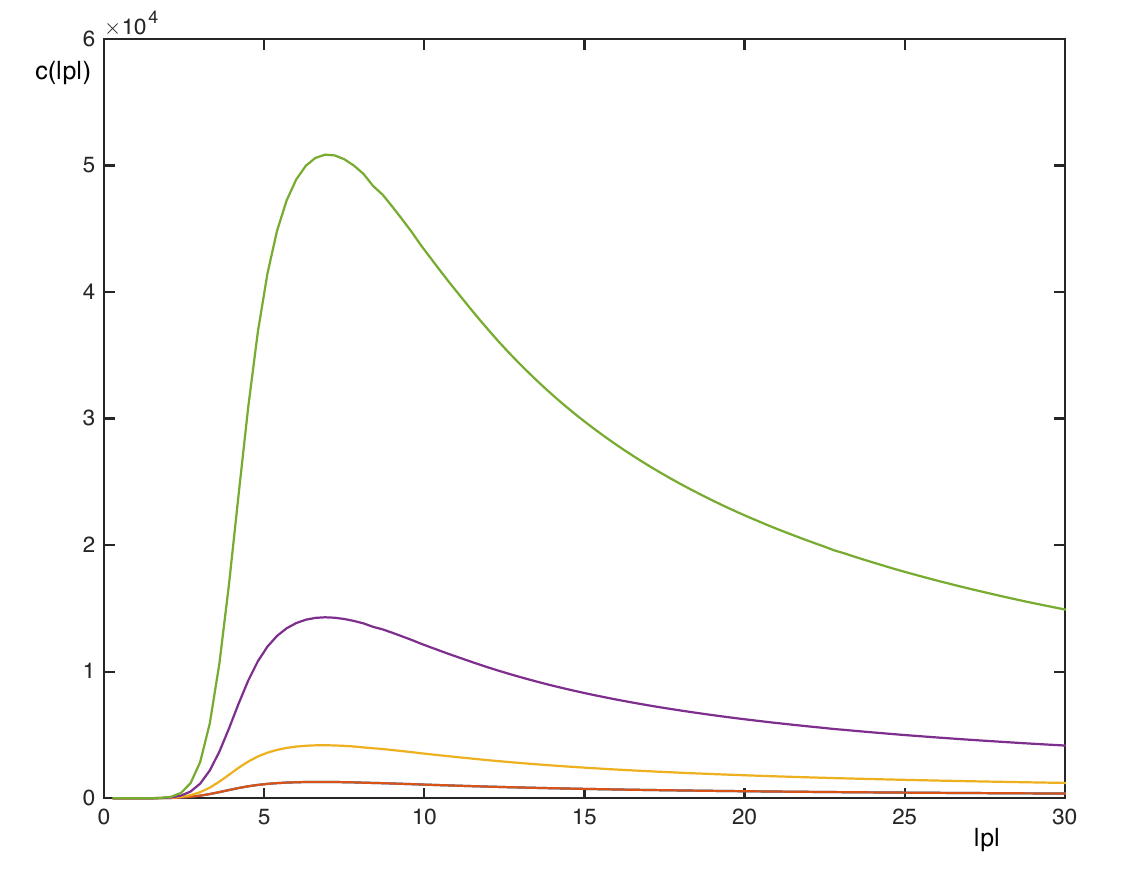}
\caption{\small Shape of the function $\mathrm{c}(P)$ in the Fermi Golden Rule \eqref{golden}  for different values of $\mu$ if the form factors $\hat{\rho}_1$ and $\hat{\rho}_2$ are gaussian.}
\end{figure}
\end{center}
\begin{proof}
We compute, for any $\varepsilon > 0$,
\begin{align*}
& \Pi_\Omega \Phi( h_0 ) ( H_0(P) - P^2 - \ii \varepsilon )^{-1}\bar \Pi_\Omega  \Phi( h_0 ) \Pi_\Omega \\
& = \Pi_\Omega a( h_0 ) ( H_0(P) - P^2 - \ii \varepsilon )^{-1}\bar \Pi_\Omega a^*( h_0 ) \Pi_\Omega \\
& = \int_{ \mathbb{R}^{3+d} } | h_0 ( \xi , k ) |^2 ( ( P - \xi )^2 - P^2 + |k| - \ii \varepsilon )^{-1} \Pi_\Omega \d k \d \xi \\
& = \int_{ \mathbb{R}^{3+d} } | k |^{2\mu} | \hat{ \rho }_1 ( | \xi | ) |^2 | \hat{\rho}_2( | k | ) |^2 ( - 2 P \cdot \xi  + \xi^2 + |k| - \ii \varepsilon )^{-1} \Pi_\Omega \d k \d \xi .
\end{align*}
Going to polar coordinates gives
\begin{align*}
& \Pi_\Omega \Phi( h_0 ) ( H_0(P) - P^2 - \ii \varepsilon )^{-1}\bar \Pi_\Omega  \Phi( h_0 ) \Pi_\Omega \\
& = 4 \pi^2 \int_{ \mathbb{R}^d \times \mathbb{R}^+ } \omega^{2 + 2\mu} | \hat{ \rho }_1 ( | \xi | ) |^2 | \hat{\rho}_2( \omega ) |^2 ( - 2 P \cdot \xi  + \xi^2 + \omega - \ii \varepsilon )^{-1} \Pi_\Omega \d \xi \d \omega .
\end{align*}
Taking the imaginary part and letting $\varepsilon \to 0$, we thus obtain
\begin{align*}
& \Pi_\Omega \Phi( h_0 ) \mathrm{Im} \big ( ( H_0(p) - P^2 - \ii 0^+ )^{-1} \bar \Pi_\Omega \big ) \Phi( h_0 ) \Pi_\Omega \\
& = 4 \pi \int_{ \mathbb{R}^d \times \mathbb{R}^+ } \omega^{2 + 2\mu} | \hat{ \rho }_1 ( | \xi | ) |^2 | \hat{\rho}_2( \omega ) |^2 \delta ( - 2 P \cdot \xi  + \xi^2 + \omega ) \Pi_\Omega \d \xi \d \omega \\
& = 4 \pi \int_{ \mathbb{R}^d } ( 2 P \cdot \xi  - \xi^2 )^{2 + 2\mu} | \hat{ \rho }_1 ( | \xi | ) |^2 | \hat{\rho}_2( 2 P \cdot \xi  - \xi^2 ) |^2 \mathds{1}_{ [0,\infty) }( 2 P \cdot \xi  - \xi^2 )  \Pi_\Omega \d \xi .
\end{align*}
The integral above vanishes for $P = 0$ but is strictly positive otherwise since we have assumed that $\mathrm{supp}( \hat{ \rho }_1 ) = \mathrm{supp}( \hat{\rho}_2 ) = [ 0 , \infty )$. This proves Lemma \ref{Fermi}.
\end{proof}

Next, we have the following limiting absorption principle for the operator $\overline{H}_{\alpha_J}(P)$.
\begin{lemma} \label{limiting}
Let  $J \subset \mathbb{R}$ be a compact interval such that $J \cap \sigma_{\mathrm{pp}}(H_0(P))= \{P^2\}$. As above we set 
$
\mathcal{S}:=\{z \in \mathbb{C} \mid \mathrm{Re}(z) \in J,  \vert \mathrm{Im}(z)  \vert \in (0,1] \}. 
$
Let $\mu > 1/2$ and $\alpha_J \ge \sup J-\inf J+1$. For all $s \in (1/2, 1]$, there is $g_c>0$ such that, for all $\vert g \vert \le g_c$ and $P \in \mathbb{R}^d$,
\begin{equation*}  
    \sup_{z\in \mathcal S} \| \langle A \rangle^{-s} (\overline{H}_{\alpha_J}(P)-z)^{-1} \langle A \rangle^{-s}\|<\infty ,
\end{equation*}
 with $\langle A \rangle = ( 1 + A^* A )^{1/2}$. Furthermore, for all $g,g' \in [-g_c,g_c]$ and $z,z' \in \mathcal{S}$,
 \begin{equation}  \label{eq:epsilon2}
\| \langle A \rangle^{-s} \big((\overline{H}'_{\alpha_J}(P)-z')^{-1} - (\overline{H}_{\alpha_J}(P)-z)^{-1} \big) \langle A \rangle^{-s}\| \leq C (\vert g-g' \vert^{s-\frac12} + \vert z-z' \vert^{s-\frac12}), 
\end{equation}
where $\overline{H}'_{\alpha_J}=H_{0}(P)+g' \Phi(h_0) + \alpha_J \Pi_\Omega$.
\end{lemma}
\begin{proof}
The proof follows resembles that of Theorem \ref{thm:LAPH(p)}, replacing $H_0(P)$ by $H_0(P) + \alpha_J \Pi_\Omega$ and observing that a Mourre estimate of the form of the hypothesis (2) of Appendix \ref{app:Mourre} holds without compact remainder (i.e. $K=0$). The main difference is the Hölder continuity in the coupling constant $g$ stated in \eqref{eq:epsilon2}. However, as explained in \cite{FaMoSk11_01}, the latter follows in the same way as the Hölder continuity in the spectral parameter $z$.
\end{proof}
%
%

\subsubsection{Proof of Theorem \ref{coi}}
We follow closely \cite{FaMoSk11_01} and do a proof by contradiction. By Theorem \ref{thm:LAPH(p)} we know that $H(P)$ does not have eigenvalue in $\mathbb{R} \setminus ( P^2 - 1 , P^2 + 1)$. Let $J = [P^2 - 1 , P^2 + 1 ]$ and let $|g|$ be sufficiently small as in the statement of Lemma \ref{limiting}. Assume that $H(P)$ has an eigenvector $\varphi_g \in \mathcal{D}( H(P) )$, $\| \varphi_g \| = 1$, with eigenvalue $\lambda_g \in J$. By definition of $\overline{H}_{\alpha_J}(P)$, one has that
\begin{equation} \label{equ1}
(\overline{H}_{\alpha_J}(P)- \lambda_g) \varphi_g= \alpha_J \Pi_\Omega \varphi_g.
\end{equation}
The limiting absorption principle stated in Lemma \ref{limiting} implies that $\lambda_g$ cannot be an eigenvalue of   $\overline{H}_{\alpha_J}(P)$ and, therefore, that $\Pi_\Omega \varphi_g \neq 0$.
Applying the operator $\Pi_\Omega (\overline{H}_{\alpha_J}(P)- \lambda_g - \ii \varepsilon)^{-1}$ on both sides of \eqref{equ1}, we deduce that 
\begin{equation*}
\Pi_\Omega \varphi_g =  \alpha_J \Pi_\Omega  (\overline{H}_{\alpha_J}(P)- \lambda_g - i \varepsilon)^{-1}\Pi_\Omega\varphi_g   - \ii \varepsilon \Pi_\Omega  (\overline{H}_{\alpha_J}(P)- \lambda_g - \ii \varepsilon)^{-1 }\varphi_g.
\end{equation*}
Using the spectral theorem, it is not difficult to verify that the second term on the right side goes to zero as $\varepsilon \to 0$. Hence
\begin{equation*}
\Pi_\Omega \varphi_g =  \lim_{\varepsilon \rightarrow 0^+} \alpha_J \Pi_\Omega  (\overline{H}_{\alpha_J}(P)- \lambda_g - \ii \varepsilon)^{-1}\Pi_\Omega\varphi_g .
\end{equation*}
Note that the limit exists by Lemma \ref{limiting}. Taking the scalar product with $\Pi_\Omega \varphi_g$ and using    the second resolvent formulae
\begin{align*}
 \big (\overline{H}_{\alpha_J}(P)- \lambda_g - \ii \varepsilon \big )^{-1}= &\big ( \overline{H}_{0,\alpha_J}(P)- \lambda_g - \ii \varepsilon \big )^{-1} \\
 &- g \big ( \overline{H}_{\alpha_J}(P)- \lambda_g - \ii \varepsilon \big )^{-1} \Phi(h_0) \big ( \overline{H}_{0,\alpha_J}(P)- \lambda_g - \ii \varepsilon \big )^{-1},  \\
 \big ( \overline{H}_{\alpha_J}(P)- \lambda_g - \ii \varepsilon \big )^{-1} =& \big ( \overline{H}_{0,\alpha_J}(P)- \lambda_g - \ii \varepsilon \big )^{-1}  \\
 &- g \big ( \overline{H}_{0,\alpha_J}(P)- \lambda_g - \ii \varepsilon \big )^{-1}  \Phi(h_0)  \big ( \overline{H}_{\alpha_J}(P)- \lambda_g - \ii \varepsilon \big )^{-1} , 
\end{align*}
where $\overline{H}_{0,\alpha_J}(P) := H_0(P) + \alpha_J \Pi_\Omega$, one arrives at 
\begin{align*}
  \| \Pi_\Omega   \varphi_g \| ^2 =& \frac{ \alpha_J \| \Pi_\Omega \varphi_g \|^2 }{P^2- \lambda_g+ \alpha_J}\\
 &  - \frac{ g^2 \alpha_J }{(P^2- \lambda_g+ \alpha_J)^2} \lim_{\varepsilon \rightarrow 0^+} \big \langle \Pi_\Omega \varphi_g \vert    \Phi(h_0) \big ( \overline{H}_{\alpha_J}(P)- \lambda_g - \ii \varepsilon \big )^{-1}    \Phi(h_0) \Pi_\Omega \varphi_g \big \rangle\\
 =& \frac{ \alpha_J  \| \Pi_\Omega \varphi_g \|^2 }{P^2- \lambda_g+ \alpha_J} \left(1 +  \frac{g^2}{P^2- \lambda_g+ \alpha_J} \lim_{\varepsilon \rightarrow 0^+} \big \langle \Omega \vert \Phi(h_0) \big ( \overline{H}_{0,\alpha_J}(P)- \lambda_g -  \ii \varepsilon \big )^{-1}  \Phi(h_0) \Omega \big \rangle \right) \\
 &   + \frac{ g^2 \alpha_J }{(P^2- \lambda_g+ \alpha_J)^2} \lim_{\varepsilon \rightarrow 0^+} \big \langle \Phi(h_0) \Pi_\Omega \varphi_g \vert  \Big [  \big ( \overline{H}_{\alpha_J}(P) - \lambda_g - \ii \varepsilon \big )^{-1} \\
 &\qquad \qquad \qquad \qquad \qquad \qquad \qquad \qquad - \big ( \overline{H}_{0,\alpha_J}(P)- \lambda_g - \ii \varepsilon \big )^{-1} \Big]   \Phi(h_0) \Pi_\Omega \varphi_g \rangle.
\end{align*}
Using Lemma \ref{limiting} together with the fact that $\Phi( h_0 ) \Omega \in \mathcal{D}( A )$, it is not difficult to verify that the limits above exist. We set 
\begin{align*}
F_{\alpha_J}(P)&:= \big \langle \Omega \vert \Phi(h_0) \big ( \overline{H}_{0,\alpha_J}(P)- \lambda_g -  \ii 0^+ \big )^{-1}  \Phi(h_0) \Omega \rangle , \\
 G_{\alpha_J}(P)&:= g^{-\frac12 } \langle  \Omega  \vert    \Phi(h_0) \left[ \big ( \overline{H}_{\alpha_J}(P)- \lambda_g - \ii 0^+ \big )^{-1} - \big ( \overline{H}_{0,\alpha_J}(p)- \lambda_g - \ii 0^{+} \big )^{-1} \right]   \Phi(h_0) \Omega \big \rangle.
\end{align*}
Since $\Pi_\Omega \Phi( h_0 ) \Pi_\Omega = 0$, we can rewrite
\begin{equation*}
F_{\alpha_J}(P) = \big \langle \Omega \vert \Phi(h_0) \big ( H_0(P)- \lambda_g -  \ii 0^+ \big )^{-1}  \Phi(h_0) \Omega \rangle =: F(P) .
\end{equation*}
Gathering all terms together and using that $ \| \Pi_\Omega   \varphi_g \| \neq 0$, we finally obtain that 
\begin{equation*}
1=  \frac{ \alpha_J }{P^2- \lambda_g+ \alpha_J} \left( 1 + g^2  \frac{1}{P^2- \lambda_g+ \alpha_J} F(P) +  g^{\frac52}  \frac{1}{ P^2- \lambda_g+ \alpha_J  } G_{\alpha_J}(P) \right).
\end{equation*}
Taking the imaginary part gives
\begin{equation} \label{impossible}
\mathrm{Im} (F(P)) = -g^{\frac12} \mathrm{Im}(G_{\alpha_J}(P) ).
\end{equation}
By Lemma \ref{Fermi},  $\mathrm{Im} (F(P)) = \mathrm{c}(P) > 0$, and by Lemma \ref{limiting}, $\mathrm{Im}(G_{\alpha_J}(P) )$ is uniformly bounded in $|g| \le g_c$ and $P$. Therefore \eqref{impossible} cannot hold and $\lambda_g$ cannot be an eigenvalue of $H(P)$.

\appendix

\section{Estimates of operators in Fock space and commutation relations}\label{app:Fock}
\subsection{Estimates of operator in Fock spaces}

We recall some standard estimates and commutation relations that have been used several times in the main text.

\subsubsection{Basic relative bounds}
Consider the subspace $\mathcal{D} ( \omega^{-1/2} ) \subset \mathrm{L}^2( \mathbb{R}^{3+d} )$, equipped with the norm
\begin{equation*}
\| h \|^2_{ \omega^{-1/2} } := \int_{ \mathbb{R}^{3+d} } ( 1 + \omega(k)^{-1} ) | h ( k , \xi ) |^2 \d k \d \xi ,
\end{equation*}
The following lemma is standard (see for instance \cite[Lemma 17]{FrGrSc01_01}).
\begin{lemma}\label{lm:appFock1}
$\quad$
\begin{itemize}[leftmargin=30pt, itemsep=5pt]
\item[(a)]  Let $f_i \in \mathrm{L}^2( \mathbb{R}^{3+d} )$, $i=1,\dots,n$. Then
\begin{align*}
\big \| a^\sharp(f_1) \cdots a^\sharp(f_n) ( N+\mathds{1} )^{- \frac{n}{2} } \big\| \le C_n \| f_1 \|_2 \dots \| f_n \|_2,   
\end{align*}
where $a^\sharp$ stands for $a$ or $a^*$.
\item[(b)] Let $f_i \in \mathcal{D} ( \omega^{-1/2} )$, $i=1,\dots,n$. Then
\begin{align*}
\big \| a^\sharp(f_1) \cdots a^\sharp(f_n) ( \d \Gamma( \omega ( k ) ) + \mathds{1} )^{- \frac{n}{2} } \big\| \le C_n  \| f_1 \|_{ \omega^{-1/2} } \dots \| f_n \|_{ \omega^{-1/2} } .
\end{align*}
\end{itemize}
\end{lemma}
We also recall the following lemma. For a proof, see for instance \cite[Section 3]{GeGeMo04_01}.
\begin{lemma}\label{lm:appFock2}
Let $b_1$ and $b_2$ be two self-adjoint operators on $\mathrm{L}^2( \mathbb{R}^{3+d} )$ such that $b_2 \ge 0$, $\mathcal{D} (b_2) \subset \mathcal{D} (b_1)$ and $\| b_1 \varphi \|_2 \le \| b_2  \varphi \|_2$ for all $\varphi \in \mathcal{D} (b_2)$. Then $\mathcal{D} ( \d\Gamma( b_2 ) ) \subset \mathcal{D}( \d\Gamma( b_1 ) )$ and we have that $\| \d \Gamma( b_1 ) \Phi \| \le \| \d \Gamma( b_2 ) \Phi \|$ for all $\Phi \in \mathcal{D} ( \d\Gamma( b_2 ) )$.
\end{lemma}
%
%

\subsubsection{Commutator estimates}
We recall the commutation relations between basic operators acting in Fock space (see \cite{DeGe99_01} for more details). As usual, the commutator $[ B_1 , B_2 ]$ of two operators $B_1$ and $B_2$ is defined as a quadratic form on $\mathcal{D}( B_1 ) \cap \mathcal{D}( B_2 )$. If the corresponding quadratic form admits an extension to a closed quadratic form on a larger domain, the associated operator is then denoted by the same symbol $[B_1, B_2 ]$.

Given two self-adjoint operators $b_1$ and $b_2$ on $\mathrm{L}^2( \mathbb{R}^{3+d} )$, we have that
\begin{align}\label{eq:comm1}
[\d \Gamma ( b_1 ) , \d \Gamma ( b_2 )]  = \d \Gamma ( [ b_1 , b_2 ] ) ,
\end{align}
Likewise, if $b$ is a self-adjoint operator on $\mathrm{L}^2( \mathbb{R}^{3+d} )$ and $f,g \in \mathrm{L}^2( \mathbb{R}^{3+d} )$, recalling the definition $\Phi ( f ) =  a^{*} ( f ) + a (f )$, we have that
\begin{align}
&[ \Phi (f) , \Phi (g) ] = 2\ii \im \langle f , g \rangle, \label{eq:comm2} \\
& [ \Phi (f) , \d \Gamma ( b ) ] = \ii \Phi ( \ii b f ) . \label{eq:comm3}
 \end{align}
The commutator of an operator of the form $\Gamma ( b_1 )$ with $\d \Gamma( b_2 )$ is given by
\begin{equation}
[ \Gamma ( b_1 ) , \d \Gamma( b_2 ) ]  = \d \Gamma( b_1 , [ b_1 , b_2 ] ) , \label{eq:comm4}
\end{equation}
where, if $c_1$, $c_2$ are two operators on $\mathrm{L}^2( \mathbb{R}^{3+d} )$, the operator $\d \Gamma( c_1 , c_2 )$ on $\mathcal{H}_{\mathrm{f}}$ is defined by its restriction to ${\otimes_s^l} \mathrm{L}^2( \mathbb{R}^{3+d} ) $ as
\begin{align*}
&\d \Gamma( c_1, c_2 )\vert_\mathbb{C} := 0 ,  \\
& \d \Gamma( c_1, c_2 )\vert_{  {\otimes_s^l} \mathrm{L}^2( \mathbb{R}^{3+d} ) } := \sum_{j=1}^{l} \underbrace{c_1 \otimes \cdots \otimes c_1 }_{j-1} \otimes c_2 \otimes \underbrace{c_1 \otimes \cdots \otimes c_1 }_{n - j} . 
\end{align*}
Finally if $b$ is a bounded self-adjoint operator on $\mathrm{L}^2( \mathbb{R}^{3+d} )$ and $f \in \mathrm{L}^2( \mathbb{R}^{3+d} )$,
\begin{align}
& [ \Gamma( b ) , a^*( f ) ] = a^*( ( b - \mathds{1} ) f ) \Gamma( b ),  \label{eq:comm5}\\
& [ \Gamma( b ) , a( f ) ] = \Gamma( b ) a( ( \mathds{1} - b ) f ) . \label{eq:comm6}
\end{align} 

The following result was used in the proof of Theorem \ref{prop:loc_spectrum}.
\begin{lemma}\label{lm:appFock3}
Let $\tilde{j}_0 \in \mathrm{C}_0^\infty( \mathbb{R} ; [ 0 , 1 ] )$ be such that $\mathrm{supp} ( \tilde{j}_0 ) \subset [0,1]$, $\tilde j_0 \equiv 1$ on $[0,1/2]$, and let $\tilde{\mathbf{j}}_0 = \tilde{j}_0( | x |/ R )$ with $R > 1$. Let $\mu > -1$, and $g \in \mathbb{R}$. For all $\varphi \in \mathcal{D} ( N )$, we have that
\begin{equation}
\big \| [ \d \Gamma( \xi_j ) , \Gamma( \tilde{\mathbf{j}}_0 ) ] \varphi \big \| \le \frac{ C }{ R } \| ( N + \mathds{1} ) \varphi \| , \quad j = 1 , 2 ,3 , \label{eq:comm7}
\end{equation}
and for all $\varphi \in \mathcal{D}( \d \Gamma( \xi )^2 ) \cap \mathcal{D} ( N^2 ) \cap \mathcal{D}( \d \Gamma( |x|^{-1} ) )$, 
\begin{equation}
\big \| [ H( 0 ) , \Gamma( \tilde{\mathbf{j}}_0 ) ] \varphi \big \| \le \frac{ C }{ R } \| ( \d\Gamma( \xi )^2 + N + \d \Gamma( |x|^{-1} ) + \mathds{1} ) \varphi \| . \label{eq:comm8}
\end{equation}
\end{lemma}
\begin{proof}
We compute by \eqref{eq:comm4}
\begin{equation*}
[ \d \Gamma( \xi_j ) , \Gamma( \tilde{\mathbf{j}}_0 ) ] = \d \Gamma ( \tilde{\mathbf{j}}_0 , [ \xi_j , \tilde{\mathbf{j}}_0 ] ) .
\end{equation*}
Since $\xi_j = - i \partial_{x_j}$ and $\tilde{\mathbf{j}}_0 = \tilde{j}_0( | x |/ R )$, we see that $[ \xi_j , \tilde{\mathbf{j}}_0 ]$ is a bounded operator in $\mathrm{L}^2( \mathbb{R}^{3+d})$ with $\| [ \xi_j , \tilde{\mathbf{j}}_0 ] \| \le C / R$. Using in addition that $\| \tilde{\mathbf{j}}_0 \| = 1$, it is not difficult to deduce that \eqref{eq:comm7} holds.

Newt we prove \eqref{eq:comm8}. Using \eqref{eq:comm4}, \eqref{eq:comm5} and \eqref{eq:comm6}, we obtain
\begin{align*}
[ H( 0 ) , \Gamma( \tilde{\mathbf{j}}_0 ) ] =& [ \d \Gamma( \xi ) , \Gamma( \tilde{\mathbf{j}}_0 ) ] \cdot \d \Gamma( \xi ) + \d\Gamma( \xi )\cdot [\d \Gamma( \xi ) , \Gamma( \tilde{\mathbf{j}}_0 ) ] \\
& + g a^*( ( \mathds{1} - \tilde{\mathbf{j}}_0 ) h_0 ) \Gamma( \tilde{\mathbf{j}}_0 ) + g \Gamma( \tilde{\mathbf{j}}_0 ) a( ( \tilde{\mathbf{j}}_0 - \mathds{1} ) h_0 ) \\
=& 2 [ \d \Gamma( \xi ) , \Gamma( \tilde{\mathbf{j}}_0 ) ] \cdot \d \Gamma( \xi ) + \sum_j [\d\Gamma( \xi_j ), [\d \Gamma( \xi_j ) , \Gamma( \tilde{\mathbf{j}}_0 ) ] ] \\
& + g a^*( ( \mathds{1} - \tilde{\mathbf{j}}_0 ) h_0 ) \Gamma( \tilde{\mathbf{j}}_0 ) + g \Gamma( \tilde{\mathbf{j}}_0 ) a( ( \tilde{\mathbf{j}}_0 - \mathds{1} ) h_0 ) .
\end{align*}
The first term is estimated using \eqref{eq:comm7}. Since in addition $N$ commutes with $\d \Gamma( \xi )$, we can write
\begin{align*}
\big \| [ \d \Gamma( \xi ) , \Gamma( \tilde{\mathbf{j}}_0 ) ] \cdot \d \Gamma( \xi ) \varphi \| \le \frac{C}{R} \sum_j \big \| (N+\mathds{1} ) \d \Gamma( \xi_j ) \varphi \big \| \le \frac{C}{R} \big \| \big ( N^2 + \d \Gamma( \xi )^2 + \mathds{1} \big ) \varphi \big \|.
\end{align*}
To estimate the second term, we have to compute the expression of the second commutator $[\d\Gamma( \xi_j ), [\d \Gamma( \xi_j ) , \Gamma( \tilde{\mathbf{j}}_0 ) ] ]$. In view of \eqref{eq:comm4}, it is not difficult to verify that
\begin{align*}
\big \| [\d\Gamma( \xi_j ), [\d \Gamma( \xi_j ) , \Gamma( \tilde{\mathbf{j}}_0 ) ] ] \varphi \| \le \frac{C}{R} \big \| \big ( \d \Gamma( |x|^{-1} ) + N^2+ \mathds{1} \big ) \varphi \big \|.
\end{align*}
Finally the last two terms are estimated using \eqref{eq:a18} and Lemma \ref{lm:appFock1}. Since $N$ commutes with $\Gamma( \tilde{\mathbf{j}}_0 )$ and since $\Gamma( \tilde{\mathbf{j}}_0 )$ is a contraction, this gives
\begin{align*}
& \big \| a^*( ( \mathds{1} - \tilde{\mathbf{j}}_0 ) h_0 ) \Gamma( \tilde{\mathbf{j}}_0 ) \varphi \big \| \le \frac{C}{R} \| ( N^{\frac12} + \mathds{1} ) \varphi \| , \quad \big \| \Gamma( \tilde{\mathbf{j}}_0 ) a( ( \tilde{\mathbf{j}}_0 - \mathds{1} ) h_0 ) \varphi \big \| \le \frac{C}{R} \| ( N^{\frac12} + \mathds{1} ) \varphi \|.
\end{align*}
Combining the previous estimates proves the lemma.
\end{proof}
%
%

\subsection{Partition of unity in Fock space}
Now we consider the operator $\check{\Gamma}( \mathbf{j} ) : \mathcal{H}_{\mathrm{f}} \to \mathcal{H}_{\mathrm{f}} \otimes \mathcal{H}_{\mathrm{f}}$ defined at the beginning of Section \ref{sec:ess_spectrum}. It follows from our choice of the maps $\mathbf{j}_0$, $\mathbf{j}_\infty$ that $\check{\Gamma}( \mathbf{j} )$ is  an isometry,
$
\check{\Gamma}( \mathbf{j} )^* \check{\Gamma}( \mathbf{j} ) = \mathds{1} .
$
 If $b$ is a self-adjoint operator on $\mathrm{L}^2( \mathbb{R}^{3+d} )$, one can show by a direct computation that
\begin{align}\label{eq:AppA2}
\d \Gamma( b ) &= \check{\Gamma}( \mathbf{j} )^* \big ( \d \Gamma( b ) \otimes \mathds{1} + \mathds{1} \otimes \d \Gamma( b ) \big ) \check{\Gamma}( \mathbf{j} ) + \frac12 \d \Gamma( [ \mathbf{j}_0 , [ \mathbf{j}_0 , b ] ] + [ \mathbf{j}_\infty , [ \mathbf{j}_\infty , b ] ] ) ,
 \end{align}
and likewise, for any $h \in \mathrm{L}^2( \mathbb{R}^{3+d} )$,
\begin{align}\label{eq:AppA3}
\Phi( h ) &= \check{\Gamma}( \mathbf{j} )^* \big ( \Phi( \mathbf{j}_0 h ) \otimes \mathds{1} + \mathds{1} \otimes \Phi( \mathbf{j}_\infty h) \big ) \check{\Gamma}( \mathbf{j} ) .
 \end{align}
We can then prove Lemma \ref{lm:decomp_in_Fock}.

\begin{proof}[Proof of Lemma \ref{lm:decomp_in_Fock}]
It follows from \eqref{eq:AppA2} and \eqref{eq:AppA3} that
\begin{align*}
H( P ) &= \big ( \check{\Gamma}( \mathbf{j} )^* \big ( - \d\Gamma( \xi ) \otimes \mathds{1} + \mathds{1} \otimes ( P -  \d\Gamma( \xi ) ) \big ) \check{\Gamma}( \mathbf{j} ) + \frac12 \d \Gamma( [ \mathbf{j}_0 , [ \mathbf{j}_0 , \xi ] ] + [ \mathbf{j}_\infty , [ \mathbf{j}_\infty , \xi ] ] ) \big )^2 \notag \\
& \quad + \check{\Gamma}( \mathbf{j} )^* \big ( \d\Gamma( \omega(k) ) \otimes \mathds{1} + \mathds{1} \otimes \d\Gamma( \omega(k) )  \big ) \check{\Gamma}( \mathbf{j} ) + \frac12 \d \Gamma( [ \mathbf{j}_0 , [ \mathbf{j}_0 , \omega(k) ] ] + [ \mathbf{j}_\infty , [ \mathbf{j}_\infty , \omega(k)  ] ] ) \notag \\
& \quad + g \check{\Gamma}( \mathbf{j} )^* \big ( \Phi( \mathbf{j}_0 h_0 ) \otimes \mathds{1} + \mathds{1} \otimes \Phi( \mathbf{j}_\infty h_0 ) \big ) \check{\Gamma}( \mathbf{j} ) . 
\end{align*}
Since $\mathbf{j}_0$ and $\mathbf{j}_\infty$ commute with $\omega(k)$, we have that $[ \mathbf{j}_0 , [ \mathbf{j}_0 , \omega(k) ] ] + [ \mathbf{j}_\infty , [ \mathbf{j}_\infty , \omega(k)  ] ] = 0$. Moreover, since $\xi = -\ii \nabla_x$ and $\mathbf{j}_0$, $\mathbf{j}_\infty$ are multiplication operator in the variable $x$, we also have that $[ \mathbf{j}_0 , [ \mathbf{j}_0 , \xi ] ] + [ \mathbf{j}_\infty , [ \mathbf{j}_\infty , \xi ] ] = 0$. Therefore the lemma is proven.
\end{proof}

\section{Absence of a ground state for $H(P)$ if $\mu \le -1/2$}\label{app:absence}

We prove Proposition~\ref{prop:nogroundstate} by adapting an argument of \cite{DeGe04_01}. The argument goes by contradiction. Suppose that $\psi(P)$ is a normalized ground state of $H(P)$, namely $H(P) \psi(P) = E_g \psi(P)$ with $\| \psi(P) \| =1$. The pull-through formula \eqref{eq:pullthrough} gives
\begin{align*}
( H ( P + \xi ) - E_g + |k| ) a( k , \xi ) \psi( P ) = - g h_0( k , \xi ) \psi( P ) ,
\end{align*}
where $h_0$ is the coupling function defined in \eqref{eq:coupling}. Since $E_g = \inf \sigma( H( P + \xi ) )$ for any $\xi \in \mathbb{R}^d$ according to (i) of Theorem \ref{thm:main}, we deduce that for a.e. $( k , \xi ) \in ( \mathbb{R}^3 \setminus \{ 0 \} ) \times \mathbb{R}^d$,
\begin{align*}
a( k , \xi ) \psi( P ) = - g h_0( k , \xi ) ( H ( P + \xi ) - E_g + |k| )^{-1} \psi( P ) .
\end{align*}
In other words, writing $\psi(P) = ( \psi(P)^{(0)} , \psi(P)^{(1)} , \dots ) \in \mathcal{H}_{\mathrm{f}}$, we have that
\begin{align}\label{eq:psip1}
\psi( P )^{(1)}( k , \xi ) = - g h_0( k , \xi ) \big ( ( H ( P + \xi ) - E_g + |k| )^{-1} \psi( P ) \big )^{(0)} ,
\end{align}
and, for all $n \in \mathbb{N}$,
\begin{align}
& \sqrt{n+1} \psi( P )^{(n+1)}( k , \xi , k_1 , \xi_1 , \dots , k_n , \xi_n ) \notag \\
&= - g h_0( k , \xi ) \big ( ( H ( P + \xi ) - E_g + |k| )^{-1} \psi( P ) \big )^{(n)} ( k_1 , \xi_1 , \dots , k_n , \xi_n ) . \label{eq:psip2}
\end{align}

From \eqref{eq:psip1} and the expression \eqref{eq:coupling} of $h_0$, we deduce that for a.e. $k \in \mathbb{R}^3 \setminus \{ 0 \}$, the map $\xi \mapsto \psi( P )^{(1)}( k , \xi )$ is continuous on $\mathbb{R}^d$. Moreover,
\begin{align*}
\psi( P )^{(1)}( k , 0 ) = - g |k|^{-1} h_0( k , 0 ) \psi( P)^{(0)} = - g |k|^{-1 + \mu} \hat{\rho}_2( |k|) \hat{\rho}_1( 0 ) \psi( P )^{(0)}  .
\end{align*}
Since $k \mapsto \psi( P )^{(1)}( k , 0 ) \in \mathrm{L}^2( \mathbb{R}^3 )$ but $k \mapsto |k|^{-1 + \mu} \hat{\rho}_2( |k|) \notin \mathrm{L}^2( \mathbb{R}^3 )$ (because $\mu \le -1/2$ and $\hat\rho_2(0)\not=0$), and since $\hat{\rho}_1( 0 ) \neq 0$, , this implies that $\psi( P )^{(0)} = 0$.

Proceeding in the same way, one verifies using \eqref{eq:psip2} with $\xi = 0$ that $\psi(P)^{(n)} = 0$ for all $n \in \mathbb{N} \cup \{ 0 \}$. This contradicts the fact that $\| \psi(P) \| = 1$.

\section{Complements to Section \ref{S4}}\label{extras}
 
\subsection{Extraction of $w_{0,0}(z,r,l)$ using one-particle states} 
\label{extrac}
We begin this appendix by explaining how to extract $w_{0,0}(z,r,l)$ from an operator $H(z)$ given as a series of Wick monomials on $\mathcal{H}$, the kernels being continuous functions in $(r,l)$.  This is of some importance, in particular to prove analyticity in $z$ of $w_{0,0}(z,r,l)$, and rotation-invariance in $\xi$-space. To do so, we just choose a sequence of one-particle states that concentrate on a point  in $(r,l)$-space (i.e. converge weakly to $\delta(r-r_0) \delta(l-l_0)$ for a given pair $(r_0,l_0)$). To do so, let $f: \mathbb{R} \rightarrow \mathbb{R}$, $h : \mathbb{R}^d \rightarrow \mathbb{R}$ be  smooth functions with compact supports and such that $\|f \|_{\mathrm{L}^2( \mathbb{R} )} = \| h \|_{ \mathrm{L}^2( \mathbb{R}^d ) } = 1$. Consider the sequence of functions $(f_n)$ and $(h_n)$, defined by 
\begin{equation*}
f_n(r)=n^{1/2} f(n(r-r_0)), \qquad h_n(l)=n^{d/2} h(n(l-l_0)).
\end{equation*}
We introduce the sequence of $1$-particle states
\begin{equation*}
\psi^{(n)}:=\int_{\mathbb{R}^{3+d}} f_n(\vert k \vert) h_n(\xi)  a^*(k,\xi) \Omega \d k \d \xi .
\end{equation*}
We show that 
\begin{equation*}
\langle \psi^{(n)} \vert H(z) \psi^{(n)} \rangle \rightarrow w_{0,0}(z,r_0,l_0) + \mathcal{E}(z)
\end{equation*}
as $n$ tends to $\infty$. Since $\psi^{(n)}$ is a one-particle state,
\begin{equation*}
\langle \psi^{(n)} \vert H(z) \psi^{(n)} \rangle = \mathcal{E}(z) + \langle \psi^{(n)}\vert  w_{0,0}(z,\d \Gamma( |k| ),\d\Gamma(\xi)) \psi^{(n)} \rangle + \langle \psi^{(n)}\vert W_{1,1}(z) \psi^{(n)} \rangle. 
\end{equation*}
Using the pull-trhough formula, one obtains that 
\begin{align*}
 \langle \psi^{(n)}\vert   & w_{0,0}(z,\d \Gamma( |k| ),\d\Gamma(\xi)) \psi^{(n)} \rangle =\int_{\mathbb{R}^{3+d}} \vert f_n(\vert k \vert)\vert ^2  \vert h_n(\xi) \vert^2 w_{0,0}(z, \vert k \vert , \xi) \d k \d \xi \\
 &= \int_{\mathbb{R}^{3+d}}  \vert f(\vert k '\vert)\vert ^2  \vert h(\xi') \vert^2 w_{0,0} \Big( z, \frac{\vert k' \vert}{n} + r_0 , \frac{\xi'}{n} + l_0 \Big) \d k' \d \xi' ,
\end{align*}
and the first limit follows from the dominated convergence theorem using that $w_{0,0}$ is continuous and that the functions $f$ and $h$ have compact supports. Similar arguments show that $\langle \psi^{(n)}\vert W_{1,1}(z) \psi^{(n)} \rangle$ converge to zero. 

Rotation invariance of $H(z)$ in $\xi-$space then clearly implies rotation invariance of the function $w_{0,0}$ in $\xi$-space. Indeed, since $H(z)$ is rotation invariant in $\xi$-space, 
\begin{equation*}
\langle \psi^{(n)}  \vert H(z)  \psi^{(n)} \rangle  = \langle \psi^{(n)} \vert  \mathcal{U}_{\mathcal{R}}^{*}  H(z)  \mathcal{U}_{\mathcal{R}} \psi^{(n)} \rangle \rightarrow w_{0,0}(z,r_0, \mathcal{R} l_0) + \mathcal{E}(z).
\end{equation*}
from which we deduce that  $w_{0,0}$ is invariant under rotations in $\xi$-space.

\subsection{Estimates on second derivatives of the resolvent}\label{app:rotation}
Let $j \in \mathbb{N} \cup \{ 0 \}$. Assuming that the estimates stated in Lemma \ref{lm:induc} hold at step $j-1$, we verify that the function 
$$(z,r,l) \mapsto R^{(j)}(z,r,l)=\frac{ \overline{\chi}^{2}_{ \rho_{j+1} }(r)}{w^{(j)}_{0,0}(z,r,l)   + \mathcal{E}^{(j)}(z) }$$
has first  derivatives in $r$ and second derivatives in $l_q$ that satisfy the bounds
\begin{align*}
\vert  \partial_{ r } R^{(j)}(z,r,l) \vert \leq C  \rho_{j+1}^{-2}, \qquad \vert  \partial_{ l_q } R^{(j)}(z,r,l) \vert \leq C  \rho_{j+1}^{-3/2}, \qquad  \vert  \partial^{2}_{l_q l_q'} R^{(j)}(z,r,l) \vert \leq C  \rho_{j+1}^{-2}
\end{align*}
for all  $z \in D^{(j+1)}$ and $(r,l) \in [3 \rho_{j+1}/4, \rho_{j+1}] \times \mathbb{R}^d$. We concentrate on the estimates concerning the partial derivatives with respect to $l_q$, the estimates concerning the partial derivative with respect to $r$ can be proven in the same way without using rotation invariance. We have that
\begin{align*}
  \partial_{ l_q } R^{(j)}(z,r,l)   & = -\frac{ \overline{\chi}^{2}_{ \rho_{j+1} }(r) \partial_{l_q} w_{0,0}^{(j)}(z,r,l)}{(w^{(j)}_{0,0}(z,r,l)   + \mathcal{E}^{(j)}(z))^2 },\\
   \partial_{ l_q l_q' } R^{(j)}(z,r,l)   & = -\frac{ \overline{\chi}^{2}_{ \rho_{j+1} }(r) \partial^{2}_{l_q l_{q'}} w_{0,0}^{(j)}(z,r,l)}{(w^{(j)}_{0,0}(z,r,l)   + \mathcal{E}^{(j)}(z))^2 } + 2 \overline{\chi}^{2}_{ \rho_{j+1} }(r) \frac{ (\partial_{l_q} w_{0,0}^{(j)}) (\partial_{l_{q'}} w_{0,0}^{(j)})(z,r,l) }{(w^{(j)}_{0,0}(z,r,l)   + \mathcal{E}^{(j)}(z))^3 }.
\end{align*}
Since $H^{(j)}$ is rotation invariant in $\xi$-space, the kernel $w^{(j)}_{0,0}$ is also rotation invariant in $\xi$-space by Appendix \ref{extrac}, and therefore we have that $(\partial_{l_q} w_{0,0}^{(j)})(z,r,(l_1,\dots,l_d)) = 0$ whenever one of the arguments $l_k$ vanishes. Hence, by \eqref{eq:AA4},
\begin{align*}
\vert \partial_{l_q} w_{0,0}^{(j)}(z,r,l) \vert  &= \vert  (\partial_{l_q} w_{0,0}^{(j)})(z,r,l) -( \partial_{l_q} w_{0,0}^{(j)})(z,r,(l_1,\dots,\underbrace{0}_q,\dots,l_d)) \vert \leq \frac{9 \vert l_q \vert }{4},\\
\vert w^{(j)}_{0,0}(z,r,l)   + \mathcal{E}^{(j)}(z) \vert & \geq \frac{3}{4}l^2 + \frac{1}{16} \rho_{j+1} \ge \frac18 | l_q | \rho_{j+1}^{\frac12}.
\end{align*}
The bounds follow by plugging these estimates in the expressions of the partial derivatives.

\subsection{Analyticity and analyticity conservation by the Feshbach-Schur map}\label{app:analyticity}
In this section we show that if the statement of Lemma \ref{lm:induc} holds at step $j$, then $H^{(j+1)}(z)- \d\Gamma(\xi)^{2}_{\mid \mathcal{H}^{(j+1)}}$ is analytic on $D^{(j+1)}$.  We first show that $W_{0,0}^{(j)}(z) - \d\Gamma(\xi)^{2}_{\mid \mathcal{H}^{(j)}}$ is analytic on $D^{(j)}$. This is a consequence of Appendix \ref{extrac}. Indeed,
\begin{align*}
 \langle \psi^{(n)}\vert   & [w_{0,0}^{(j)}(z,\d \Gamma( |k| ),\d\Gamma(\xi))- \d \Gamma(\xi)^2] \psi^{(n)} \rangle \\
 &= \int_{\mathbb{R}^{3+d}} \vert f(\vert k '\vert)\vert ^2  \vert h(\xi') \vert^2 [w_{0,0}^{(j)} \Big( z, \frac{\vert k' \vert}{n} + r_0 , \frac{\xi'}{n} + l_0 \Big) -  (\frac{\xi'}{n} + l_0)^2] \d k' \d \xi' ,
\end{align*}
converges to $w_{0,0}^{(j)} ( z,  r_0 , l_0 ) -   l_0^2$ as $n$ tends to infinity.  This convergence is \textit{uniform} in $z \in D^{(j)}$. Indeed, using the property $\mathcal{C}^{1,2}(\rho)$ of $w_{0,0}$, one gets  that 
\begin{align*}
w_{0,0}^{(j)}(z,r',l')- l'^2 & -w_{0,0}^{(j)}(z,r,l)+ l^2=\int_{r}^{r'} (\partial_{r''} w_{0,0}^{(j)})(z,r'',l') \d r''  \\
&+ \sum_{i=1}^{d} \int_{l_i}^{l'_i} [(\partial_{l''_i} w_{0,0}^{(j)})(z,r, (l_1,\dots,l_{i-1},l''_i,l'_{i+1},\dots,l'_d))-2 l''_i] \d l''_i
\end{align*}
Using \eqref{eq:AA2}, it follows that 
\begin{equation*}
\vert w_{0,0}^{(j)}(z,r',l')- l'^2  -w_{0,0}^{(j)}(z,r,l)+ l^2 \vert \leq C \|(r,l)-(r',l')\|,
\end{equation*}
for all $(r,l)$ and $(r',l')$, \textit{uniformly} in $z$.  This shows that the convergence is uniform in $z$, and, therefore, since a similar result holds for the convergence of  $\langle \psi^{(n)}\vert W^{(j)}_{1,1}(z) \psi^{(n)} \rangle$ towards zero, we deduce that $z \mapsto w_{0,0}^{(j)}(z,r,l)- l^2 + \mathcal{E}^{(j)}(z)$ is analytic on $D^{(j)}$. The analyticity of the bounded operator valued function $z \mapsto W_{0,0}^{(j)}(z)- \d\Gamma(\xi)^{2}_{\mid \mathcal{H}^{(j)}} + \mathcal{E}^{(j)}(z)$ then follows from Morera's theorem; see \cite{FaFrSc14_01} for similar calculations. One can verify that $z \mapsto W_{\geq 1}^{(j)}(z)$ is analytic as well. Using the definition of the Feshbach-Schur map in \eqref{eq:def-fesh}, it is then clear that $z \mapsto H^{(j+1)}(z)- \d\Gamma(\xi)^{2}_{\mid \mathcal{H}^{(j+1)}}$ is analytic on $D^{(j+1)}$.  Hence, analyticity is preserved at each iteration step, provided  that it holds for the first decimation step.

\section{Abstract results}\label{app:Mourre}

In this section we recall results from Mourre's conjugate operator theory in an abstract framework. Our main concern is the Fermi Golden Rule criterion established in \cite{FaMoSk11_01} in the context of the so-called singular Mourre's theory introduced by Georgescu, G{\'e}rard and M{\o}ller in \cite{GeGeMo04_01,GeGeMo04_02}. Here we only recall the main results that we used in Section \ref{section:Mourre}. For more details and explanations, the reader is invited to consult \cite{GeGeMo04_01} and \cite{FaMoSk11_01}.

We consider a complex, separable Hilbert space $\mathcal{H}$ and two self-adjoint operators $H$ and $M$ on $\mathcal{H}$, with $M \ge 0$. We consider in addition a symmetric operator $R$ relatively $H$-bounded, and we define $H' := M+R$ on $\mathcal{D}(M) \cap \mathcal{D}(H)$. We set
$
\mathcal{G} = \mathcal{D}( M^{\frac{1}{2}} )\cap \mathcal{D}( |H|^{\frac{1}{2}} ),
$
equipped with the norm of the intersection topology:
\begin{align*}
\| \varphi \|^2_{ \mathcal{G} } :=  \big \| M^{\frac{1}{2}} \varphi \big \|^2_{ \mathcal{H} } +\big \| |H|^{\frac{1}{2}} \varphi \big \|^2_{ \mathcal{H} } + \| \varphi \|^2_{ \mathcal{H} }. 
\end{align*}
Setting
\begin{align*}
\| \varphi \|_{ \mathcal{G}^* } :=  \big \| ( M + | H | + \mathds{1} )^{-\frac{1}{2}} \varphi \big \|_{ \mathcal{H} } ,
\end{align*}
one verifies that $\mathcal{G}^*$, the dual space of $\mathcal{G}$, identifies with the completion of $\mathcal{H}$ with respect to $\| \cdot \|_{ \mathcal{G}^* }$. The (possibly unbounded) operators $H$ and $M$ on $\mathcal{H}$ may then be seen as bounded operators from $\mathcal{G}$ to $\mathcal{G}^*$. The corresponding operators will be denoted by the same symbols.

In our setting the ``conjugate operator'' $A$ is closed and maximal symmetric on $\mathcal{H}$, with $\dim \mathrm{Ker} (A^* - \ii) = 0$. This implies that $A$ is the generator of a $\mathrm{C}_0$-semigroup of isometries $\{ W_t \}_{t \ge 0}$, in the sense that
\begin{equation*}
\mathcal{D} (A) = \big \{ \varphi \in \mathcal{H} ,\lim_{t \to 0^+} ( \ii t )^{-1}
( W_t \varphi - \varphi ) \text{ exists} \big \}, \quad  A \varphi = \lim_{t \to 0^+} ( \ii t )^{-1}
( W_t \varphi - \varphi ).
\end{equation*} 

Let $\{ W_{1,t} \}$ be a $\mathrm{C}_0$-semigroup on a Hilbert space $\mathcal{H}_1$ with generator $A_1$, and  $\{W_{2,t} \}$ be a $\mathrm{C}_0$-semigroup on a Hilbert space $\mathcal{H}_2$ with generator $A_2$. We recall that the class $\mathrm{C}^1(A_1 ; A_2)$ is defined as the set of bounded operators $B \in \mathcal{B}( \mathcal{H}_1 ; \mathcal{H}_2) $ satisfying
\begin{equation*}
\| W_{2,t} B - B W_{1,t} \|_{\mathcal{B}( \mathcal{H}_1 ; \mathcal{H}_2 )} \le C t , \quad 0 \le t \le 1,
\end{equation*}
for some positive constant $C$. One can verify (see \cite{GeGeMo04_01}) that $B$ is of class $\mathrm{C}^1( A_1 ; A_2 )$ if and only if the quadratic form $_2[B,\ii A]_1$ defined on $\mathcal{D}(A_2^*) \times \mathcal{D}(A_1)$ by 
\begin{equation*}
\langle \varphi,  {_2}[B,\ii A]_1 \psi \rangle = \ii \langle B^* \varphi ,  A_1 \psi \rangle - \ii \langle A_2^* \varphi , B \psi \rangle,
\end{equation*}
extends to a bounded quadratic form on $\mathcal{H}_2 \times \mathcal{H}_1$. Hence $_2[B,\ii A]_1$ is associated to a bounded operator $[B,\ii A]^0 \in \mathcal{B}( \mathcal{H}_1 ; \mathcal{H}_2 )$. We then have that
\begin{equation*}
[B,\ii A]^0  \psi = \lim_{t\to0^+}  t^{-1} [ B W_{1,t} - W_{2,t} B ] \psi,
\end{equation*}
for all $\psi \in \mathcal{H}_1$ (see \cite{GeGeMo04_01}). Similarly, the class $\mathrm{C}^2( A_1 ; A_2 )$ is defined as the set of operators $B \in \mathrm{C}^1( A_1 ; A_2 )$ such that $[B,\ii A]^0 \in \mathrm{C}^1( A_1 ; A_2 )$.

We remark that, in the particular case where $\mathcal{H}_1 = \mathcal{H}_2 = \mathcal{H}$, and if $A$ is a self-adjoint operator on $\mathcal{H}$, the well-known class $\mathrm{C}^1(A)$ corresponds to the set of self-adjoint operators $B$ on $\mathcal{H}$ such that there exists $z\in\mathbb{C} \setminus \mathbb{R}$ such that $(B-z)^{-1}\in\mathrm{C}^1( A ; A )$.

With these notations we can state the main hypotheses of \cite{GeGeMo04_01} and \cite{FaMoSk11_01}.
\begin{itemize}
\item[(1)] $H \in \mathrm{C}^1( M )$ and $[ H , \ii M ]^0$ is relatively $H$-bounded.
\item[(2)] There exists an interval $I \subset \mathbb{R}$ such that, for all $\lambda \in I$, there exist constants $\mathrm{c}_0>0$, $C_1 \in \mathbb{R}$ and a function $\chi_\lambda \in \mathrm{C}_0^\infty( \mathbb{R})$ with $0\le \chi_\lambda \le 1$ and $\chi_\lambda = 1$ in a neighborhood of $\lambda$, such that
\begin{equation*}
 H' \ge \mathrm{c}_0 \mathds{1} - C_1 \chi_\lambda^\perp(H)^2 \langle H \rangle - K,
\end{equation*}
in the sense of quadratic forms on $\mathcal{D}(H) \cap \mathcal{D}( M )$, where $\chi_\lambda^\perp(H) = \mathds{1} - \chi_\lambda(H)$, $\langle H \rangle = ( 1 + H^2 )^{1/2}$, and $K$ is a compact operator on $\mathcal{H}$.
\item[(3)] For all $t > 0$, $W_t$ and $W_t^*$ preserve $\mathcal{G}$, and we have that
\begin{equation*}
\sup_{0<t<1} \| W_t^\sharp \varphi \|_\mathcal{G} <  \infty, 
\end{equation*}
for all $\varphi \in \mathcal{G}$, where $W_t^\sharp$ stands for $W_t$ or $W_t^*$. The generator of the $\mathrm{C}_0$-semigroup $W_t |_\mathcal{G}$ is denoted by $A_\mathcal{G}$ and the generator of the $\mathrm{C}_0$-semigroup given as  the extension of $W_t$ to $\mathcal{G}^*$ is denoted by $A_{\mathcal{G}^*}$.
\item[(4)] $H\in \mathrm{C}^2(
  A_\mathcal{G};A_{\mathcal{G}^*})$ and, for all $\varphi \in \mathcal{D}(H) \cap \mathcal{D}( M )$, 
$
  H' \varphi = [ H , \ii A ]^0 \varphi.
$
\end{itemize}
Under these hypotheses, the following theorem is proven in \cite{GeGeMo04_01}.
\begin{theorem} \label{thm:LAP} 
Suppose that Hypotheses (1)--(4) above hold. Let $J \subset I$ be a compact interval such that $\sigma_{\mathrm{pp}}(H) \cap J=  \emptyset$ and let 
$
S=\{ z \in \mathbb{C} , \mathrm{Re}(z) \in J , 0 < | \mathrm{Im}(z) | \leq 1\}.
$
Then 
\begin{equation*} 
    \sup_{z\in S} \| \langle A \rangle^{-s} (H-z)^{-1} \langle A \rangle^{-s}\|<\infty ,
\end{equation*}
for any $1/2<s\le 1$, with $\langle A \rangle = ( 1 + A^* A )^{1/2}$. Moreover the function $z\to   \langle A \rangle^{-s} (H-z)^{-1} \langle A \rangle^{-s} $ is uniformly H\"older continuous of order $s-1/2$ in $S$. In particular, the spectrum of $H$ in $J$ is purely absolutely continuous, and the following local decay property holds:
\begin{equation*}
\big \| \langle A \rangle^{-s} e^{ - \ii t H } \chi( H ) \langle A \rangle^{-s} \big \| \lesssim t^{ - s + \frac12 } , \quad t \to \infty ,
\end{equation*}
for any $1/2 < s \le 1$ and $\chi \in \mathrm{C}_0^\infty( J ; \mathbb{R} )$.
\end{theorem}

\bibliographystyle{amsplain}
\providecommand{\bysame}{\leavevmode\hbox to3em{\hrulefill}\thinspace}
\providecommand{\MR}{\relax\ifhmode\unskip\space\fi MR }
\providecommand{\MRhref}[2]{%
  \href{http://www.ams.org/mathscinet-getitem?mr=#1}{#2}
}
\providecommand{\href}[2]{#2}

\end{document}

%% file: modele.tex
\begin{figure}[here]
\begin{center}
\begin{tikzpicture}

      \fill[color=gray!25]  (7.2,0) ellipse (0.15 and 0.32);
      \draw[decorate,decoration={snake, segment length=0.5mm,amplitude=0.3mm}]   (7.2,0) ellipse (0.15 and 0.32);

    \fill[color=gray!25]  (5.2,0) ellipse (0.25 and 0.55);
      \draw[decorate,decoration={snake, segment length=0.5mm,amplitude=0.3mm}]  (5.2,0)  ellipse (0.25 and 0.55);

          \fill[color=gray!25]  (3.2,0) ellipse (0.35 and 0.75);
      \draw[decorate,decoration={snake, segment length=0.5mm,amplitude=0.3mm}]  (3.2,0) ellipse (0.35 and 0.75);

          \fill[color=gray!25]  (1.2,0) ellipse (0.45 and 0.95);
      \draw[decorate,decoration={snake, segment length=0.5mm,amplitude=0.3mm}]  (1.2,0) ellipse (0.45 and 0.95);

  \draw[->](7.2,0.4)--(7.2,0.57); 
     \draw[->](7.2,-0.35)--(7.2,-0.52);         
 \draw[->](7.4,0.1)--(7.6,0.2); 
     \draw[->](7,-0.02)--(6.8,-0.1);

     \draw[->](5.2,0.6)--(5.2,0.77); 
     \draw[->](5.2,-0.55)--(5.2,-0.72);         
 \draw[->](5.5,0.12)--(5.7,0.22); 
     \draw[->](4.9,-0.02)--(4.7,-0.1);

   \draw[->](3.2,0.8)--(3.2,0.97); 
     \draw[->](3.2,-0.75)--(3.2,-0.92);         
 \draw[->](3.6,0.15)--(3.8,0.25); 
     \draw[->](2.8,-0.05)--(2.6,-0.13);

\draw[->](1.2,1)--(1.2,1.17); 
     \draw[->](1.2,-0.95)--(1.2,-1.12);         
 \draw[->](1.7,0.18)--(1.9,0.28); 
     \draw[->](0.7,-0.07)--(0.5,-0.15);

  \draw[->](0,0)--(12,0); 
   \draw (12,0) node[below] { $x$};   
    \fill[ color=black] (10,0) circle (0.08) ;
     \draw (10,0) node[below] { $q(t)$};   
  
   \draw (0.5,-1.5) node[right] { \footnotesize $\mathbb{R}^3$};   
    \draw (2.5,-1.5) node[right] { \footnotesize  $\mathbb{R}^3$};   
    \draw (4.5,-1.5) node[right] {  \footnotesize  $\mathbb{R}^3$};    
    \draw (6.5,-1.5) node[right] { \footnotesize  $\mathbb{R}^3$};

   \draw[-,dashed](0.5,-2)--(2,-1);  
     \draw[-,dashed](0.5,1.5)--(2,2.5);  
     \draw[-,dashed](0.5,-2)--(0.5,1.5);  
     \draw[-,dashed](2,2.5)--(2,-1);

     \draw[-,dashed](2.5,-2)--(4,-1);  
     \draw[-,dashed](2.5,1.5)--(4,2.5);  
     \draw[-,dashed](2.5,-2)--(2.5,1.5);  
     \draw[-,dashed](4,2.5)--(4,-1);

        \draw[-,dashed](4.5,-2)--(6,-1);  
     \draw[-,dashed](4.5,1.5)--(6,2.5);  
     \draw[-,dashed](4.5,-2)--(4.5,1.5);  
     \draw[-,dashed](6,2.5)--(6,-1);

        \draw[-,dashed](6.5,-2)--(8,-1);  
     \draw[-,dashed](6.5,1.5)--(8,2.5);  
     \draw[-,dashed](6.5,-2)--(6.5,1.5);  
     \draw[-,dashed](8,2.5)--(8,-1);

  \draw[-,dotted](10,0)--(1,1);
    \draw[-,dotted](10,0)--(1,-1);

\end{tikzpicture}
\end{center}
\caption{ \small  The particle moves through vibrating membranes (real scalar fields) and looses its energy in these independent membranes. This energy cannot be recovered by the particle if the phase velocity $c$ of the vibrating waves is sufficiently large. }

\end{figure}

%% file: spectre.tex
\begin{figure}[H]
\begin{center}
\begin{tikzpicture}

 \fill [gray!30] (0,0) rectangle (2,4);
 \fill [gray!30] (9,-0.2) rectangle (11,4);
 \draw[->](0,0)--(3,0);
  \draw[->](9,0)--(12,0); 
  
   \draw[->](0,0)--(0,5);
  \draw[->](9,-0.2)--(9,5);

 \draw [domain=0:2]  plot (\x,\x* \x);
 
  \draw [domain=9:11,dotted]  plot (\x,\x*\x-18 *\x +81);
  
   \fill[ color=black] (0,0) circle (0.05) ;
   
      \fill[ color=black] (9,-0.2) circle (0.05) ;
  
    \draw (1,-0.1) node[below] {\small$ g=0$};
 \draw (10,-0.1) node[below] {\small$ g \neq 0$};

 \draw (11.8,0) node[below] {\small$ \vert P\vert$};
 \draw (2.8,0) node[below] {\small$ \vert P\vert$};

   \draw (0,5) node[left] {\small $ \sigma(H(P))$};
 \draw (9,5) node[left] {\small $ \sigma(H(P))$};

\end{tikzpicture}
\end{center}
\caption{ \small Spectrum of $H(P)$ if $\mu > 1/2$ and $g=0$ (left) or $g \neq 0$ (right). The grey shaded region  represents the absolutely continuous part of the spectrum. If $g=0$, the infimum of the spectrum is equal to $0$ for all $P$ and it is an eigenvalue if and only if $P=0$;  for all $P \neq 0$, $P^2$ is an eigenvalue of $H(P)$ embedded in the continuous spectrum. If $g\neq 0$, the infimum of the spectrum, $E_g$, is negative for all $P$, and independent of $P$. It is an eigenvalue if $P=0$; for $P \neq 0$ (more precisely $|P| \in ( \nu_1 , \nu_2 )$), the spectrum is purely absolutely continuous. }

\end{figure}